\documentclass[twocolumn]{aastex631}
\usepackage[figuresright]{rotating}
\usepackage{longtable}

\begin{document}

\title{VLBI Imaging of Parsec-scale Radio Structures in Nearby Low-luminosity AGN}

\correspondingauthor{Xiaopeng Cheng}
\email{xcheng0808@gmail.com}
\correspondingauthor{Tao An}
\email{antao@shao.ac.cn}

\author[0000-0003-4407-9868]{Xiaopeng Cheng}
\affiliation{Korea Astronomy and Space Science Institute, 776 Daedeok-daero, Yuseong-gu, Daejeon 34055, Korea}
\author[0000-0003-4341-0029]{Tao An}
\affiliation{Shanghai Astronomical Observatory, CAS, 80 Nandan Road, Shanghai 200030, China}
\affiliation{Xinjiang Astronomical Observatory, Chinese Academy of Sciences, Urumqi 830011, China}
\affiliation{University of Chinese Academy of Sciences, 19A Yuquanlu, Beijing 100049, China}
\author[0000-0003-3389-6838]{Willem A. Baan}
\affiliation{Xinjiang Astronomical Observatory, Chinese Academy of Sciences, Urumqi 830011, China}
\affiliation{Netherlands Institute for Radio Astronomy ASTRON, NL-7991 PD Dwingeloo, the Netherlands}
\author[0000-0002-1824-0411]{Ranieri D. Baldi}
\affiliation{INAF - Istituto di Radioastronomia, Via P. Gobetti 101, I-40129 Bologna, Italy}
\author[0000-0001-7361-0246]{David R. A. Williams-Baldwin}
\affiliation{Jodrell Bank Centre for Astrophysics, School of Physics and Astronomy, The University of Manchester, Manchester, M13 9PL, UK}
\author[0000-0002-4148-8378]{Bong Won Sohn}
\affiliation{Korea Astronomy and Space Science Institute, 776 Daedeok-daero, Yuseong-gu, Daejeon 34055, Korea}
\affiliation{University of Science and Technology, Gajeong-ro 217, Yuseong-gu, Daejeon 34113, Republic of Korea}
\author[0000-0002-5544-2354]{Robert Beswick}
\affiliation{Jodrell Bank Centre for Astrophysics, School of Physics and Astronomy, The University of Manchester, Manchester, M13 9PL, UK}
\author{Ian M. McHardy}
\affiliation{School of Physics and Astronomy, University of Southampton, Southampton, SO17 1BJ, UK}



\begin{abstract}

We report the results of high-resolution 5 GHz Very Long Baseline Array and European VLBI Network observations of 36 nearby galaxies, an extension of the Legacy e-MERLIN Multi-band Imaging of Nearby Galaxies (LeMMINGs) survey. Our sample includes 21 low ionization nuclear emission regions (LINERs), 4 Seyferts, 3 absorption line galaxies (ALGs), and 8 HII galaxies. We achieved an unprecedented detection rate, successfully imaging 23 out of 36 sources with a detection threshold of $\sim$20 $\mu$Jy beam$^{-1}$. The radio sizes are typically of $\leq$ 5 pc. Core identification was achieved in 16 sources, while 7 others were identified as core candidates. Radio luminosities of the sample range from 10$\rm ^{34}$ to 10$\rm ^{38}$ erg s$^{-1}$. Our analysis reveals a predominance of compact core structures, with ten sources exhibiting a one-sided core jet morphology and NGC 2146 exhibiting a rare two-sided jet structure. The study advances our understanding of the compactness of radio sources at various scales, indicating a core-dominated nature in all but one galaxy NGC2655. We find moderate to strong correlations between radio luminosity, black hole mass, optical [O III] line luminosity, and hard X-ray luminosity, suggesting a common active galactic nucleus (AGN) core origin. These results provide new insights into the fundamental plane of black hole activity and support the role of the synchrotron process in Low-luminosity AGN (LLAGN) radio emission.

\end{abstract}


\section{Introduction} \label{sec:intro}

Low-luminosity active galactic nuclei \citep[LLAGNs, bolometric luminosity L$\rm _{bol}$ $\leq$ 10$\rm ^{42}$ erg s $\rm ^{-1}$; e.g.][]{2008ARA&A..46..475H} are found at the center of half of the nearby galaxies and encompass low-luminosity Seyferts, low ionization nuclear emission regions \citep[LINERs;]{1980A&A....87..152H}, transition nuclei (spectral properties intermediate between those of LINERs and HII regions), HII galaxies, and absorption-line galaxies (ALGs, no significant optical emission lines) \citep{1997ApJS..112..315H}. The spectral energy distribution (SED) of LLAGNs is very prominent in the radio and infrared bands and lacks the canonical optical-UV ``big blue bump" \citep{1996PASP..108..637H,1999ApJ...516..672H,1999ApJ...525L..89Q}. The power source of LLAGNs is still a matter of debate \citep{2019NatAs...3..387P}. Hunting for LLAGNs can help to probe the cosmological evolutionary models for SMBHs and galaxies and complete the census of local SMBH activity at the low end of the luminosity function \citep{2004ApJ...600..580G,2017MNRAS.468.4782B}. Understanding the properties of LLAGNs is thus key for comprehensively testing models for the coevolution of SMBHs and galaxies across cosmic time.

The main physical processes that dominate the radio emission in LLAGNs are thermal radiation from H II regions and non-thermal synchrotron emission associated with a powerful starburst, an AGN, or both \citep{2019NatAs...3..387P}. Thermal radiation from H II regions, due to star formation (SF), and non-thermal radiation from supernova remnants (SNRs)  show diffuse, low surface brightness (T$\rm _{b} < 10 ^{4} - 10^{5}$ K) radio morphology \citep{1992ARA&A..30..575C,2006MNRAS.369.1221B,2019NatAs...3..387P,2021A&ARv..29....2P,2022MNRAS.515.5758M}. In contrast, AGN often exhibit compact, bright non-thermal radio cores (T$\rm _{b} > 10 ^{7}$ K), sometimes associated with relativistic jets. However,  non-thermal synchrotron radiation with brightness temperature between 10$\rm ^{5}$ and 10$\rm ^{7}$ K can be explained by different types of sources, e.g., a young radio supernova (RSN), an AGN, or a combination of AGN and compact transient source \citep{2009A&A...499L..17B}. A key way to discriminate between these possibilities is to probe the size of the radio emission on sub-parsec scales.
For AGN-driven radio emission, if the non-thermal synchrotron radiation from LLAGN arises from a radiatively inefficient accretion flow \citep[RIAF;][]{1994ApJ...428L..13N,2014ARA&A..52..529Y}, their radio sources should be compact, only tens of Schwarzschild radii in size \citep[e.g., Sgr A*][]{1995Natur.374..623N,2022ApJ...930L..12E}. On the other hand, if the radio emission is dominated by a relativistic jet, it would show a core-jet morphology similar to that seen in more powerful radio-loud AGN \citep[e.g.,][]{2007A&A...464..553K,2014ApJ...787...62M}.

Several low-resolution (arcsec resolution) radio surveys of nearby galaxies have been carried out with the Very Large Array (VLA) at 1.4 - 15 GHz \citep[e.g.,][] {1989MNRAS.240..591S,1989ApJ...343..659U,2000ApJS..129...93F,2002ApJS..142..223F,2000ApJ...542..186N,2002A&A...392...53N,2001ApJS..133...77H,2001ApJ...558..561U,2006AJ....132..546G,2018A&A...616A.152S} and Multi-Element Radio-Linked Interferometer Network (MERLIN) at 5 GHz \citep{2006A&A...451...71F}. These surveys targeted a complete magnitude-limited (B$\rm _{T} \leq$ 12.5 mag) sample of bright nearby galaxies from the Palomar survey \citep[][]{1995ApJS...98..477H,1997ApJS..112..315H} and found compact cores in over half of low-luminosity Seyfert galaxies and LINERs \citep{2002A&A...392...53N}, and in more than 25 per cent of transition objects \citep{2002ApJS..142..223F,2018A&A...616A.152S}. 
However, the low resolution of these surveys made it difficult to pinpoint the origin of the radio emission and distinguish between diffuse thermal emission from star formation or supernova remnants and compact non-thermal emission from an AGN.

Very Long Baseline Interferometry (VLBI) provides the highest angular resolution,  revealing the fine-scale structures of compact radio sources on milliarcsecond (mas) scales. This spatial resolution allows VLBI observations to localize non-thermal radio emission to the nucleus on parsec and sub-parsec scales. Several cm-wavelength VLBI observations of LLAGNs have been conducted \citep[e.g.,][]{2000ApJ...542..197F,2001ApJ...562L.133U,2002A&A...392...53N,2004ApJ...603...42A,2004A&A...418..429F,2009ApJ...706L.260G,2013MNRAS.432.1138P}, imaging more than 60 active galaxies on parsec-scales to sub-parsec scales. These studies targeted relatively radio-bright galaxies with flat/inverted spectra. As a result, most of these sources show compact and high-brightness temperature cores in the VLA or e-MERLIN images \citep{2002A&A...392...53N}. Only a few HII galaxies were detected in VLBI, suggesting that their arcsecond-scale radio cores seen by VLA/e-MERLIN likely arise from star formation or supernova remnants rather than AGN \citep{2021MNRAS.500.4749B}. An unbiased VLBI survey of a flux-limited LLAGN sample is still needed to elucidate the nature of their nuclear radio emission.

More recently, \citet{2018MNRAS.476.3478B,2021MNRAS.500.4749B} performed a large high-resolution ($<$ 0.2 arcsec) and high sensitivity ($\sim$70 $\mu$Jy beam$\rm ^{-1}$) 1.5 GHz radio survey of nearby 280 active and quiescent Palomar galaxies ($\delta > 20^{\circ}$ and distance $<$ 100 Mpc) with the enhanced Multi-Element Radio-Linked Interferometer Network (e-MERLIN) array as part of the Legacy e-MERLIN Multiband Imaging of Nearby Galaxies Survey (LeMMINGs). They detected radio emissions from 125 of the 280 galaxies (a detection rate of 44.6\%) with flux densities down to 0.25 mJy. For 106 of the 125 sources, the radio cores were identified with typical sizes $\leq$ 100 pc. Conversely for the remaining 19 sources, the radio emission was not associated with the central optical nucleus. Thirty-one sources show clear jet structures, known as jetted galaxies. LINERs show elongated core-brightened radio structures while Seyferts reveal the highest fraction of symmetric morphologies. It is interesting that 47 of the 140 HII galaxies were detected and seven galaxies show core-jet structures similar to LINERs and Seyferts \citep{2021MNRAS.500.4749B}, suggesting the presence of an active nucleus \citep{1993ApJ...405L...9L}. C-band e-MERLIN observations (50 mas of angular resolution) the whole LeMMINGs sample have already been completed (Williams-Baldwin et al., 2025, in prep.) and the data are used in this work.
However, higher resolution observations on milliarcsecond scales with VLBI are desirable to further investigate the nature of the detected radio emission.

To comprehensively study the radio emission in LLAGNs, a sub-mJy VLBI survey of the Palomar sample (inactive galaxies) is needed, including both active and inactive galaxies.
We have selected 36 galaxies from the LeMMINGs sample \citep{2018MNRAS.476.3478B,2021MNRAS.500.4749B} for follow-up observations with the Very Long Baseline Array (VLBA) and European VLBI Network (EVN) at 5 GHz. This flux-limited sample allows us to investigate the parsec-scale radio properties of LLAGNs down to sub-mJy levels and provide new constraints on the origin of their nuclear radio emission.

In this paper, we present the first high-resolution ($\sim$1 mas) and high sensitivity ($\sim$20 $\mu$Jy beam$\rm ^{-1}$) phase referencing VLBI results of 36 nearby LLAGNs with the VLBA and EVN at 5 GHz and discuss the physical nature of the radio emission in sub-parsec scales. We describe the sample selection, VLBI observations, and the data reduction in Section \ref{sec:obs} and present the imaging results in Section \ref{sec:result}. Section \ref{sec:disscussion} contains the correlations between radio luminosity and optical luminosity, X-ray nuclear luminosity, and black hole (BH) mass, and a discussion of the possible origins of radio emission in these sources. Throughout this paper, the standard $\Lambda$CDM cosmological model with H$_{0}$ = 71~km\,s$^{-1}$\,Mpc$^{-1}$, $\Omega_{\rm M} = 0.27$, and $\Omega_{\Lambda} = 0.73$ is adopted.

\section{Sample Selection} \label{sec:sample}

We have selected a sample of 36 galaxies from the recent LeMMINGs survey \citep{2018MNRAS.476.3478B,2021MNRAS.500.4749B} for follow-up VLBI observations, see Table \ref{tab:information}. The parent LeMMINGs sample consists of 280 nearby galaxies from the Palomar survey with J2000 declination $\rm \delta$ $>$ $20^{\circ}$ and  B$\rm _{T}$  $\leq$ 12.5.
Throughout this survey, e-MERLIN detected radio emission in 125 targets (typically $<$ a few mJy) on scales of a few tens of parsecs to kpc scales. Cross-matching of the LeMMINGs sample with the existing VLBI databases identified 29 detected sources: (i) 16 sources show a typical core-jet structure with a projected linear size between 0.1 to 20 pc; (ii) 13 sources show an unresolved core due to the limited resolution and sensitivity ($\sim 80 \mu$Jy/beam) \citep[e.g.,][]{2004A&A...418..429F,2005A&A...435..521N}. 
This indicates that half of these sources have a parsec-scale fine structure revealed by the flux-density-limited VLBI survey ($\rm S_{VLBI} >$ 0.45 mJy).

Among the remaining 96 sources in the LeMMINGs sample without previous VLBI observations, 48 sources have flux density higher than 1 mJy, including 17 weak sources (5 HII galaxies, 2 Seyferts, and 10 LINERs) with only an unresolved radio core and 31 bright sources (11 LINERs, 2 Seyferts, 3 ALGs, and 15 HII galaxies) with jetted morphologies. 
We finally selected 36 sources, including 21 LINERs, 4 Seyferts, 3 ALGs with total flux density $>$ 1 mJy, and 8 HII galaxies with total flux density $>$ 3 mJy. 
18 galaxies with flux $>$3 mJy were observed with the VLBA at 5 GHz during Feb-Mar 2022. The other 18 were observed with the EVN at 5 GHz from Jan-Dec 2022 (see Table \ref{tab:obs}). This flux-limited VLBI sample allows us to probe the sub-parsec scale radio properties of LLAGN.

\section{Observations and data reduction} \label{sec:obs}

\subsection{Observations} 
\label{sec:observation}

Observations of the LLAGN sample were conducted for 20 hours with the VLBA and 45 hours with the EVN.
Based on the same motivation of deep imaging, the EVN and VLBA observations had similar schedules. The observations were made at the C band (5 GHz) in 2022. To ensure the success of the observations and optimize the $uv$ coverage of each source, we split the VLBA and EVN samples into 4 groups, respectively. Due to the weakness of the targets, we used phase-referencing observations with a cycle time of 3.5 mins: 40 s for the phase calibrator and 2.8 min for the target. The angular separations between the phase calibrator and target are smaller than 2 deg, see columns 9 and 10 of Table \ref{tab:information}. Each source was observed for 1 to 2.5 h with an effective on-source time of 0.5 to 2 h approximately. The correlation positions for the targets, obtained from the results of e-MERLIN observations \citep{2018MNRAS.476.3478B,2021MNRAS.500.4749B}, are listed in columns 3 and 4 of Table \ref{tab:information}.

The EVN observations (project codes of EC082A to EC082F; PI: X.-P. Cheng) were carried out at 5 GHz during the e-EVN sessions from 2022 January 18 to 2022 December 6. Nine to Fourteen EVN antennas were used with a 2-bit sampling at an aggregate data rate of 2048 Mbps (2 dual polarization, 8 consecutive 32 MHz channels). Due to the limited network bandwidths, the Russian stations (SV, Zc, and BD) used the 1024 Mbps experiment setup: 2 dual-polarization and 4 consecutive 32 MHz channels. There are 5 e-MERLIN antennas participating in the observations with a 512 Mbps experiment setup: 2 dual polarization and 2 consecutive 32 MHz channels. The detailed information is summarized in Table \ref{tab:obs}. The data correlation was done with the EVN software correlator SFXC \citep{2015ExA....39..259K} at JIVE (Joint Institute for VLBI ERIC), Dwingeloo, the Netherlands, with 0.5-MHz frequency resolution and 1-s integration time.

The VLBA observations (project codes of BA152A to BA152D; PI: Tao An) were carried out from 2022 March 8 to March 16. All 10 VLBA antennas were used with a 2-bit sampling at an aggregate data rate of 4096 Mbps (2 dual polarization, 4 consecutive 128 MHz channels). The detailed information is summarized in Table \ref{tab:obs}.
The data were correlated using the DiFX software correlator \citep{2007PASP..119..318D,2011PASP..123..275D} at the Array Operations Center in Socorro (New Mexico, U.S.) with 256 frequency points per channel and 1-s integration time.

\subsection{Data Reduction}
\label{data reduction}

The data were calibrated using the National Radio Astronomy Observatory (NRAO) Astronomical Image Processing System \citep[AIPS version 31DEC19,][]{2003ASSL..285..109G} software package, following the general EVN and VLBA data calibration strategy. We removed the visibility data of side channels because of the low amplitude while loading the raw data into AIPS, and flagged out off-source data and some bad data (typically due to the inclement weather). Before proceeding further, we ran the task ACCOR to remove the sampler bias. A priori amplitude calibration was carried out with the system temperatures and the antenna gain curves measured at each station during the observations. The dispersive delays caused by the ionosphere were corrected according to the maps of total electron content provided by the Global Positioning System (GPS) satellite observations. Phase errors due to the antenna parallactic angle variations were also removed. The stations Pie Town (PT) and Effelsberg (EF) were chosen as the reference antennas during the VLBA and EVN data calibration process. The instrumental single-band delays and phase offsets were corrected using a 2-min scan of the calibrator's data. After these calibrations, we ran the global fringe fitting on the phase-referencing calibrators with a 40 s solution interval and a point source model \citep{1995ASPC...82..189C} by combining all the channels and applied the solutions to targets by interpolation. The bandpass calibrations were performed in the final step.

We first imaged the phase-referencing sources. The imaging procedure was performed through a number of iterations of model fitting with a group of point-source models, and the self-calibration in DIFMAP \citep{1994BAAS...26..987S}. We reran the fringe fitting and amplitude calibration in AIPS with the source model made in DIFMAP and transferred the solutions to the data of the targets by linear interpolation. Finally, we obtained the calibrated visibility data of the targets.

We imaged all the faint targets without any self-calibration in DIFMAP. We first checked the data using natural weighting within the area of at least 1000 $\times$ 1000 mas around the target position from the e-MERLIN result and set a firm detection of the source with a signal-to-ratio (SNR) larger than 7$\sigma$. For the sources with SNR between 5 and 7, we checked the SNR in lower-resolution data by using the UVTAPER parameter, ranging between 25 and 80 M$\lambda$, to confirm or refute a real detection. This parameter specifies the widths of the Gaussian function in U and V directions to weigh down the long-baseline data points. NGC 3735 is resolved into 3 
weak components ($\sim$ 5 $\sigma$) in the full-resolution image. The combination ($\leq$ 7$\sigma$) of the three components is detected in the low-resolution map with a size of 14.29 mas (2.6 pc) and flux density of 0.5 mJy.
For the remaining 13 undetected sources, the SNRs are below 5 in both the full-resolution and lower-resolution data, we further searched for potential radio emission at the positions reported by the Gaia astrometry \citep{2020yCat.1350....0G} and Panoramic Survey Telescope and Rapid Response System \citep[Pan-STARRS1;][]{2016arXiv161205560C} but did not detect any of these sources.

\subsection{Model Fitting}

The visibility data of the detected sources were fitted with circular Gaussian components in DIFMAP using model fitting to quantitatively describe the emission structure. We derived the positions from fitting Gaussian models in the image plane in AIPS. 
The derived total flux densities range between 0.2 and 4.1 mJy. The model-fitting parameters are reported in Table \ref{tab:image}.

We used the results of the model fitting to calculate the brightness temperatures of all the components. The  source’s rest frame brightness temperature T$_{\rm b}$ of the emission region can be calculated using the following formula:
\begin{equation}
  T_{\rm b} = 1.22\times10^{\rm 12}\frac{S}{ \nu^{2}d^{2}}(1+z),
\end{equation}
where $S$ is the flux density of the component in Jy, $z$ is the redshift, $\nu$ is the observing frequency in GHz, and $d$ is the fitted Gaussian size (FWHM) of the component in mas. For unresolved sources, the estimated brightness temperatures should be considered lower limits. 
The estimated core brightness temperatures, listed in column 10 of Table \ref{tab:image}, are in the range $\sim$ $10\rm ^{5} - 10^{9}$ K.

\section{Results} \label{sec:result}

\subsection{Detection Rates}

Of the 36 sources observed, twenty-three sources were detected and imaged with VLBI for the first time, resulting in an overall detection rate of 64\%. The radio positions for the detected sources are limited by the positional accuracy of the phase calibrators, which is typically 0.1--3.5 mas, and by the accuracy of the Gaussian fit to the source brightness distribution, which depends on the SNR of the source detection and is typically 0.1--0.6 mas. The overall accuracy is generally $<1$ mas, except for NGC 2985 which has a larger uncertainty of about 4 mas because of the 3.5 mas position error of the phase calibrator.

The detection rates are 14/21 ($\sim$67\%) for LINERs, 3/4 (75\%) for Seyferts,  3/8 (37.5\%) for HII galaxies, and 3/3 (100\%) for  ALGs. The higher detection rate for LINERs and Seyferts than that in HII galaxies suggests weaker AGN contributions in HII galaxies \citep{2018MNRAS.476.3478B}. All three ALGs are detected in our observations, suggesting that their radio emission is clearly AGN-powered. 
We note that the detection rates for the LINERs and Seyferts are consistent with the LeMMINGs survey detection rates, but the detection rates for the H II and ALGs are higher than the LeMMINGs survey detection rates \citep{2018MNRAS.476.3478B,2021MNRAS.500.4749B}, since we reach deeper sensitivity limits of $\geq$ 5$\sigma$ level ($\approx$100$\mu$Jy). The non-detection of 13 sources indicates that the radio emission detected by the VLA and e-MERLIN is not from a compact source at the mas scale resolution. Here, we provide 5$\sigma$ upper limits in Table \ref{tab:image}.

\subsection{Parsec-scale Radio Morphology }

Figure \ref{fig:my_label} presents the full- and low-resolution naturally weighted total intensity radio maps for the 23 detected sources. These sources display a large variety of morphologies on parsec scales, ranging from compact unresolved single components to extended and complex features. The image parameters including source name, peak intensity, and rms noise level are given in the upper left corner of the full-resolution map, and the parameters of uv-taper, peak intensity, and the rms noise are given in the left corner of the uv-tapered map. In all of the images, the lowest contour represents 3 times the off-source image noise level, with successive contours increasing by factors of 2. The red and yellow crosses indicate the optical positions reported by Gaia astrometry \citep{2020yCat.1350....0G} and Pan-STARRS1 \citep{2016arXiv161205560C} with the uncertainties. The positions of the VLBI peak are marked as white crosses in both full- and low-resolution images for each source. As the uncertainties of the VLBI results are too small ($<$ 1 mas), we just add a large size of 2 mas for the white crosses. Table \ref{tab:image} lists the image parameters for all the sources: source name, VLBI positions, beam size, peak intensity, and off-source rms noise.

The identification of the core component is crucial for characterizing the parsec-scale morphology and investigating the origin of the radio emission. The detailed descriptions of the individual sources, including the identification of the core and the reliability of faint features are discussed in Section \ref{notes}. The radio cores are identified based on compactness, brightness, and proximity to optical positions. 

The 23 sources display three morphological classes: single component, one-sided core-jet, and two-sided core-jet. The presence of prominent compact radio emission qualifies all of the sources detected as AGN candidates. A single unresolved core is detected in 52\% (12/23) of the sources. In 48\% (11/23) of the resolved sources, ten sources show one-sided core-jet structures, and one source, NGC 2146, shows a core and two-sided jet structure.

Twelve sources (NGC 507, NGC 2300, NGC 2841, NGC 3245, NGC 3414, NGC 3735, NGC 3982, NGC 4036, NGC 4102, NGC 5005, NGC 5485, and NGC 6340) show a single compact core in our observations.
The 1.5 GHz e-MERLIN observations show three types of radio morphologies in large scales: four sources (NGC 3735, NGC 3982, NGC 5485, and NGC 6340) showing compact cores; two H II galaxies (NGC 3245 and NGC 4102) showing very complex structures (star-formation dominated); the remaining 6 sources showing core-jet or twin-jet structures.
The VLBI positions of 3 sources (NGC 507, NGC 2841, and NGC 3414) are almost consistent with the optical positions. For the other 9 sources, the Gaia-VLBI position differences range from 23 - 1000 mas (1.8 - 95 pc), suggesting that the true nuclei in these sources are possibly heavily obscured at optical wavelengths \citep{1998ApJ...496..133B}.
As the radio cores of 5 sources (NGC 3735, NGC 3982, NGC 4036, NGC 4102, and NGC 5005) are more than 6.4 pc away from the optical peak and not bright enough, we cannot confirm the AGN cores from the current data. The other 7 sources with very compact structures ($<$ 0.05 - 1.1 pc), high brightness temperatures (5 $\times$ 10$^{6}$ K - 2 $\times$ 10$^{7}$ K), and close to the optical peak allow them to be identified as the AGN cores. 

Ten sources (NGC 410, NGC 777, NGC 1161, NGC 2985, NGC 2655, NGC 3348, NGC 3884, NGC 4736, NGC 4750, and NGC 5548) display a one-sided core-jet structure. The radio emission is dominated by the core, the brightest and most compact component. The position of the optical nucleus from the catalog of the Gaia and Pan-STARRS1 is much more consistent with the radio core.
Six sources show core-jet, twin jets, or complex structures in the 1.5 GHz e-MERLIN observations, while 4 LINER sources (NGC 1161, NGC 2985, NGC 3884, and NGC 4750) show compact cores.

NGC 2146 shows a bright central component straddled between two weaker symmetric features, suggesting a two-sided core-jet structure, along the northwest-southeast direction with an extent of about 20 mas (1.2 pc). 
The core is identified based on the very compact size (0.06 pc) and high brightness temperature (4.2 $\times$ 10$^{7}$ K), typical of non-thermal origin. The separation between the optical position and our VLBI position is about 1.7 arcsec (100 pc). The 1.5-GHz e-MERLIN observation detected a core-jet structure in this region, previously identified as a radio supernova in the study by \citet{2000A&A...358...95T}. However, the observations in our current study --- particularly the compactness and high brightness temperature of the core, coupled with the symmetric nature of the surrounding features --- strongly support the interpretation that the central component of NGC 2146 is indeed an AGN.

\subsection{Source Compactness}

The morphology of LLAGNs can be quantified with compactness, a feature discernible only through high-resolution VLBI observations.  Compactness serves as a quantitative measure of the extent to which radio emissions are confined to the central regions of galaxies and is crucial for understanding the nature of these emissions, be they from star-forming regions, active galactic nuclei, or other phenomena.  The resolution of the NVSS at 1.4 GHz \citep{1998AJ....115.1693C} is about 45 arcsec and the e-MERLIN at 1.5 GHz is about 0.2 arcsec \citep{2018MNRAS.476.3478B}. So the ratio  $\rm S_{e-MERLIN}/S_{NVSS}$ can be used as a rough indicator of the compactness on arcsecond scales. The resolution of the e-MERLIN at 5 GHz is about 50 mas and the resolution of our VLBI observations is about 2 mas. The ratio of $\rm S_{VLBI}/S_{e-MERLIN}$ can be used as a rough indicator of the compactness on mas scales. The preliminary fluxes have been extracted from the 5 GHz e-MERLIN data and will be published in a forthcoming work (Williams-Baldwin et al., 2025, in prep.).

The compactness on arcsecond scales derived as $\rm S_{e-MERLIN}/S_{NVSS}$ shows a wide range between 0.4\% and 144\%. Figure \ref{fig:compact}a shows the compactness on arcsecond scales as a function of NVSS flux density. NGC 5485 has a ratio larger than 1. The higher e-MERLIN flux density could be due to variability with a variation factor higher than 1.44. We used the Pearson correlation test to reveal the correlation between the compactness of arcsecond and NVSS flux density. That gives a correlation with $\tau = -0.7$ and $P < 0.0001$, where $\tau = 0$ means no correlation, $\tau = \pm 1$ means strong correlation and $P$ gives the probability of no correlation. We found a strong anti-correlation between the compactness of arcsecond scales and NVSS flux density, suggesting that all the sources show a similar contribution of radio emission from the compact sub-arcsecond scales but different origins of extended emission on arcsecond scales. 

Figure \ref{fig:compact}b shows the relation between the NVSS flux density and the compactness on mas scales derived as $\rm S_{VLBI}/S_{e-MERLIN}$, which varies in the range of 0.04 -- 2.17. Nine sources have a ratio larger than 1. For the 23 detected sources, the median value is 0.77, indicating that a substantial fraction of the 5 GHz e-MERLIN emission is from the pc scale core region. NGC 2655 has a relatively low ratio, $\rm S_{VLBI}/S_{e-MERLIN}$ = $\sim$0.04. This implies that NGC 2655 is lobe-dominated in arcsec scales, a scaled-down version of FR II morphology  (Williams-Baldwin et al., 2025, in prep.). We did not find any correlation or anti-correlation between the NVSS flux density and the mas scale compactness. In Figure \ref{fig:compact}c, we plot the dependence of the mas scale compactness on the arcsecond sale compactness. Our results confirm the similar trend of no significant correlation between the mas scale compactness with the large-scale emission.

Most of our undetected sources (11/13) were also not detected by the e-MERLIN at 5 GHz (Williams-Baldwin et al., 2025, in prep.), indicating that the 1.5 GHz low-brightness radio emission in these sources is from the star-forming activities, previous nuclear starburst or inactive SMBHs, rather than active, accreting SMBHs \citep{2018MNRAS.476.3478B,2021MNRAS.500.4749B}.

\subsection{Notes on Individual Galaxies}\label{notes}

In this section, we generally concentrate on the radio properties of each source with the flux density and morphology obtained from the NRAO VLA Sky Survey \citep[NVSS,][]{1998AJ....115.1693C} at 1.4 GHz with an angular resolution of $45''$ and an rms of 0.45 mJy, the new e-MERLIN 1.5 GHz \citep[LeMMINGs,][]{2018MNRAS.476.3478B,2021MNRAS.500.4749B}, the archive and our VLBI results, and as well as other available radio data.
The optical positions reported by Gaia astrometry \citep{2020yCat.1350....0G} and Pan-STARRS1 \citep{2016arXiv161205560C} are also discussed for the radio core identifications. Since most of these nearby galaxies are spatially resolved and show low surface brightness of $\sim$ 18 - 20 mag in the optical V-band images, their localization accuracy is strongly affected and poor. The identification of the radio cores is mainly based on compactness, brightness, specific morphology, and proximity to the optical center.
\\

\textbf{\large NGC 410.} This is an X-ray bright elliptical galaxy.
NVSS detected a 5.8 mJy point source.
The 1.5-GHz e-MERLIN observations show an extended core-jet structure with a total flux density of 3.62 mJy. 
The source was not detected with the VLBA at 5 GHz before \citep{2004A&A...418..429F}, which gives an upper limit of 0.40 mJy (5$\sigma$).
Our 5-GHz VLBA observations detected two radio components with a total flux density of 0.56 mJy, consistent with the position of the core region in e-MERLIN.
The Gaia position has a separation of 44 mas (16.2 pc) with respect to the VLBI peak position.
The counterpart in the Pan-STARRS1 catalog is relatively closer (25 mas) to our VLBI position, but not consistent with the Gaia position within 3$\sigma$. 
Because of the optical faintness with a mean magnitude of m$\rm _{g}$ = 20.2, both the multi-epoch full-sky Gaia astrometry and Pan-STARRS1 give a large position uncertainty of 16.5 and 5.4 mas, respectively.
Considering the VLBI position uncertainty of about 0.5 mas, the position of VLBI is consistent with both optical positions within 5$\sigma$ error, indicating the true position of the nucleus.
The more compact (0.64 pc) and brighter (5.8 $\times$ 10$^{6}$ K) component in the north is identified as the core. 
The flux densities of the core and the jet component are 0.35 and 0.21 mJy, respectively.
Our image also reveals a core-jet structure of $\sim$ 10 mas ($\sim$ 3.6 pc) along the northeast-southwest direction.

\textbf{\large NGC 507.}
This is a massive elliptical galaxy. 
NVSS detected a double radio source straddling the galaxy along the northwest-southeast direction with a total flux density of 61.7 mJy.
In the higher resolution ($\sim$3.5 arcsec) VLA observations at 1.4 GHz, the source shows a weak core and very diffuse and faint extended emission \citep{1986A&AS...64..135P}.
The 1.5-GHz e-MERLIN observations reveal a core-jet structure in the middle of the NVSS image with a total flux density of 1.5 mJy but consistent with the core position in \citet{1986A&AS...64..135P}, indicating a weak core and double jets structure obtained in VLA. A large fraction of the extended radio emission is resolved in e-MERLIN.
The very faint and diffuse radio emission, possibly related to a previous outburst, is also detected with the LOFAR and uGMRT at low frequencies \citep{2022A&A...661A..92B}. 
The 5-GHz EVN observations only detected one compact component in the southeast of the core position of e-MERLIN with an angular separation of about 1.5 arcsec. The flux density of the component is 1.22 mJy. No optical nucleus in Gaia is found within 18 arcsec. We searched the Pan-STARRS1 catalog and found a counterpart very close to our VLBI position with a separation of 26 mas (8.4 pc) to the VLBI peak, consistent with each other within 1$\sigma$ error.
We also checked the 2-Micron All Sky Survey \citep[2MASS;][]{2006AJ....131.1163S} and found an infra-red counterpart that is much consistent with our VLBI position (separation $<$ 0.04 arcsec).
The compact size ($<$ 1 pc), high brightness temperature (5.1 $\times$ 10$^{6}$ K), and close to the optical and infra-red nucleus make it the radio core.
The radio structure that appears in e-MERLIN images should be the pc-kpc scale lobe, indicating that the jet is subluminal nearer to the core region and it eventually interacts with the surrounding interstellar medium (ISM), edge-brightening radio morphology.

\textbf{\large NGC 777.} 
This is an X-ray bright elliptical galaxy, hosting a type 2 nucleus.
NVSS detected a 7.4 mJy point source.
\citet{2001ApJS..133...77H} detected an unresolved core using VLA at 6 cm and 20 cm with flux densities of 2.63 and 3.19 mJy, respectively.
The 1.5-GHz e-MERLIN observations reveal a complex structure with a core flux density of 0.59 mJy.
The source was not detected with the VLBA at 5 GHz before \citep{2005ApJ...627..674A}, which gives an upper limit of 0.15 mJy (5$\sigma$). 
Our 5-GHz EVN observations detected two components with total flux densities of 0.45 mJy, consistent with the core position in e-MERLIN.
In the low-resolution, the image shows an extended emission structure of $\sim$12 mas (4 pc) along the east-west direction.
The Gaia position has a separation of 22 mas (7.6 pc) with respect to the VLBI peak position. 
The counterpart in Pan-STARRS1 is in the east (30 mas) of our VLBI position but is not consistent with the Gaia position. 
Because of the optical faintness with a mean magnitude of m$\rm _{g}$ = 20.0, the multi-epoch full-sky Gaia astrometry and Pan-STARRS1 only give a position uncertainty of 16.2 and 13.1 mas. 
Considering the VLBI position uncertainty of about 0.5 mas, the positions of VLBI and optical are consistent within 3$\sigma$ error, indicating the true position of the nucleus.
The East component is identified as the core because of its very compact size (0.26 pc) and high brightness temperature ($\sim$ 10$^{7}$ K).

\textbf{\large NGC 1161.}
This is classified as an S0 galaxy, hosting a LINER nucleus. 
NVSS detected a 4.9 mJy point source.
The 1.5-GHz e-MERLIN observations show a 3.13 mJy compact core.
The source was not detected with the VLBA at 5 GHz before \citep{2004A&A...418..429F}, which gives an upper limit of 0.40 mJy (5$\sigma$).
Our 5-GHz VLBA observations detected two components with flux densities of 3.78 and 0.32 mJy, respectively. 
No optical counterpart is found within 0.3 arcsec (50 pc) in the Pan-STARRS1 catalog.
The Gaia position has a separation of 22.6 mas (2.9 pc) with respect to the west component.
This component has a very compact size (0.07 pc) and high brightness temperature of $\rm 10^{8.88}$ K  and is identified as the core.
The source shows a core-jet structure along the northwest-southeast direction. 
The Gaia position in the upstream direction jet should be due to the dominant impact of the accretion disk on the Gaia coordinates and by the effects of the parsec-scale radio jet.

\textbf{\large NGC 2146.}
This is a nearby luminous infrared galaxy \citep[LIRG,][]{2003AJ....126.1607S}.
NVSS detected an unresolved point source of 1074.5 mJy.
The high-resolution 5-GHz continuum observation of this starburst galaxy made with VLA and e-MERLIN shows that no evidence of a flat radio core was found \citep{2000A&A...358...95T}, suggesting a very strong burst star formation.
The 1.5-GHz e-MERLIN observation detected a core-jet structure that was classified as a radio supernova before with a total flux density of 13.7 mJy.
This is the only source showing a two-sided core-jet structure in our 5 GHz VLBI observations along the northeast-southwest direction with a total flux density of 1.71 mJy.
The optical counterparts from Gaia and Pan-STARRS1 catalogs are about 1.7 arcsec (100 pc) away from the radio position.
The very compact size (1.14 mas $\approx$ 0.07 pc) and high brightness temperature (T$_{\rm b} >$ 10$^{\rm 7.62}$ K) suggest that the central component is possibly the true nucleus.
The reason for the position offset between optical and radio is that the true optical nucleus may be heavily obscured.
However, we can not fully exclude the possibility of SNR due to the lack of spectral index and variability. Further observations with multifrequency and multiepoch will shed light on the nature of the source.

\textbf{\large NGC 2300.}
This is classified as an S0 galaxy.
NVSS detected a 2.9 mJy point source.
The 1.5-GHz e-MERLIN observations reveal a core-jet structure along the northeast-southwest direction with a total flux density of 1.5 mJy.
The 5-GHz EVN observations only detected a point component, consistent with the e-MERLIN position, with a flux density of 0.11 mJy. 
No optical nucleus in Gaia is found within 10 arcsec from the VLBI position.
We searched the Pan-STARRS1 catalog and found a counterpart of about 1004 mas ($\sim$95 pc) in the east of our VLBI position.
The G-band optical image of Pan-STARRS1 shows a resolved low surface brightness source with a field size of 20 $\times$ 20 arcsec.
Thus, its optical centroid can be located by the Pan-STARRS1 survey but has a large systematic error of about 35 mas (1$\sigma$).
The very compact size (0.12 pc) and high brightness temperature (T$\rm _{b} > 10^{6.78}$ K) suggest that the component is dominated by the non-thermal synchrotron emission.
In addition, Hubble Space Telescope (HST) observations show no sign of dust and the low far-infrared (FIR) flux density upper limit also suggests the lack of the ISM in this galaxy, supporting that the radio source is powered by an AGN \citep{1989ApJS...70..329K,2004A&A...416...41X}.
Thus, we think the component should be the AGN core.

\textbf{\large NGC 2655.}
This is a nearby early-type spiral galaxy, hosting a LINER nucleus.
NVSS detected a 123.2 mJy point source.
\citet{2001ApJS..133...77H} detected a compact core and two-sided jetlike features with VLA at 6 cm and 20 cm.
The 1.5-GHz e-MERLIN observations reveal an S-shaped jet structure along the northeast-southwest direction with a total flux density of 88 mJy.
The source was not detected with the VLBA at 5 GHz before \citep{2002A&A...392...53N}, which gives an upper limit of 0.76 mJy (5$\sigma$).
Our 5-GHz VLBA observations detected two components, consistent with the e-MERLIN position, with a total flux density of 1.18 mJy.
The very compact size (0.32 pc) and high brightness temperature (T$\rm _{b} > 10^{6.60}$ K) of the north component suggest that the radio emission is associated with the non-thermal synchrotron emission.
No optical nucleus in Gaia is found within 10 arcsec away from the VLBI position. 
The counterpart in the Pan-STARRS1 catalog is about 330 mas (31.4 pc) in the west of our VLBI position.
The optical HST V-band image reveals a strong dust lane, suggesting that the true nucleus is heavily obscured in optical \citep{1998ApJ...496..133B}.
This may be the reason for the large position offset (330 mas) between optical and radio.
Thus, we identified the compact and bright north component as the radio core and the source shows a core-jet structure along the northeast-southwest direction, the same direction as the large-scale structures.

\textbf{\large NGC 2841.}
This is a well-known Sb spiral galaxy, hosting a LINER nucleus.
NVSS detected a two-sided core-jet structure along the northwest-southeast direction with a total flux density of 35.9 mJy.
The 1.5-GHz e-MERLIN observations also reveal a twin jet structure along the same direction with a total flux density of 1.15 mJy.
The 5-GHz EVN observations only detected a compact component, consistent with the e-MERLIN position, with a flux density of 0.97 mJy.
The Gaia position has a separation of 6.9 mas (0.3 pc) with respect to the VLBI position. 
No optical counterpart is found within 0.3 arcsec (13 pc) in the Pan-STARRS1 catalog.
Considering the uncertainty of VLBI position about 0.6 mas and 20 mas of Gaia position, the positions of VLBI and Gaia are consistent within 1$\sigma$ error.
This component is identified as the core because of its very compact size (0.05 pc) and high brightness temperature of 3.2 $\times$ 10 $^{7}$ K.

\textbf{\large NGC 2985.}
This is a nearby early-type spiral galaxy, hosting a LINER nucleus.
NVSS detected a 44.1 mJy point source.
At 1.49 GHz, \citet{1987ApJS...65..485C} measured an extended source with a total flux of 61.9 mJy.
The 1.5-GHz e-MERLIN observations reveal a slightly resolved core-jet structure along the northeast-southwest direction with a total flux density of 1.18 mJy.
The 5-GHz EVN observations show a bright component and a faint component located 10 mas to the southeast, consistent with the e-MERLIN position, with a total flux density of 1.46 mJy.
The Gaia position has a separation of 5.4 mas (0.44 pc) with respect to the VLBI position.
The counterpart in the Pan-STARRS1 catalog is about 823 mas (67.1 pc) in the southwest of our VLBI position but with a very large systematic error of about 2 arcsec. We dropped this data because of the large errors.
The very compact size (0.07 pc), high brightness temperature (7.6 $\rm \times 10^{7} $ K), and close to the optical nucleus make the bright component of the radio core.
The Gaia position in the upstream direction jet should be due to the dominant impact of the accretion disk on the Gaia coordinates and by the effects of the parsec-scale radio jet.

\textbf{\large NGC 3245.}
This is a nearby H II galaxy.
NVSS detected a 7.2 mJy point source.
The 1.5-GHz e-MERLIN observations show a complex structure with a total flux density of 2.6 mJy.
The source was not detected with the VLBA at 5 GHz before \citep{2004A&A...418..429F}, which gives an upper limit of 0.50 mJy (5$\sigma$).
Our 5-GHz EVN observations detected one compact component with a flux density of 0.96 mJy. 
The Gaia position has a separation of 62.3 mas (5.0 pc) with respect to the VLBI position.
The counterpart in the Pan-STARRS1 catalog is about 45 mas in the east of the VLBI position, consistent with the Gaia and VLBI position within 1 $\sigma$ of the large systematic error of 66 mas.
The HST V-band image reveals a nuclear dust disk while the larger-scale bulge morphology is classical, potentially hosting a lens \citep{1991rc3..book.....D}.
This suggests that the true nucleus is heavily obscured in optical.
This component has the highest brightness temperature of $\rm > 10^{9.04}$ K in our sample and very compact size (0.015 pc) and is identified as the radio core.

\textbf{\large NGC 3348.}
This is an E0 galaxy.
NVSS detected an 8.3 mJy point source.
The 1.5-GHz e-MERLIN observations show a complex structure with a triple components source of total flux density of 2.2 mJy.
Our 5-GHz EVN observation also detected three components with a total flux density of 1.62 mJy, consistent with the e-MERLIN core position. 
The Gaia position has a separation of 71 mas (13 pc) with respect to the VLBI position.
The Pan-STARRS1 position has a separation of 57 mas (10.4 pc) but is inconsistent with the Gaia position within 3 $\sigma$. 
All three components are on the same side of the Gaia or Pan-STARRS1 position.
Because of the optical faintness with a mean magnitude of m$\rm _{g}$ = 19.2, both the multi-epoch full-sky Gaia astrometry and Pan-STARRS1 give a large position uncertainty of 9.1 and 23.3 mas, respectively.
Considering the VLBI position uncertainty of about 0.5 mas, the position of VLBI is consistent with both optical positions within 5$\sigma$ error, indicating the true position of the nucleus.
With the current result, we cannot fully confirm the core-jet structure and rule out other possibilities, such as a resolved jet \citep[e.g., 3C 216:][]{2013MNRAS.433.1161A} or nuclear shocks likely formed by the episodic ejection dominated in the VLBI image \citep[e.g., NGC 4395:][]{2022MNRAS.514.6215Y}.
Lower and multiple frequencies observations are required to reveal the radio structure.
The brightest component has a brightness temperature of $\rm 5.5 \times 10^{6}$ K with a size of 3.5 mas (0.6 pc) and is identified as the AGN core candidate.

\textbf{\large NGC 3414}
This is a peculiar S0 galaxy, the outer disc is nearly face-on, and the inner disk has a higher ellipticity and perhaps a central bar \citep{2012MNRAS.427..790S}.
The low far-infrared flux density suggests that the radio source is powered by an AGN \citep{1989ApJS...70..329K}.
NVSS detected a 4.7 mJy point source.
The 1.5-GHz e-MERLIN observations show a twin jet structure with a flux density of 2.1 mJy.
Our 5-GHz EVN observations detected one compact component with a flux density of 1.92 mJy. 
The Gaia position has a separation of 7.3 mas (0.7 pc) with respect to the VLBI position.
The Pan-STARRS1 position has a separation of 14.2 mas (1.4 pc) but has a large systemic error of about 6.4 mas (1 $\sigma$).
The high brightness temperature of $\rm > 10^{7.15}$ K, very compact size (0.2 pc), and close to the optical nucleus allow the component to be identified as the core.

\textbf{\large NGC 3735.}
This is a low-luminosity Seyfert 2 galaxy.
NVSS detected an 87.6 mJy point source.
The 1.5-GHz e-MERLIN observations only detected a core with a flux density of 1.05 mJy.
In our 5 GHz EVN observation, three weak components ($\sim$ 5$\sigma$) are detected in the full-resolution map and the combination ($> \rm 7 \sigma$) of the three components is detected in the low-resolution map with a size of 14.29 mas (2.6 pc) and flux density of 0.5 mJy. The VLBI position is consistent with the position of the e-MERLIN result but $\sim$ 880 mas (160 pc) and 1.8 arcsec (341 pc) away from the Gaia and Pan-STARRS1 positions.
This component has a relatively low brightness temperature of 1.3 $\rm \times 10^{5}$ K, a large size (2.6 pc).
Based on the current result, we cannot identify the component as a young RSN or radio nucleus.
We make the component the AGN core candidate or a nuclear star cluster.
Deeper observations at low frequencies are needed to confirm this.

\textbf{\large NGC 3884.}
This is a spiral Galaxy in the Leo constellation.
NVSS detected a 14.9 mJy point source.
The 1.5-GHz e-MERLIN observations show a 4.2 mJy compact core.
Our 5-GHz VLBA observations detected two components with flux densities of 3.96 and 0.35 mJy. 
The Gaia position has a separation of 5.8 mas (2.6 pc) with respect to the south component.
The Pan-STARRS1 position has a separation of 14.3 mas (6.4 pc) but has a large systemic error of about 15 mas (1 $\sigma$).
The VLBI position is consistent with the Gaia and Pan-STARRS1 positions within 5 $\sigma$.
This component has a compact size (0.45 pc) and high brightness temperature of $\rm > 10^{8.30}$ K and is identified as the core.
The source shows a core-jet structure along the southwest-northeast direction.

\textbf{\large NGC 3982}
This is a low-luminosity Seyfert 2 galaxy.
NVSS detected a 57.7 mJy point source.
\citet{2001ApJS..133...77H} detected an unresolved core using VLA at 6 cm and 20 cm with flux densities of 1.79 and 3.56 mJy.
The 1.5-GHz e-MERLIN observations detected a 4.4 mJy compact core.
Our 5-GHz VLBA observations also only detected one compact component with a flux density of 0.44 mJy, consistent with the core position in e-MERLIN.
The Gaia and Pan-STARRS1 positions are consistent with each other within 1 $\sigma$, a separation of about 79 mas (6.4 pc) with respect to the VLBI position.
The component has a high brightness temperature of 5.5 $\rm \times 10^{6}$ K and a compact size of 0.16 pc, suggesting a non-thermal synchrotron emission.
We make the component the AGN core candidate.
Deeper observations at low frequencies are needed to confirm this.

\textbf{\large NGC 4036}
This is a S0 galaxy.
NVSS detected an 11.9 mJy compact point source.
The 1.5-GHz e-MERLIN observations detected a twin jet structure along the east-west direction with a total flux density of 8.0 mJy.
Our 5-GHz VLBA observations only detected a 0.41 mJy compact component, consistent with the core position in e-MERLIN.
The Gaia position has a separation of 104.7 mas (10.7 pc) in the east of the VLBI position.
The counterpart in the Pan-STARRS1 catalog is about 24 mas in the northwest of the VLBI position, consistent with the VLBI position within 3 $\sigma$.
The optical observations reveal an irregular dust lane in the disk, suggesting that the true nucleus is heavily obscured \citep{1998ApJ...496..133B}.
This may be the reason for the large position offset (105 mas) between Gaia and VLBI.
Thus, we identified this compact (0.27 pc) and bright (3 $\rm \times 10^{6}$ K) component as the AGN core candidate. 
Deeper observations at low frequencies are needed to confirm this.

\textbf{\large NGC 4102}
This is a nearby starburst galaxy, hosting a LINER nucleus.
NVSS detected a 273.4 mJy compact point source.
The 1.5-GHz e-MERLIN observations detected several extended star-forming regions and showed a very complex structure with a total flux density of 68.1 mJy.
The 5-GHz VLBA observations only detected a compact component, consistent with the e-MERLIN position but $\sim$ 1000 mas (61 pc) away from the optical nucleus reported by Gaia and Pan-STARRS1, with a flux density of 0.20 mJy. 
The brightness temperature of 2.1 $\rm \times 10^{6}$ K confirms that this is a non-thermal component, but not high enough to confirm the AGN core ($\rm > 10^{7-8} K$).
Based on the current result, we cannot identify the component as a young supernova or radio nucleus.
We make the component the AGN core candidate.
Deeper observations at low frequencies are needed to confirm this.

\textbf{\large NGC 4736}
This is an early-type spiral galaxy with a LINER 2 nucleus.
NVSS detected a 265 mJy point source.
\citep{1994ApJ...421..122T} detected a flat compact nuclear source at 2 cm and 6 cm with the VLA, suggesting the presence of an AGN.
The 1.5-GHz e-MERLIN observations show a compact core and two weak components with a total flux density of 5.85 mJy.
Our 5-GHz VLBA observations detected two components with a total flux density of 0.71 mJy, consistent with the core position in e-MERLIN. 
No optical nucleus in Gaia is found within 1 arcsec away from the VLBI position.
We searched the Pan-STARRS1 catalog and found a counterpart close to our VLBI position with a separation of 170 mas (3.4 pc).
Because of the large error (1 $\sigma$ = 110 mas) in the Pan-STARRS1 catalog, the positions between VLBI and Pan-STARRS1 are consistent with each other within the 2$\sigma$ error.
The very compact size ($<$ 0.07 pc), high brightness temperature (2 $\times$ 10$^{6}$ K), and close to the optical nucleus make it the core.
The source shows a core-jet structure along the southeast-northwest direction.

\textbf{\large NGC 4750}
This is a gas-rich Spiral galaxy, hosting a LINER nucleus.
NVSS detected a core-jet structure with a total flux density of 35.6 mJy.
The 1.5-GHz e-MERLIN observations show a 1.2 mJy compact core.
Our 5-GHz EVN observations detected two components with a total flux density of 0.54 mJy, consistent with the core position in e-MERLIN. 
The closet optical nucleus in Gaia is about 180 mas (18.4 pc) away from the VLBI position. We searched the Pan-STARRS1 catalog and found a counterpart closer to our VLBI position with a separation of 145 mas (14.8 pc).
Both optical positions are in the east of the VLBI position and consistent with each other.
The NIR and HST images found no evidence of star formation but a strong dust lane in the central region, suggesting that the true nucleus is heavily obscured in optical \citep{2002AJ....123..159C}.
This may be the reason for the large position offset between optical and radio.
Thus, we identified this compact (0.46 pc) and bright (1.1 $\rm \times 10^{6}$ K) component as the core candidate. 
The source shows a core-jet structure along the east-west direction.
Deeper observations at low frequencies are needed to confirm this.

\textbf{\large NGC 5005}
This is a spiral galaxy, hosting a LINER nucleus.
NVSS detected a slightly resolved source in an east-west direction with a total flux density of 182.7 mJy. 
The 1.5-GHz e-MERLIN observations reveal a double-lobed jet structure along the northwest-southeast direction with a total flux density of 42.3 mJy.
The 5-GHz VLBA image only detected a 0.29 mJy point component, consistent with the e-MERLIN position.
The optical nuclei in Gaia and Pan-STARRS1 are found about 500 mas (30 pc) away from the VLBI position. 
The reason for the position offset between optical and radio is that the optical nucleus is heavily obscured by the clumpy, dusty starburst region \citep{1998ApJ...496..133B}.
Based on the very compact size (0.12 pc), high brightness temperature (T$\rm _{b} > 10^{6.28}$ K), and the morphology in kpc scales, we identified the VLBI component is the AGN core candidate.
Deeper observations at low frequencies are needed to confirm this.

\textbf{\large NGC 5485}
This is an early-type spiral galaxy, hosting a LINER nucleus.
NVSS did not detect the source which gives an upper limit of 0.9 mJy.
The 1.5-GHz e-MERLIN observations detected a 1.3 mJy compact core.
The 5-GHz EVN observations also only detected a point component, consistent with the e-MERLIN position, with a flux density of 0.28 mJy.
The optical nuclei in Gaia and Pan-STARRS1 are found about 196 mas (24 pc) and 111 mas (14 pc) away from the VLBI position.
Based on the very compact size (0.1 pc) and high brightness temperature (T$\rm _{b} > 10^{7.28}$ K), we identify the component as the radio core.
The reason for the position offset between optical and radio is that the true optical nucleus may be heavily obscured \citep{1988AJ.....95..422E}.

\textbf{\large NGC 5548}
This is classified as a Seyfert 1.5 galaxy. 
NVSS detected a 29.1 mJy point source.
\citet{2001ApJS..133...77H} detected a classic triple linear structure at 6 cm and 20 cm with the VLA, consisting of a compact core and oppositely directed lobes.
The 1.5-GHz e-MERLIN observations show a core-jet structure with a total flux density of 5.0 mJy.
Our 5-GHz VLBA observations detected two components with flux densities of 0.80 and 0.35 mJy, consistent with the core position in e-MERLIN. 
The Gaia position has a separation of 0.7 mas (0.23 pc) with respect to the north component, consistent with each other within 1 $\sigma$.
We searched the Pan-STARRS1 catalog and found a counterpart of about 32 mas (10.5 pc) in the southeast of our VLBI position but with a very large systematic error of about 20 mas. We dropped this data because of the large errors.
This component has a very compact size (0.2 pc), high brightness temperature of $\rm > 10^{8.04}$ K, and is very close to the optical nucleus, suggesting the radio core.
The source shows a core-jet structure along the northeast-southwest direction.

\textbf{\large NGC 6340}
This source was not detected by the NVSS.
\citet{2018A&A...616A.152S} detected a point source of 0.61 mJy with the VLA at 15 GHz.
The 1.5-GHz e-MERLIN observations detected a compact core with a flux density of 1.22 mJy.
The 5-GHz EVN observations detected a point component, consistent with the e-MERLIN position, with a flux density of 0.36 mJy.
The optical nucleus in Gaia is found about 23 mas (1.8 pc) away from the VLBI position.
We searched the Pan-STARRS1 catalog and found a counterpart of about 42 mas (3.3 pc) in the south of our VLBI position a very large systematic error of about 26 mas, consistent with the Gaia position within 1$\sigma$.
Based on the very compact size (0.23 pc), high brightness temperature (T$\rm _{b} > 10^{6.36}$ K), and close to the optical nucleus, we identify the component as the radio core.

\section{Parameter correlations and discussion} \label{sec:disscussion}

In this section, we analyze correlations between the radio luminosity and the central SMBH mass, optical [O III] line luminosity, and X-ray luminosity. In addition, we also investigate the Fundamental Plane relation to constrain the accretion and jet physics in LLAGNs. As we mainly focus on our first VLBI detection results in this paper, we only present the correlations for the full VLBI sample of LeMMINGs galaxies in the following. The detailed correlation and scientific results for each optical class will be reported in a forthcoming paper, combining all the archival VLBI LLAGNs data in the literature. This will significantly enlarge the ALGs and HII galaxies samples and make statistical comparisons between the types of sources more precise.

We note that we used the total VLBI radio luminosity at 5 GHz because the core is not fully confirmed in one-third of the detected sources. To reduce errors due to differences in observation arrays and data quality, we adopt a uniform (10\%) uncertainty of radio flux in our study. The correlations of total and core luminosities show consistent results. These correlations will provide useful constraints when investigating the origin of radio emission in the nuclear region. We also include the 29 sources from the literature (listed in Table \ref{archival}) that were in the LeMMINGs sample and previously observed with VLBI in the following correlations, and hereafter refer to these as the archival sample. One of these galaxies, NGC 3079, is a well-known lobe-dominated AGN, and we therefore exclude it from the following fitting. In total, our correlation and linear regression analyses include 50 detected sources and 13 upper limits ($5\sigma$) of the undetected sources that will provide a comprehensive view of the parameter space. The LINMIX\footnote{A PYTHON module can be obtained from \url{https://linmix.readthedocs.io/en/latest/index.html}} package is used for the following fittings which include the upper limits of the undetected sources. There are a total of 6 H II galaxies (NGC 2146, NGC 3245, and NGC 4102 in our sample, NGC 2782, NGC 3504, and NGC 3665 in the archival sample) detected and imaged. We compared the distributions of H II galaxies with those of LINERs and Seyferts across the following correlations.



Here we test the correlations between the radio luminosity in VLBI observations and the BH mass. The VLBI observations enable a more accurate measure of the radio emission from only the accretion inflow and/or the sub-parsec scale of the jet. We adopt the BH mass from \citet{2021MNRAS.508.2019B} that estimated the values using the stellar velocity dispersions $\sigma$ measured from optical spectra \citep{2009ApJS..183....1H} and the empirical M$_{\rm BH}$–$\sigma$ relation from \citet{2002ApJ...574..740T}. The typical uncertainty of the BH mass is 0.5 dex.

Figure \ref{fig:BH} shows the distribution of the radio luminosity as a function of BH mass. A positive trend between the radio luminosity and BH mass is present, both for the archival sample (data points with gray crosses) and our sample. We note that the LINERs and Seyferts generally have larger BH masses than the H II galaxies. To evaluate the significance of the correlation, we used the Pearson correlation test between the two variables. We estimate the Pearson correlation coefficient for the whole sample. We find a
strong correlation with a Pearson correlation coefficient of 0.72 and a no correlation probability significance of 10$^{-11}$. We use the LINMIX package to include upper limits in the best fit:
\begin{equation}
\log L_{\rm R}=1.24(\pm0.15)\log M_{\rm BH}+27.12(\pm 1.20). 
\end{equation}
The regression analysis result is also visualized in Figure \ref{fig:BH}.


\citet{2018MNRAS.476.3478B,2021MNRAS.508.2019B} reported two 'broken' scaling relations between L$\rm _{core}$ and M$\rm _{BH}$ with distinct slopes: L$\rm_{\rm R} \propto L_{\rm BH}^{0.61\pm0.33}$ for non-jetted star-forming galaxies and L$\rm_{\rm R} \propto L_{\rm BH}^{1.65\pm0.25}$ for active galaxies. The break occurs at M$\rm _{BH} \sim 10^{6.5} M_{\odot}$. 
With only 4 SMBH masses less massive than 10$\rm ^{6.5} M_{\odot}$ in our whole sample, we have not sought to separate the galaxies into these two groups. We used a single L$\rm _{tot}$--M$\rm _{BH}$ relation. We find our fit gives L$\rm_{\rm R} \propto L_{\rm BH}^{1.24\pm0.15}$, consistent with the results of active galaxies within the errors, see Figure \ref{fig:BH}. 
Our study further supports that the radio emissions in active galaxies are AGN-driven.

It is interesting that the six H II galaxies which have M$\rm _{BH} > 10^{7}$ M$_{\odot}$ follow the sequence of the LINERs and Seyferts. This suggests that the radio emissions detected from these H~II galaxies, typically through VLBI observations, are predominantly influenced by active SMBHs, aligning them more closely with the characteristics of LINERs and Seyferts rather than being outliers in this correlation.

The [OIII], a forbidden emission line $\lambda$5007\,\AA\, coming from the narrow line region, is thought to be an isotropic indicator and should, therefore, be representative of the intrinsic power of the central engine.  We collected the [O III] luminosities (not corrected for extinction) from \citet{1997ApJS..112..315H}. Based on the typical uncertainty of [O III] intensity about 10\% - 20\% in this line \citep{1997ApJS..112..315H}, we adopt 20\% as the uncertainty of [O III] luminosity in our study. For the 3 ALGs in our sample, the [O III] luminosities are not available. In Figure \ref{fig:OIII}(a), we plot the distribution of radio luminosity as a function of optical [O III] line luminosities. We find the Seyferts have the highest [O III] luminosities, while H II galaxies have the lowest luminosity. LINERs are in the middle between the other two classes. 
A weak correlation is found at 0.38 with no correlation probability as less than 0.002 and the best fit for the whole sample using LINMIX package gives  
\begin{equation}
\log L_{\rm R}=0.46(\pm0.14)\log L_{\rm [O III]}+18.92(\pm 5.61). 
\end{equation}


Our high-resolution VLBI observations probe the jet base close to the AGN engine on the sub-parsec scales, which should be more sensitive to the BH-accretion properties than the VLA or e-MERLIN at arcsec resolutions. However, we find weaker radio-[O III] correlations than previous work with eMERLIN \citep{2021MNRAS.508.2019B}. One possibility is that a substantial contribution to the [O III] line emission is from kpc scales. Another possibility is that the [O III] line is heavily obscured by dust in our LLAGNs. The broad distribution of radio and optical luminosities in our sample, which comprises sources in various evolutionary stages, might also diminish the likelihood of observing a strong correlation. The high-resolution HST nuclear luminosity will be studied in the forthcoming paper. Nevertheless, the current correlations suggest that the probability for a galaxy to be a radio source increases with its [O III] luminosity.

To study the possible correlation between radio luminosity and Eddington ratio, we adopt the Eddington ratio from \citet{2021MNRAS.508.2019B} who estimated the values using the 3500 $\rm \times L_{\rm [OIII]}$ as the bolometric AGN luminosity. Figure \ref{fig:OIII}(b) shows the distribution of Eddington ratios as a function of VLBI 5 GHz total radio luminosity, we do not find any statistically significant correlation for our whole sample. This result is consistent with the e-MERLIN results \citep{2021MNRAS.508.2019B}.


In AGN, the hard 2-10 keV emission arises within the innermost region around the SMBH, while the soft band below 2 keV is usually from the diffuse extended region. In order to characterize the correlation between the radio and X-ray luminosity in the whole VLBI sample, we searched and adopted the luminosity data at 2-10 keV band from \citet{2022MNRAS.510.4909W}, which gives the X-ray data derived from $Chandra$ observations coincident within 2 arcsec of the nucleus. In our sample, six sources were not detected in $Chandra$ observations. We searched and found 4 sources that were observed and detected with $XMM-Newton$. Overall, only two sources (1 ALG NGC 3348 and 1 LINER 6340) in our sample and two sources (1 HII galaxy NGC 3504 and 1 LINER NGC 7217) in the archival sample were not observed or detected in X-ray now. In Figure \ref{fig:Xray}, the X-ray luminosity is plotted as a function of radio luminosity at 5 GHz to explore the intricate relationship between these two bands of emission. Similar to Figure \ref{fig:BH}, the distribution shows that the radio detection rate increases with increasing X-ray luminosity. While the X-ray luminosities range over 6 orders of magnitude (see Figure \ref{fig:Xray}, there are clear distinctions in the different optical types of sources. The Seyferts galaxies tend to have higher X-ray luminosities than LINERs.  There is a large distribution in the LINERs X-ray luminosity of order of 4 decades.  ALGs and H II galaxies tend to fall within the same mass bins as the LINERs but exhibit slightly lower radio and X-ray luminosities.

The ratio of radio to X-ray luminosity, L$\rm_{R}$/L$\rm_{X}$, is a critical metric representing the properties of accretion and ejection in compact objects. The large scatter of L$\rm_{R}$/L$\rm_{X}$ in Figure \ref{fig:Xray} implies that the origin of the emission in the radio and X-ray bands is significantly different. The distribution value is 10$\rm ^{-5}$ for radio-quiet quasars (RQQs) and coronally active stars \citep{2008MNRAS.390..847L}, suggesting coronal origin of both radio and X-ray emission in RQQs. However, recently \citet{2023MNRAS.tmp.2534W,2023MNRAS.525..164C,2024arXiv241007889C} found the L$\rm_{R}$/L$\rm_{X}$ values ranging from 10$\rm ^{-2}$ to 10$\rm ^{-6}$ is higher than 10$\rm ^{-5}$, implying jet-dominated radio emission in their RQQs. Higher L$\rm_{R}$/L$\rm_{X}$ $>$ 10$\rm ^{-5}$ in radio-loud AGNs are found \citep[e.g.,][]{2013MNRAS.432.1138P,2019NatAs...3..387P,2019MNRAS.488.4317B}, indicating the jet-dominated nature of their radio emission. 
The L$\rm_{R}$/L$\rm_{X}$ of most sources in the whole sample ranges from 10$\rm ^{-1}$ to 10$\rm ^{-5}$, with the best fitting result of 10$\rm^{-3.34\pm0.18}$ (cyan dashed line in Figure \ref{fig:Xray}). This suggests that the sub-parsec scale radio components in LLAGNs are likely associated with jet/outflow structures. We note that the Seyferts typically follow L$\rm_{R}$/L$\rm_{X}$ $\simeq$ 10$\rm ^{-5}$, suggesting a similar origin of radio and X-ray emission as RQQs \citep{2021MNRAS.508.2019B}. The detected LINERs, ALGs, and H II galaxies mostly follow L$\rm_{R}$/L$\rm_{X}$ $\simeq$ 10$\rm ^{-3}$, which indicates the jet-dominated nature of the radio emission in these detected three types of sources. It is crucial to note that the radio and X-ray observations were not performed simultaneously. Therefore, a deeper understanding would benefit from simultaneous deep multi-frequency radio and X-ray observations to accurately characterize the relationship between these emissions.

With the highly significant correlation results of radio luminosity with BH mass and X-ray luminosity, one might expect that the relationship between the X-ray luminosity, radio luminosity, and BH mass, which is known as the Fundamental Plane of black hole activity, would be tightly correlated. 
If the radio emission at 5 GHz in our VLBI observations is from the sub-parsec scale jets and hard X-ray emission is ADAF or jet in origin, the correlation in our sample should be more pronounced in the very vicinity of SMBHs down to sub-parsec scales.  
In Figure \ref{fig:FMP}, we find a tight correlation using the LINMIX package including upper limits among the 5 GHz VLBI total luminosity, the 2-10 keV luminosity, and the BH mass, with the multiple linear regression fit function as follows:
\begin{equation}
\log L_{\rm R}=\rm \xi_{RX}\log L_{\rm X} + \rm \xi_{RM}\log L_{\rm M} + \rm b_{R}
\end{equation}
We obtain $\rm \xi_{RX}$ = 0.52($\pm$0.08), $\rm \xi_{RM}$ = 1.25($\pm$0.15), $\rm b_{R}$ = 5.85($\pm$4.35) with a dispersion of $\rm \sigma_{VLBI}$ = 0.72. 

Several previous studies have also selected the nearby LLAGNs from the Palomar sample and observed them by using VLA or e-MERLIN at 1.4 to 15 GHz in sub-arcsec resolution \citep{2015MNRAS.453.3447D,2018A&A...616A.152S,2019ApJ...871...80G,2024A&A...689A.327W}. They found the $\rm \xi_{RX} \sim$ 0.44-0.53, consistent with our results within errors. These results suggest that the compact parsec-scale (sub-arcsec scale) radio emission is more correlated with the X-ray luminosity and black hole mass. 

\section{Summary and Conclusions} 

We have conducted a VLBI imaging study of a large sample of 36 nearby galaxies selected from the LeMMINGs sample at sub-mJy levels. Twenty-three galaxies were detected. The detection rates are 14/21 ($\sim$67\%) for LINERs, 3/4 (75\%) for Seyferts,  3/8 (37.5\%) for HII galaxies, and 3/3 (100\%) for ALGs. We identify the radio core in 16 sources and the core candidate in seven other sources. Our results show that about two-thirds of the nearby galaxies host faint AGNs, as revealed by the presence of compact radio cores. In our sample, ten sources show a compact core, twelve sources show a one-sided jet structure and one H II galaxy (NGC 2146) shows a two-sided core-jet structure in sub-parsec scales. Combining our results with the previous VLBI measurements of LLAGN, the main conclusions are summarized as follows: 
\begin{trivlist}

\item{1.} A strong correlation was found between radio luminosity and BH mass. The linear regression fits show a steep slope, L$\rm_{\rm R} \propto L_{\rm BH}^{1.24\pm0.15}$, in the whole sample, suggesting that the radio emission in these sources is AGN-driven which is tightly linked with the SMBHs. 

\item{2.} A weak correlation was found between radio luminosity and [O III] line luminosity. The possibility is that the [O III] emission line is from kpc scales. There is no significant correlation found between radio luminosity and the Eddington ratio, suggesting the sources are in various evolutionary stages. 

\item{3.} 
The L$\rm _{R}$/L$\rm _{X}$ ranges from 10$^{-1}$ to 10$^{-5}$  with a best fit of 10$^{-3.34}$, suggesting that the sub-parsec scale radio components in LLAGNs are associated with jet/outflow structures. Seyferts typically follow L$\rm_{R}$/L$\rm_{X}$ $\simeq$ 10$\rm ^{-5}$, suggesting a similar origin of radio and X-ray emission as RQQs. While the other three optical classes (LINERs, ALGs, and H II galaxies) mostly follow L$\rm_{R}$/L$\rm_{X}$ $\simeq$ 10$\rm ^{-3}$, which indicates the jet-dominated nature of the radio emission.

\item{4.} A tight correlation was found for the radio luminosity, BH mass, and 2-10 keV luminosity, known as the Fundamental Plane of black hole activity. The radio and X-ray correlation coefficient $\rm \xi_{RX}$ is $\sim$ 0.52 for our VLBI sample.
The compact parsec scale (sub-parsec scale) radio emission is strongly correlated with X-ray luminosity and black hole mass.

\end{trivlist}

\begin{acknowledgments}
This work is supported by the Brain Pool Program through the National Research Foundation of Korea (NRF) funded by the Ministry of Science and ICT (2019H1D3A1A01102564, RS-2024-00407499) and the National Key R\&D Programme of China (2018YFA0404603) and the Chinese Academy of Sciences (CAS, 114231KYSB20170003).
The European VLBI Network (EVN) is a joint facility of independent European, African, Asian, and North American radio astronomy institutes. Scientific results from data presented in this publication are derived from the following EVN project code: EC082. 
e-MERLIN is a National Facility operated by the University of Manchester at Jodrell Bank Observatory on behalf of STFC, part of UK Research and Innovation.
The VLBA observations were sponsored by Shanghai Astronomical Observatory through an MoU with the NRAO (Project code: BA152). 
The Very Long Baseline Array is a facility of the National Science Foundation operated under cooperative agreement by Associated Universities, Inc. 
\end{acknowledgments}
%
\vspace{5mm}
\facilities{VLBA, EVN, e-MERLIN.}
\software{AIPS \citep{2003ASSL..285..109G}, 
                    DiFX \citep{2007PASP..119..318D}, 
                    DIFMAP \citep{1994BAAS...26..987S},
                    LINMIX \citep{2007ApJ...665.1489K}}
\bibliography{sample631}{}
\bibliographystyle{aasjournal}



\begin{figure}
    \centering
    \includegraphics[height=3.5cm]{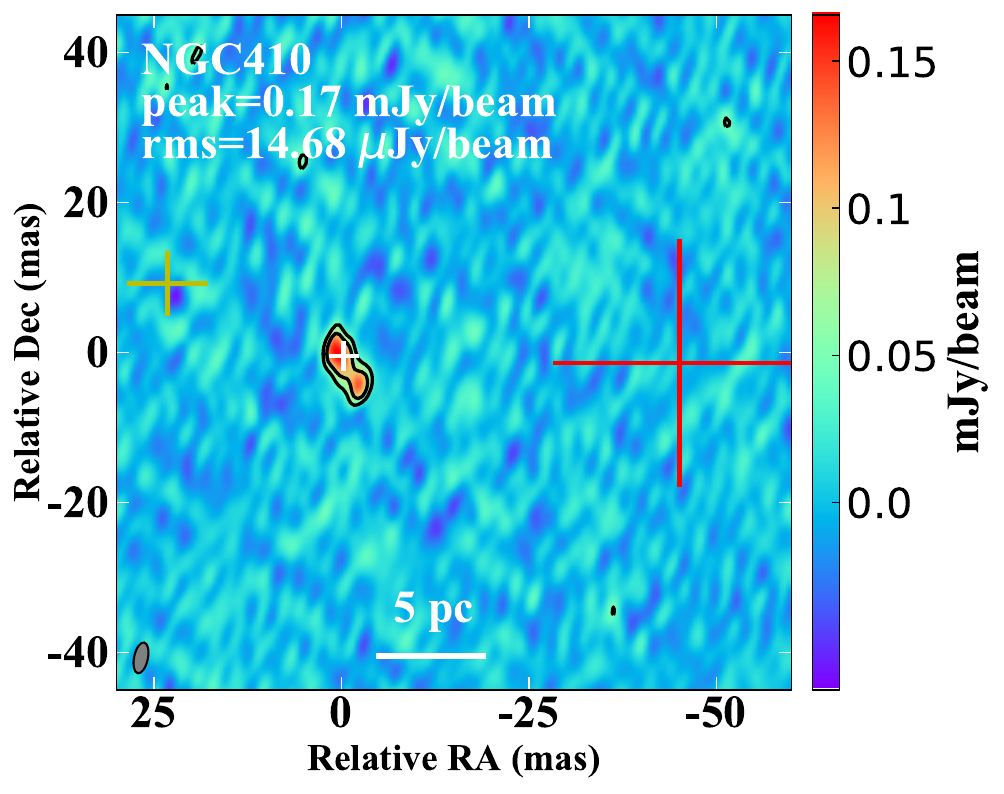}
    \includegraphics[height=3.5cm]{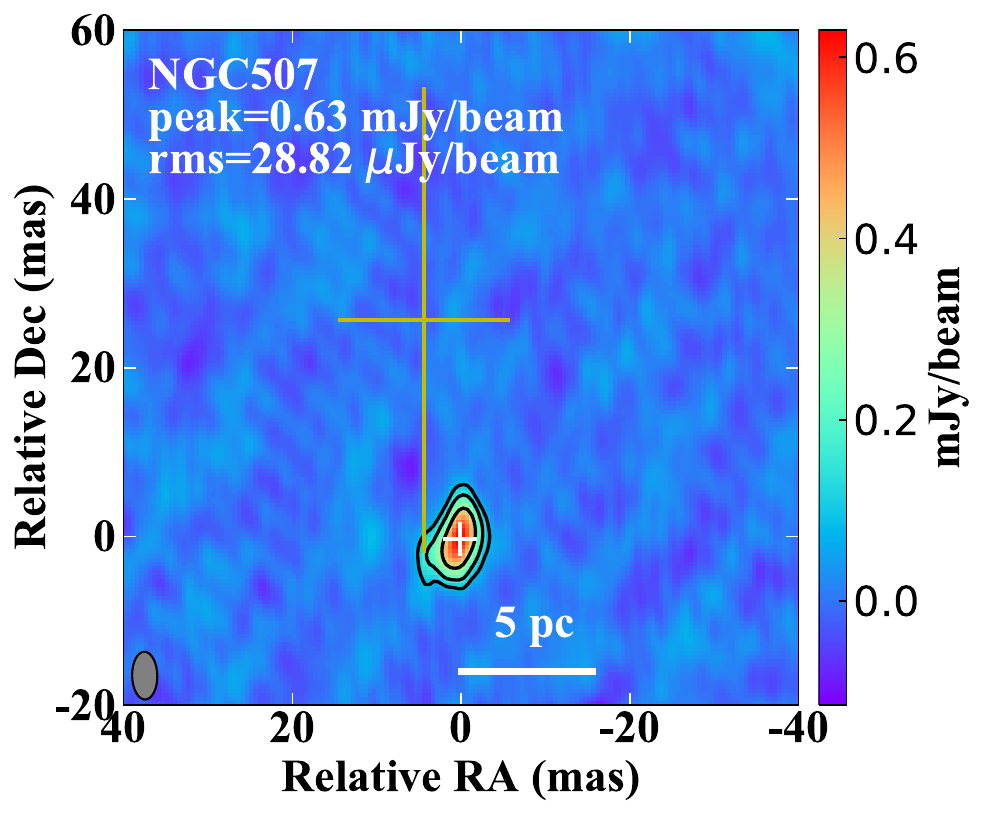}
    \includegraphics[height=3.5cm]{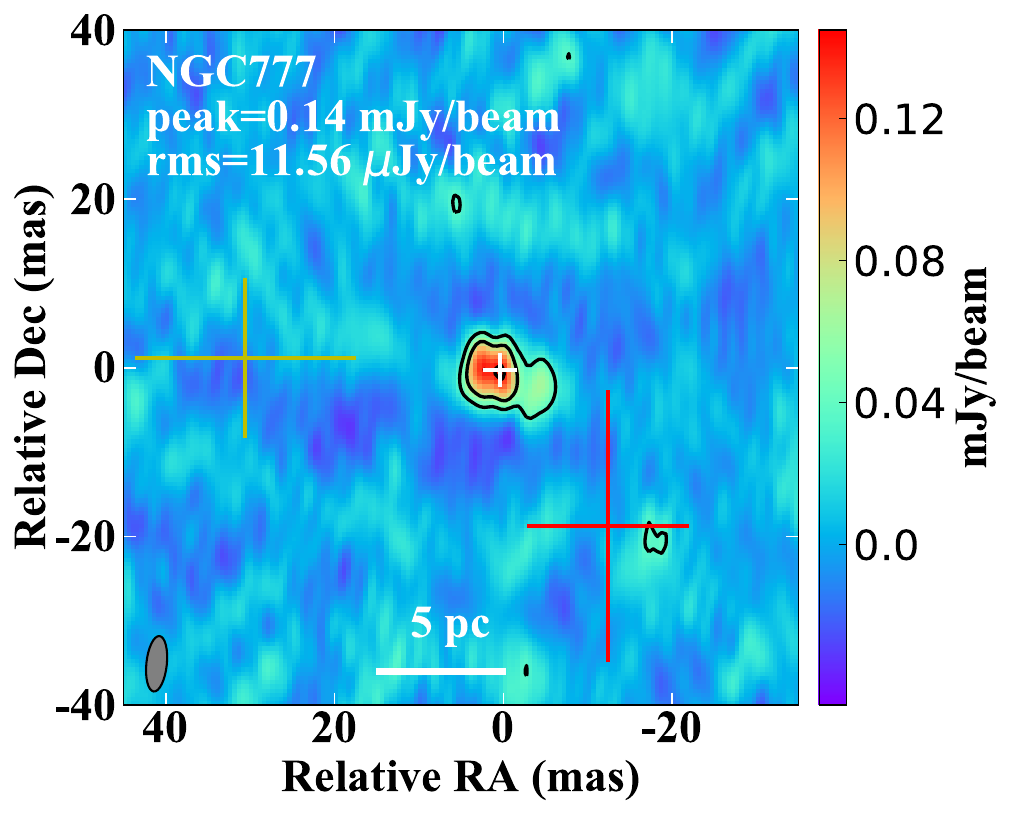}    
    \includegraphics[height=3.5cm]{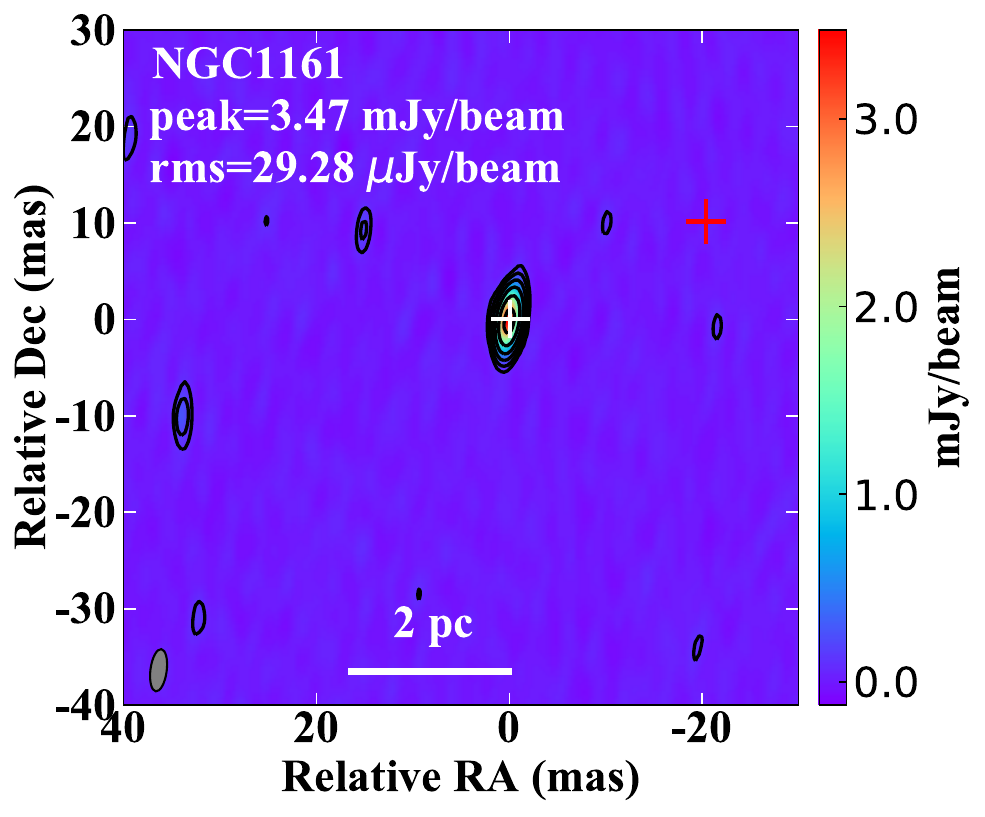}    
    \\
    \includegraphics[height=3.5cm]{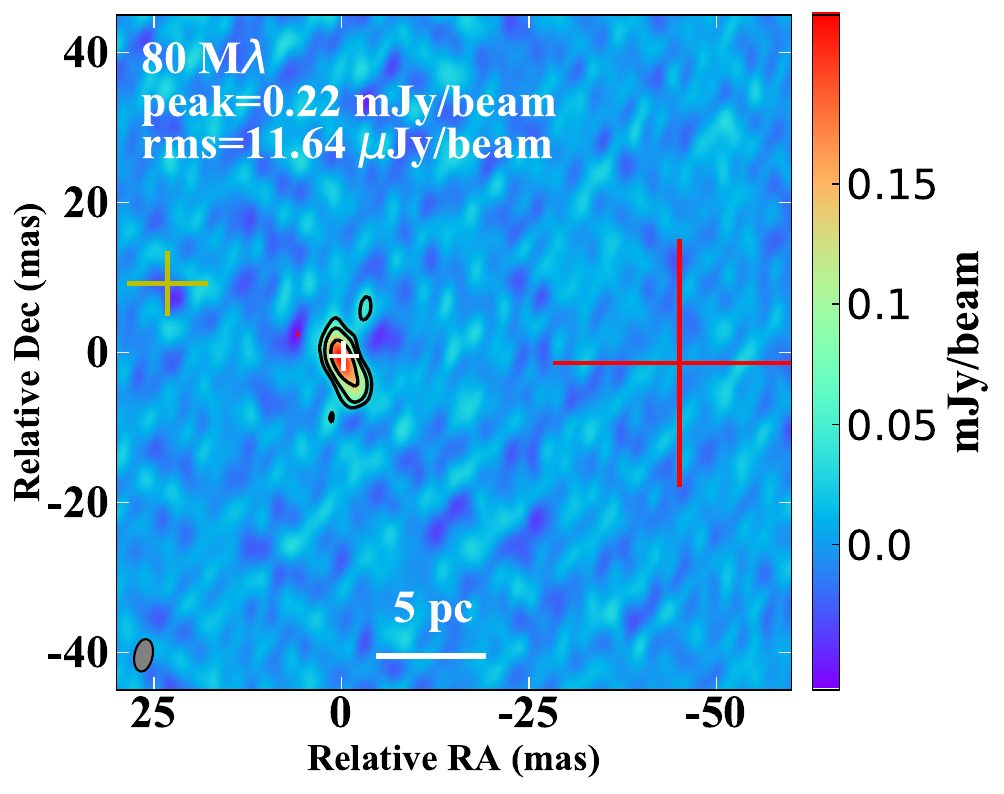}
    \includegraphics[height=3.5cm]{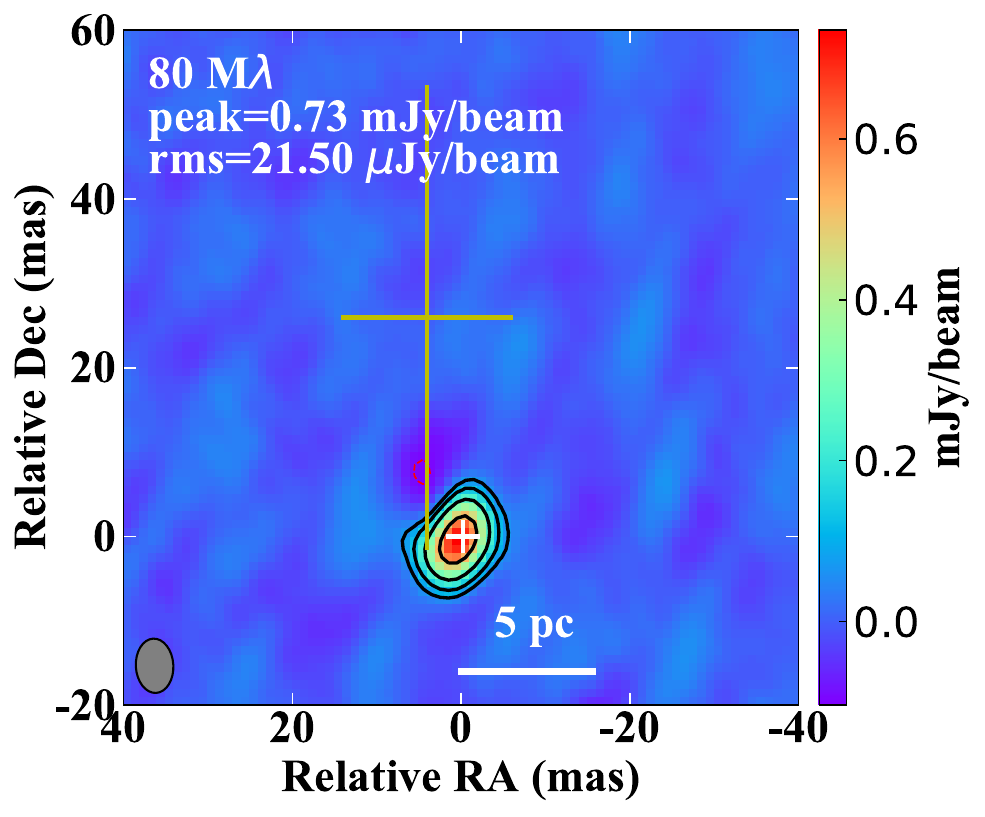}
    \includegraphics[height=3.5cm]{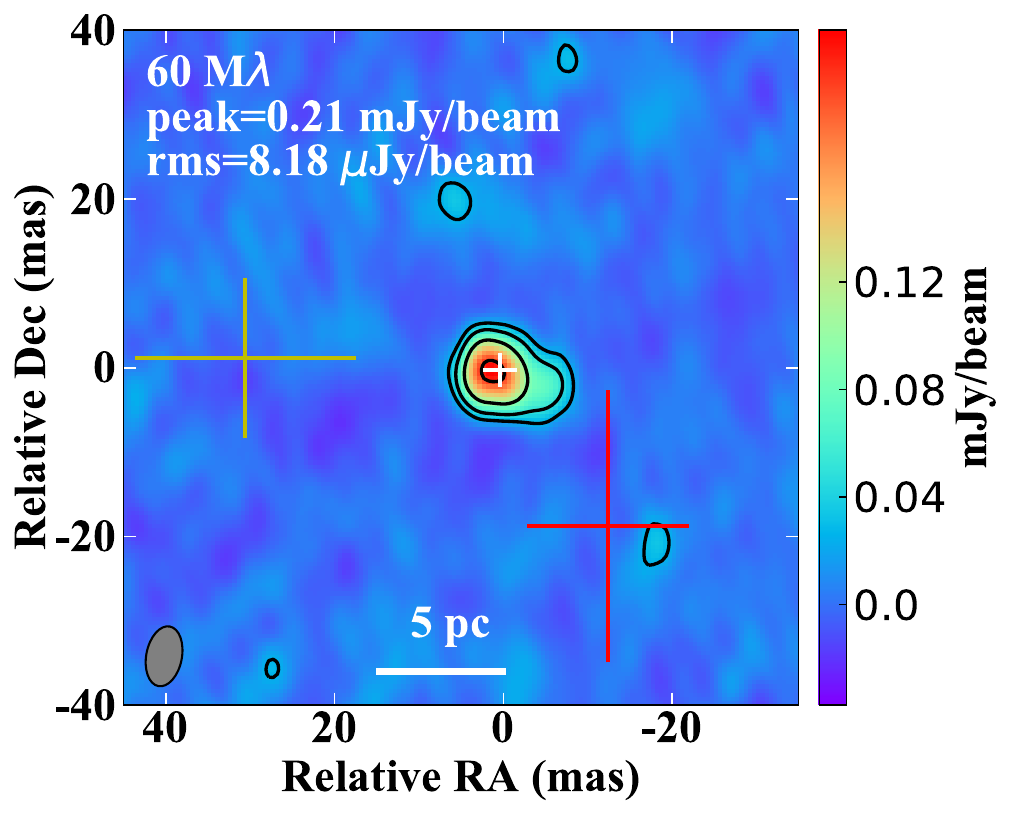}
    \includegraphics[height=3.5cm]{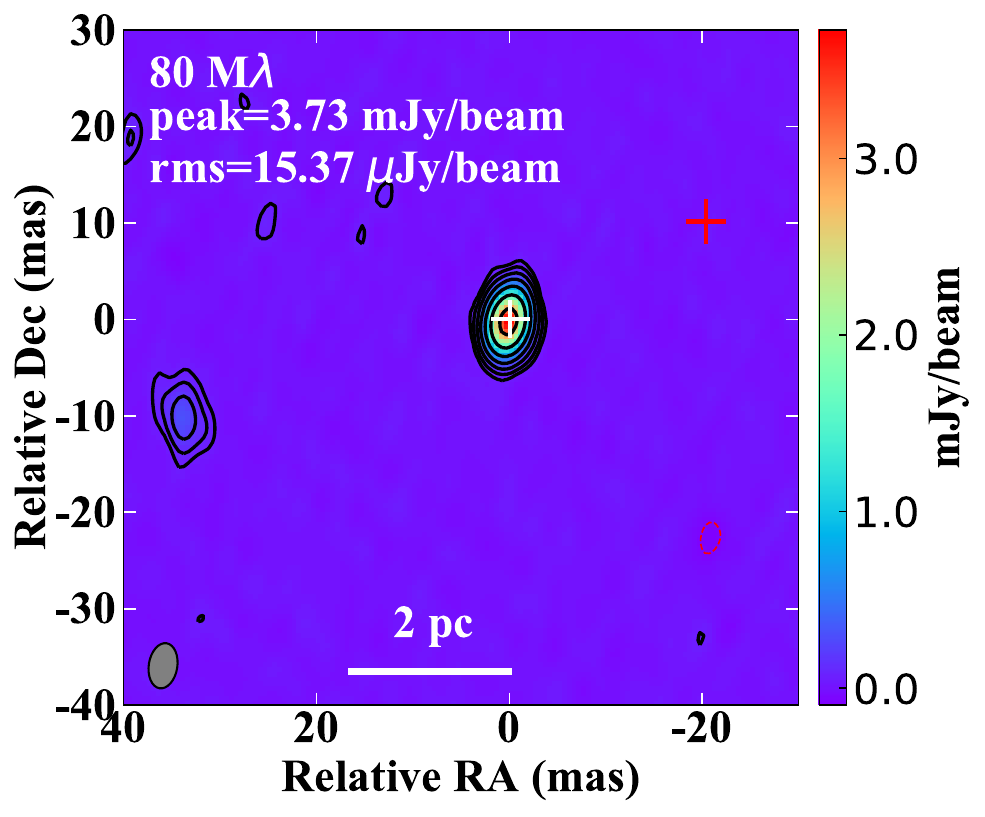}    
    \\
    \includegraphics[height=3.5cm]{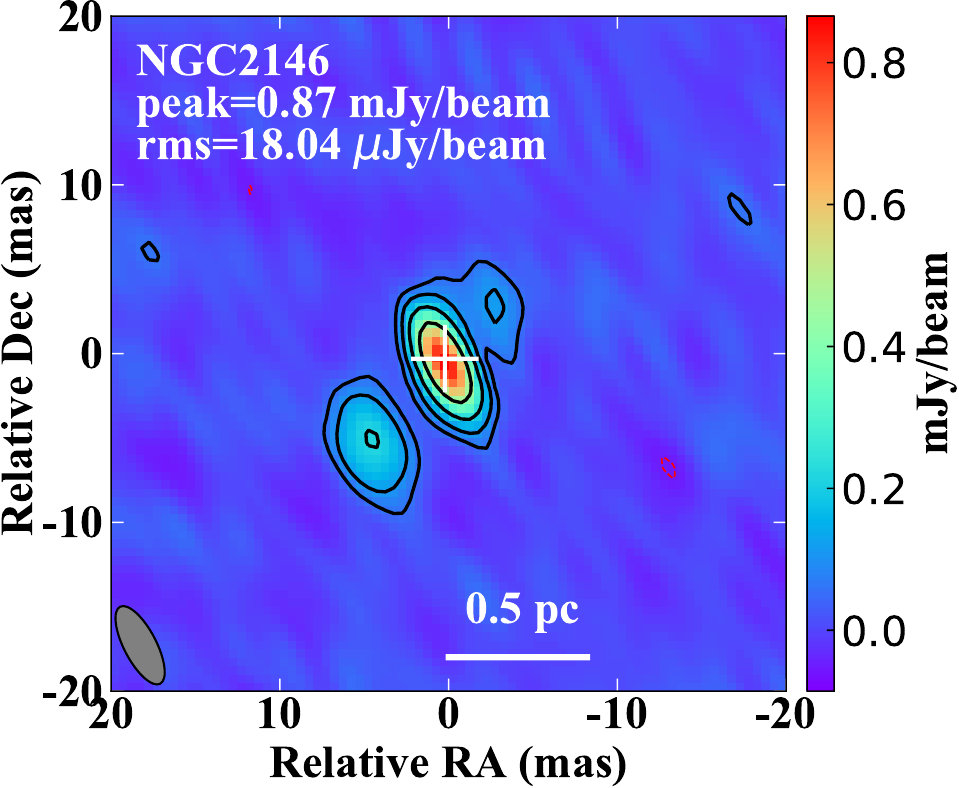}
    \includegraphics[height=3.5cm]{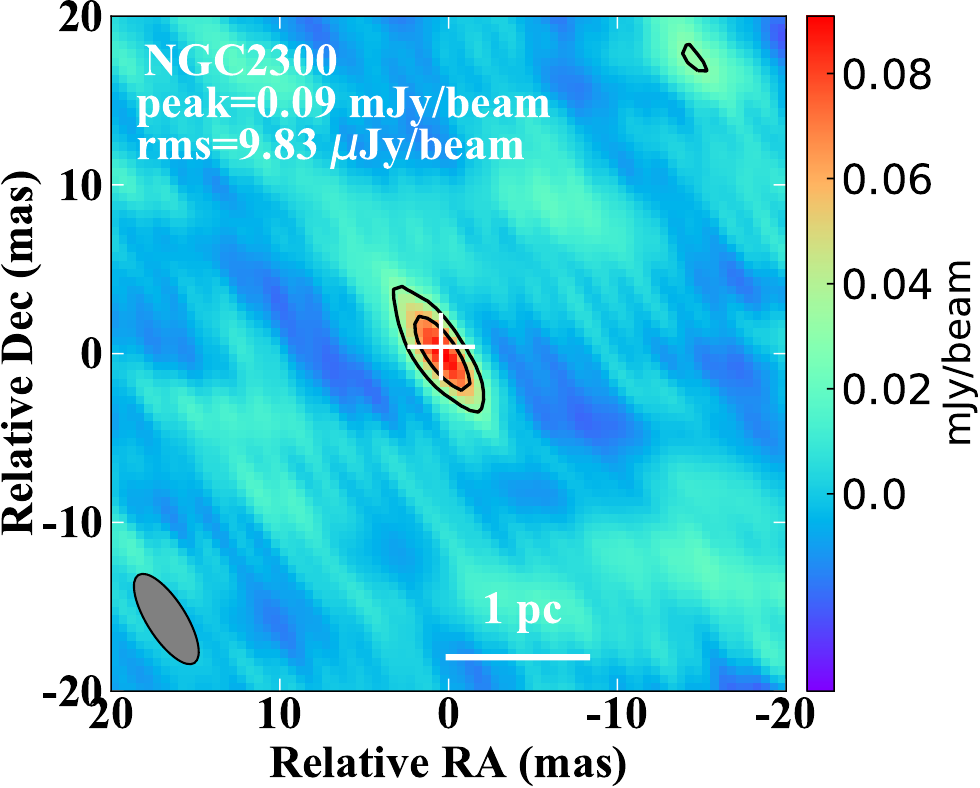}
    \includegraphics[height=3.5cm]{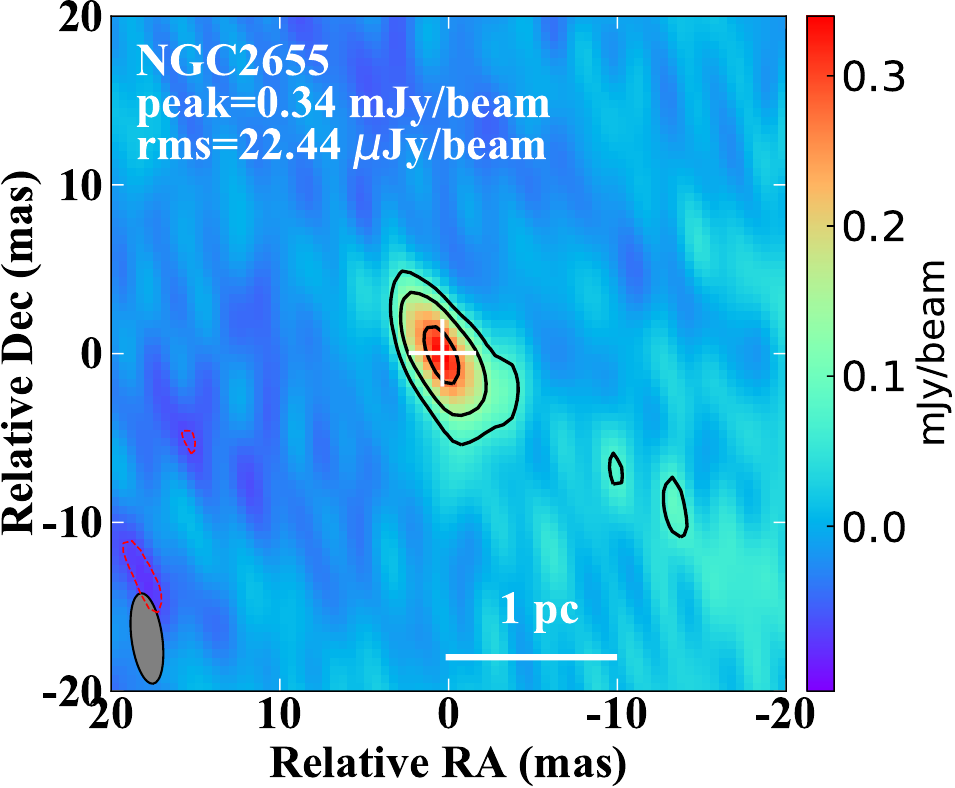}
    \includegraphics[height=3.5cm]{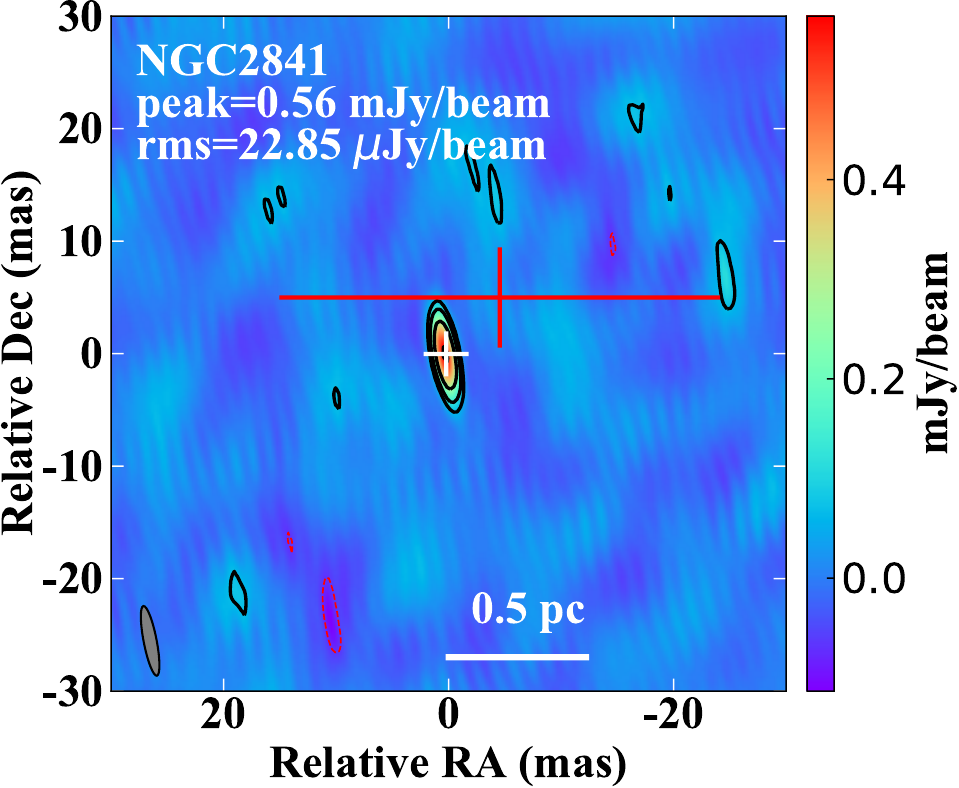}
    \\
    \includegraphics[height=3.5cm]{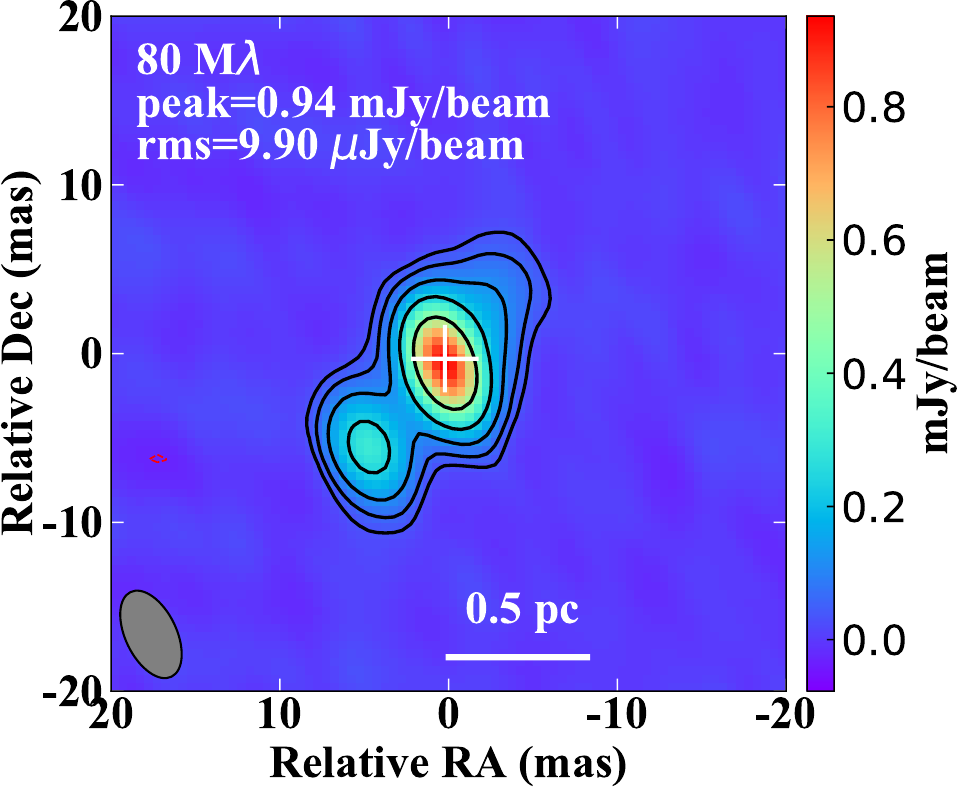}
    \includegraphics[height=3.5cm]{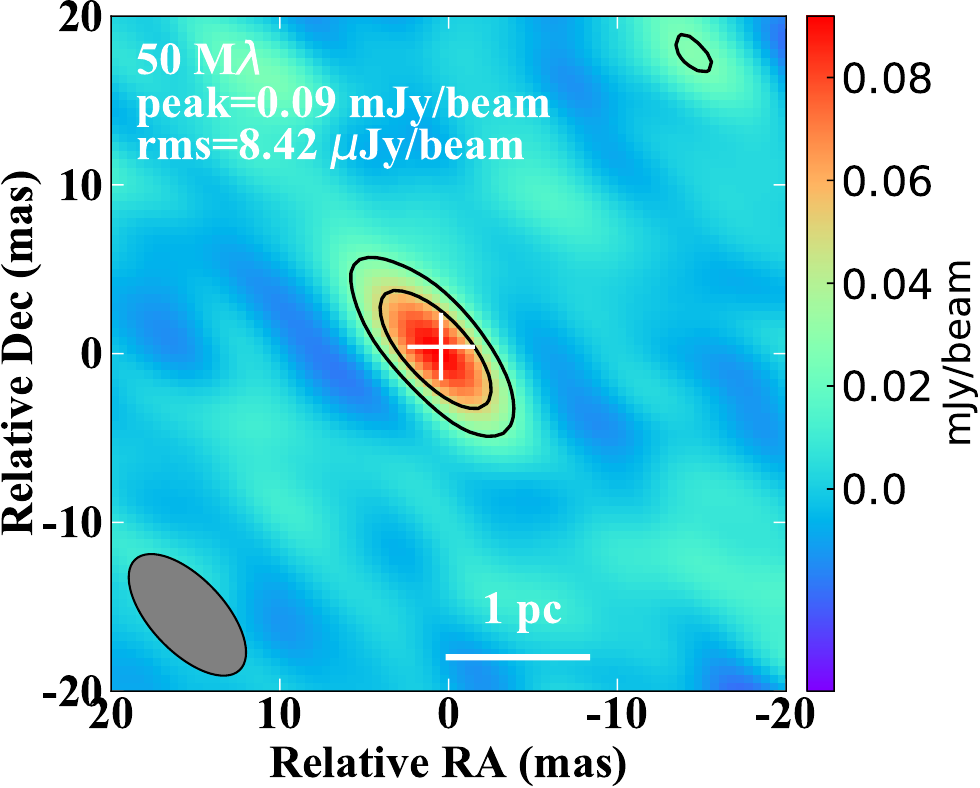}
    \includegraphics[height=3.5cm]{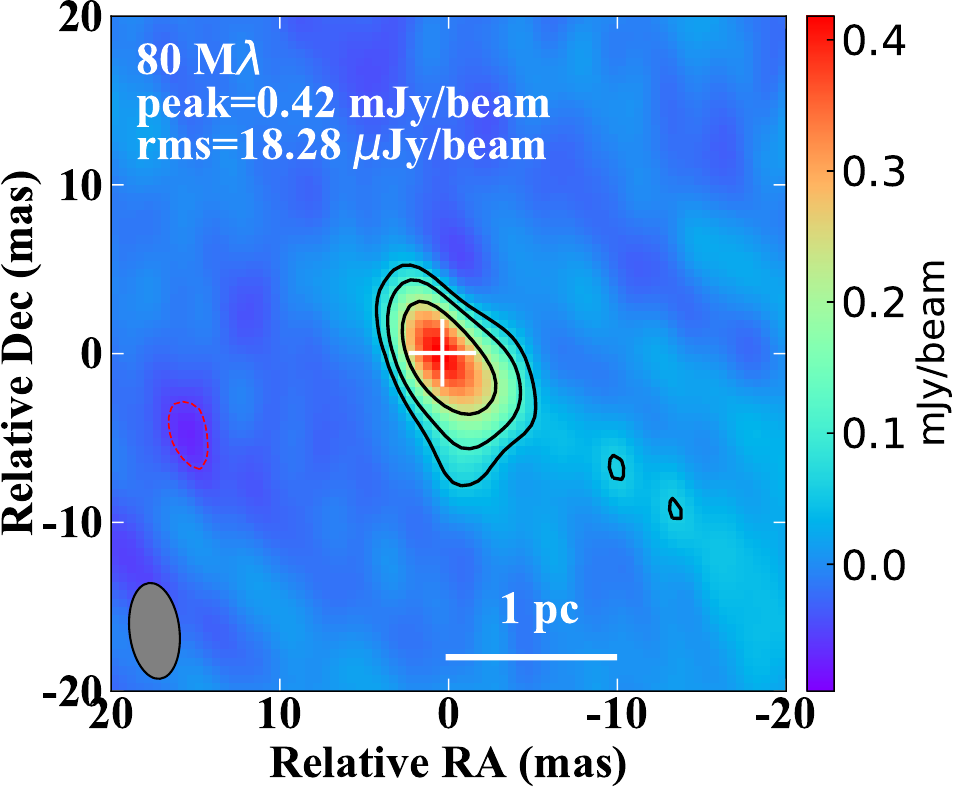}
    \includegraphics[height=3.5cm]{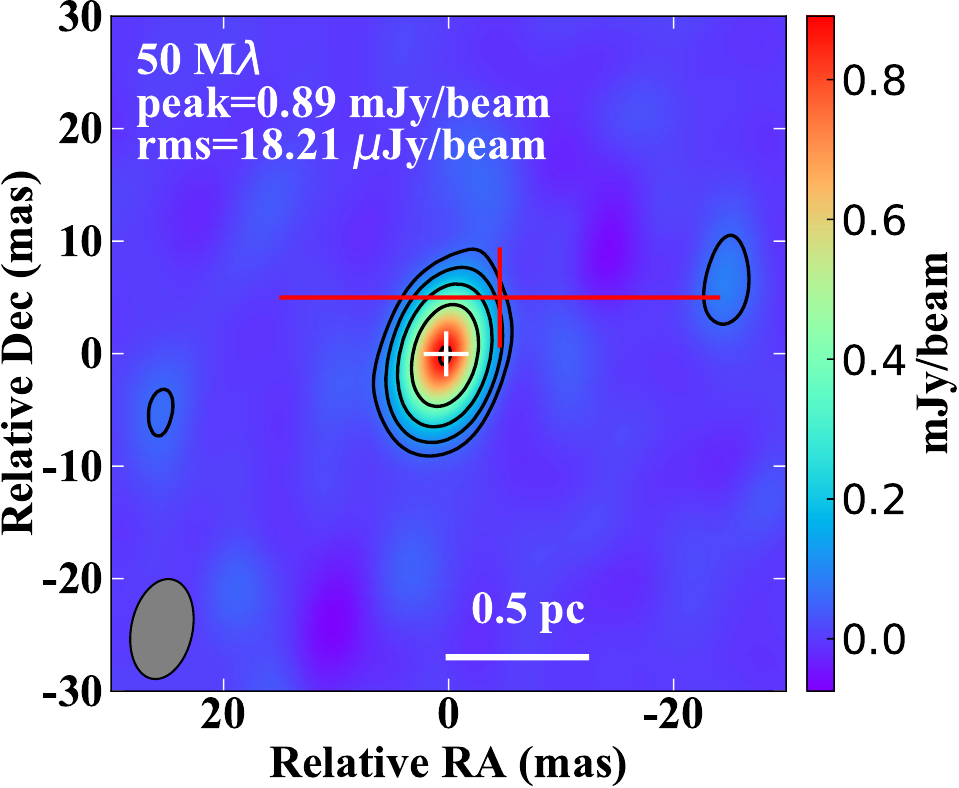}
    \\
    \includegraphics[height=3.5cm]{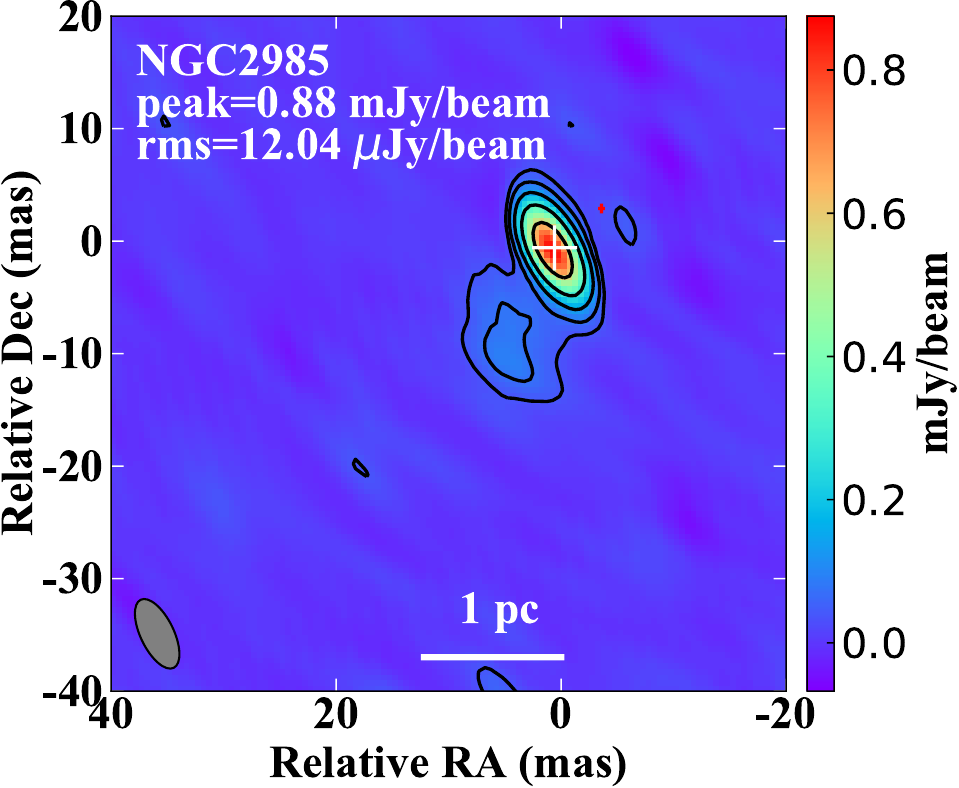}
    \includegraphics[height=3.5cm]{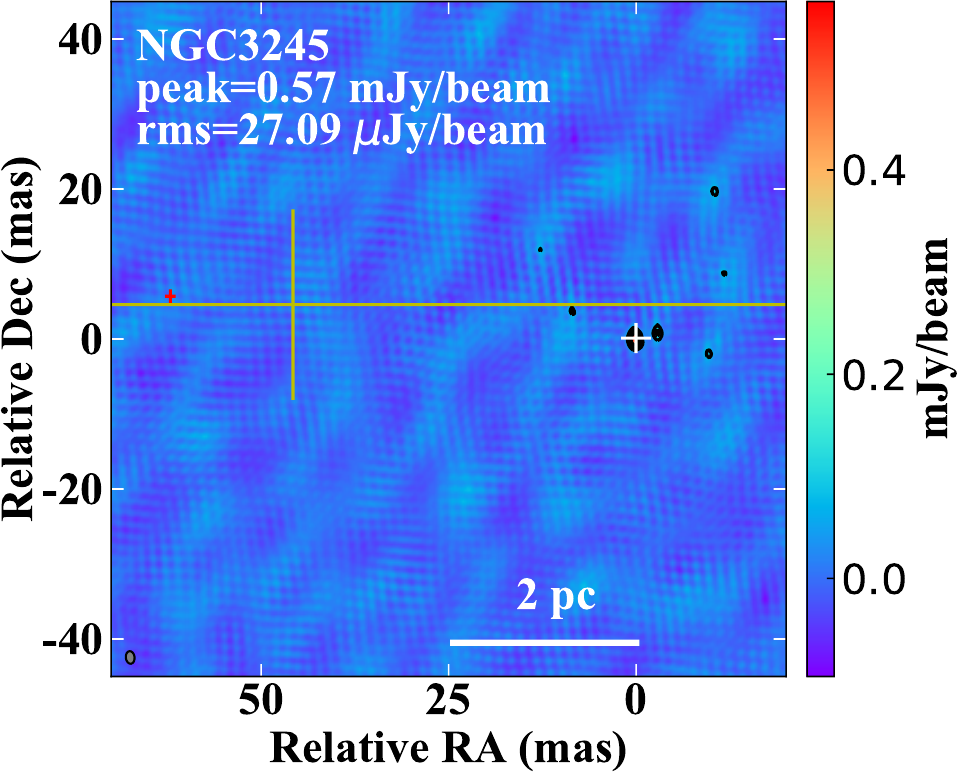}
    \includegraphics[height=3.5cm]{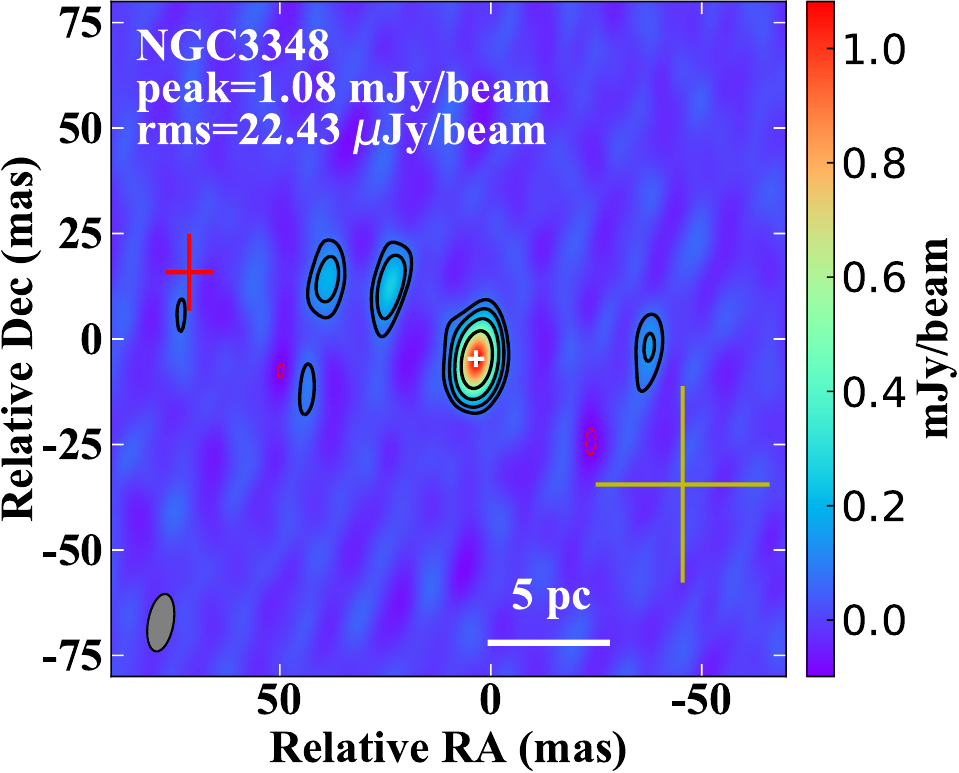}
    \includegraphics[height=3.5cm]{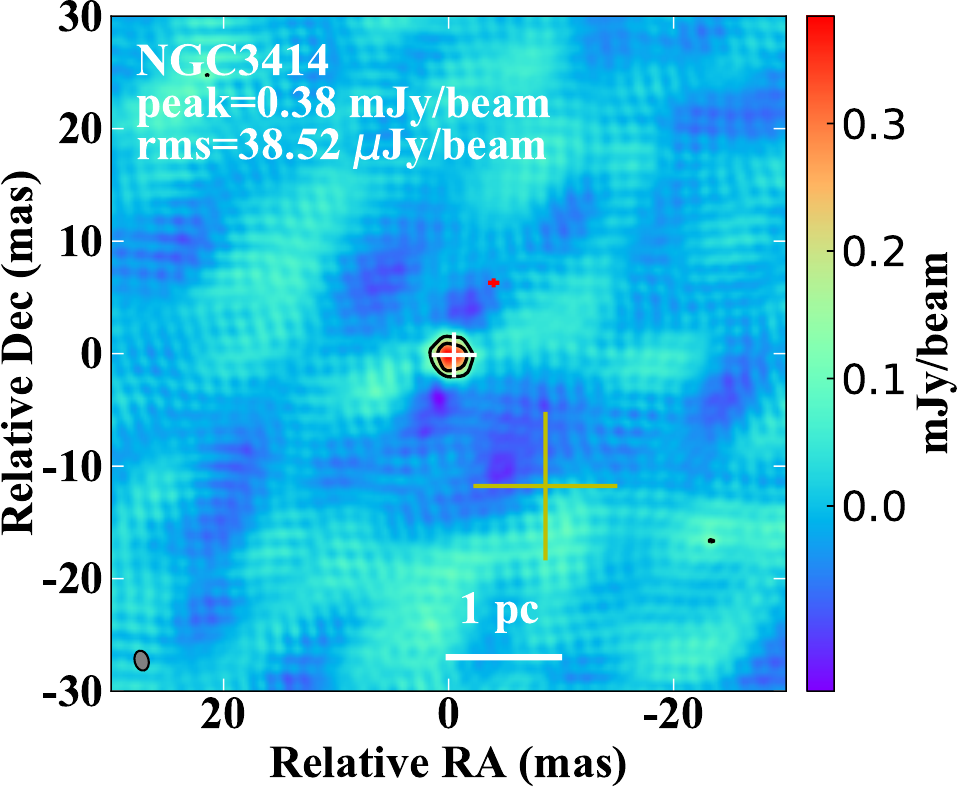}
    \\
    \includegraphics[height=3.5cm]{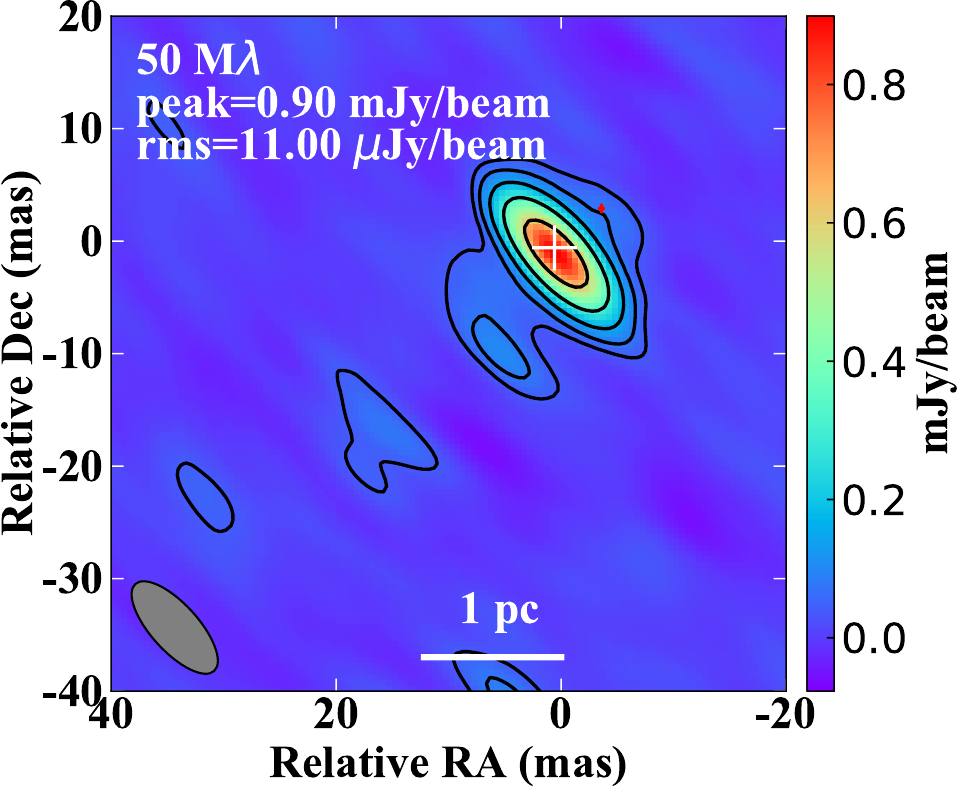}
    \includegraphics[height=3.5cm]{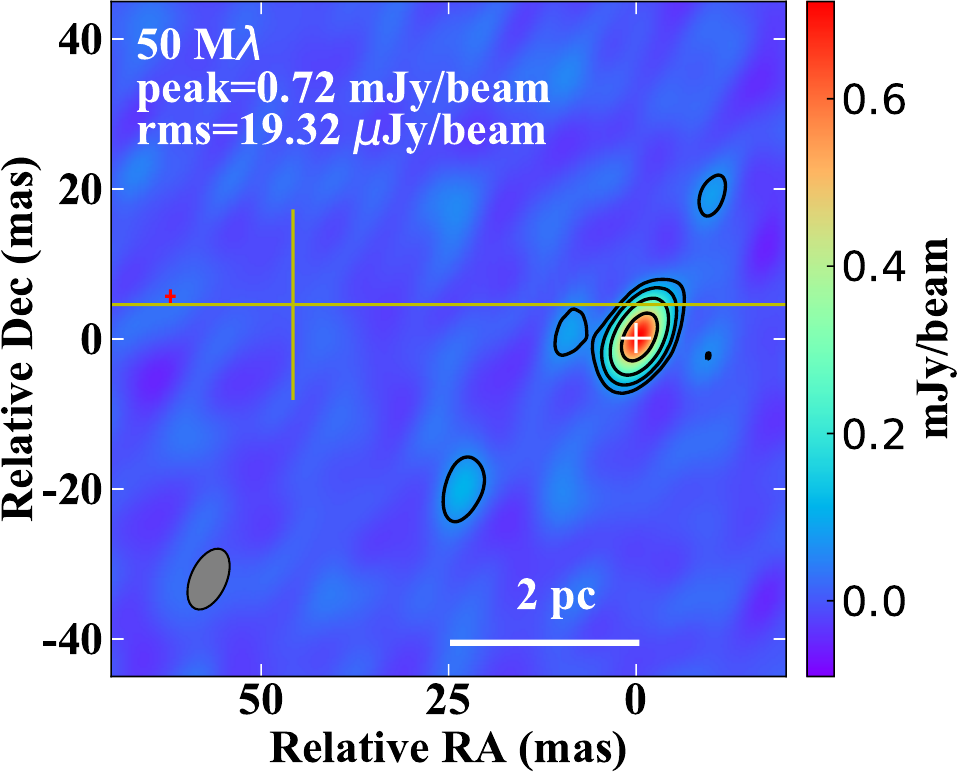}
    \includegraphics[height=3.5cm]{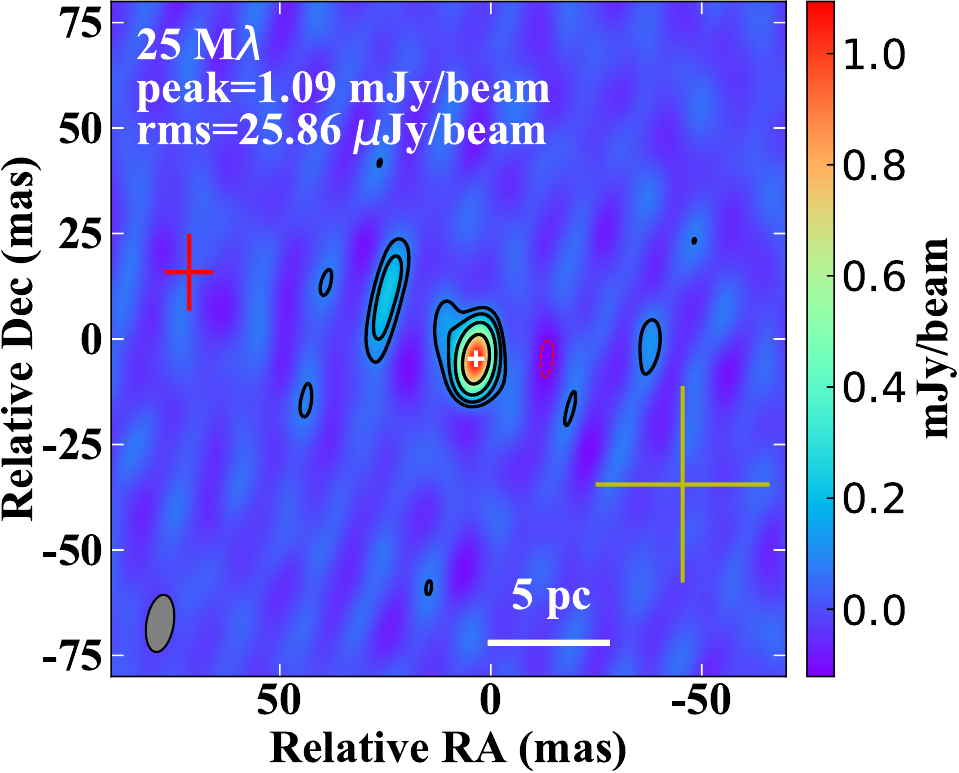}
    \includegraphics[height=3.5cm]{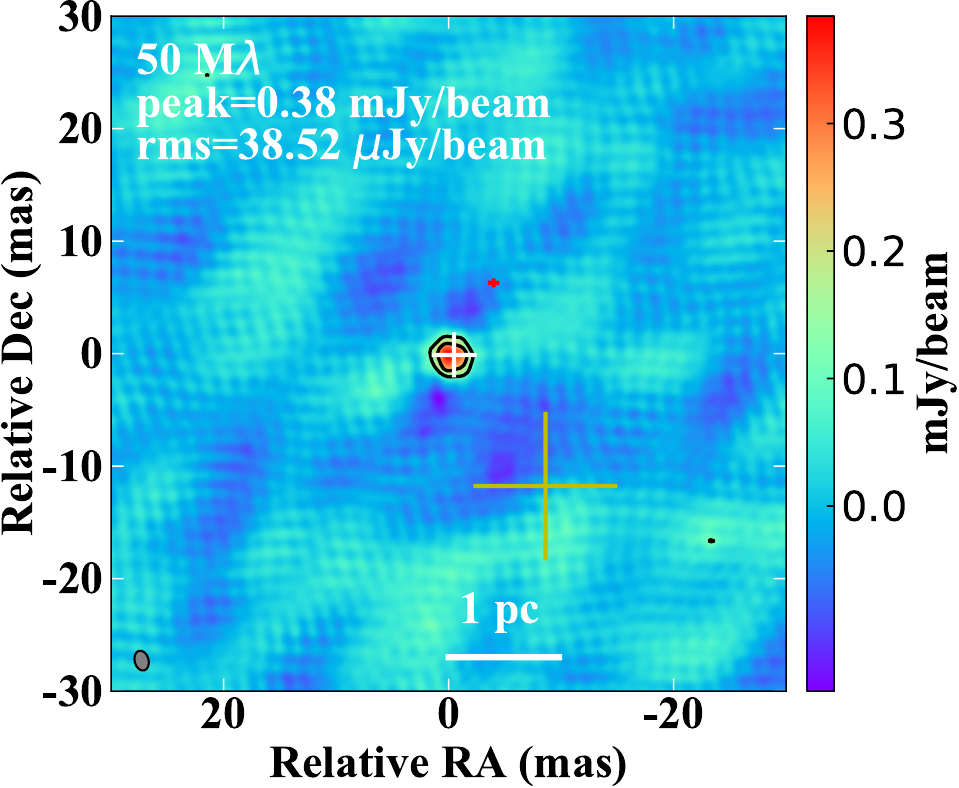}
    \\
    \caption{VLBI images of 23 detected LLAGNs at 5 GHz. For each source, two panels are shown. The upper panel shows the full-resolution map, while the lower panel shows the low-resolution map with uv-tapered scale.  The source name, peak intensity, and rms noise level for each source are shown in the upper left corner of each full-resolution map. The uv-tapered scale, peak intensity, and rms noise level for each source are also shown in the upper left corner of each low-resolution map. The lowest contour in all the images is at 3 $\sigma$ level and increases by factors of $-1$, 1, 2, 4, ..., 64. The restoring beam is presented as an ellipse on the bottom left corner of each map. The white cross indicates the VLBI peak position, while the yellow and red crosses indicate the optical positions reported by Gaia astrometry \citep{2020yCat.1350....0G} and Pan-STARRS1 \citep{2016arXiv161205560C}. The sizes of yellow and red crosses represent the 1 $\sigma$ uncertainty. As the uncertainties of VLBI positions are too small, we simply used the cross sizes of 2 mas for white crosses and the size has no special meaning. The image parameters are presented in Table \ref{tab:image}.}
    \label{fig:my_label}
\end{figure}
\addtocounter{figure}{-1}
\begin{figure}
\centering
    \includegraphics[height=3.5cm]{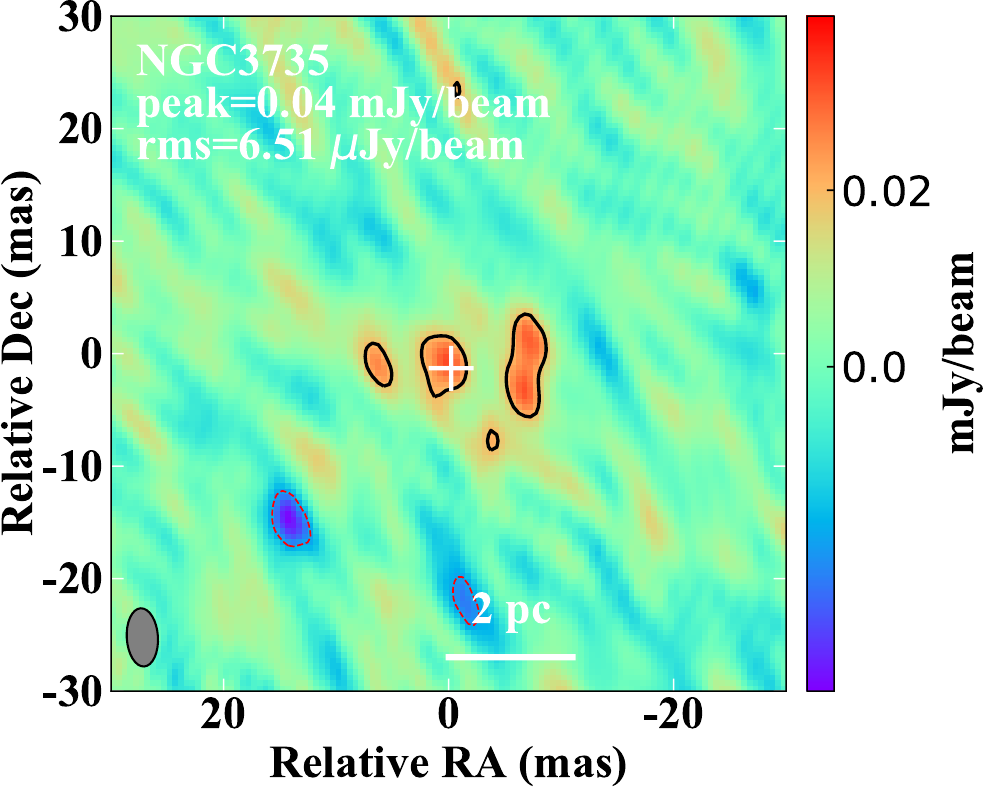}
    \includegraphics[height=3.5cm]{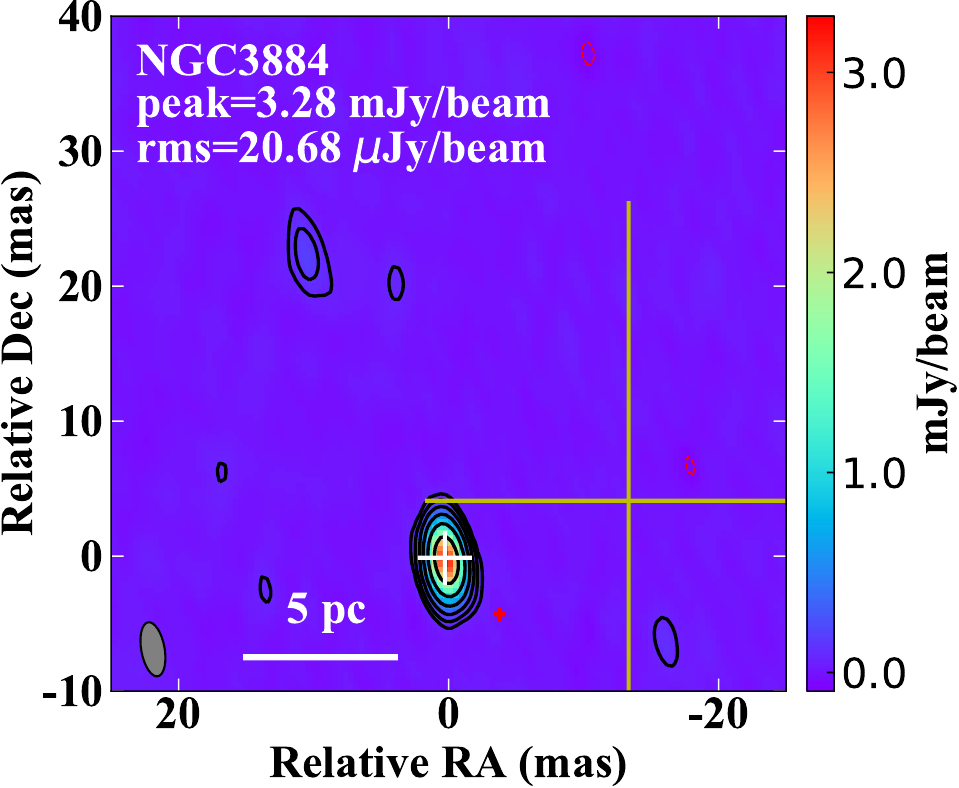}
    \includegraphics[height=3.5cm]{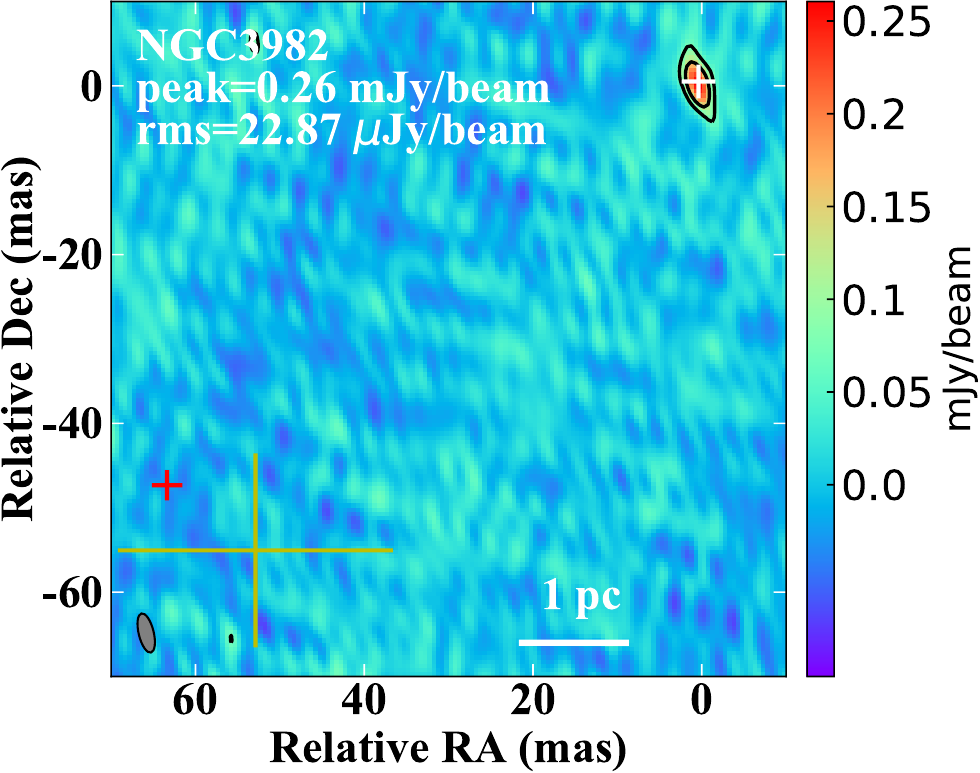}
    \includegraphics[height=3.5cm]{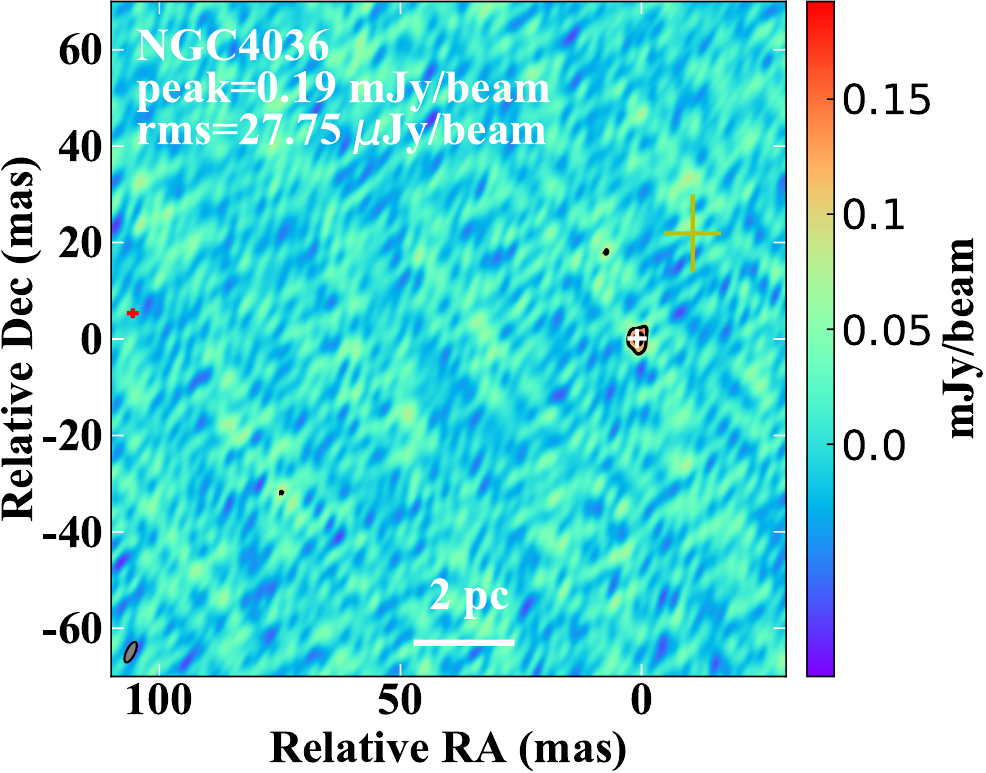}
    \\
    \includegraphics[height=3.5cm]{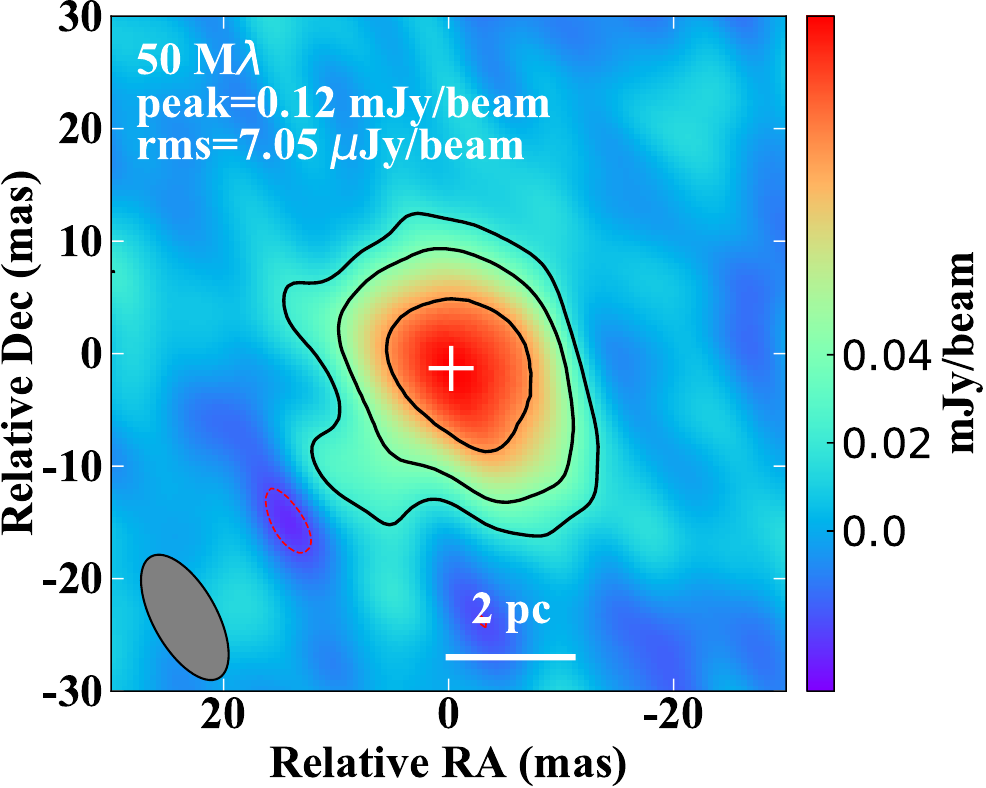}
    \includegraphics[height=3.5cm]{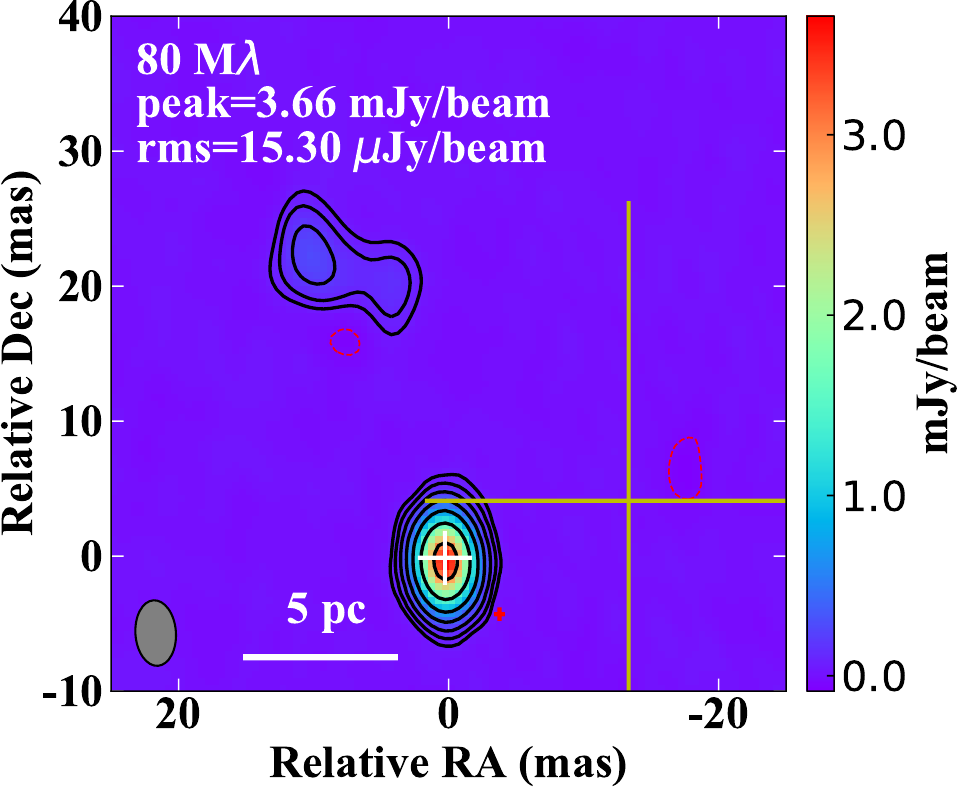}
    \includegraphics[height=3.5cm]{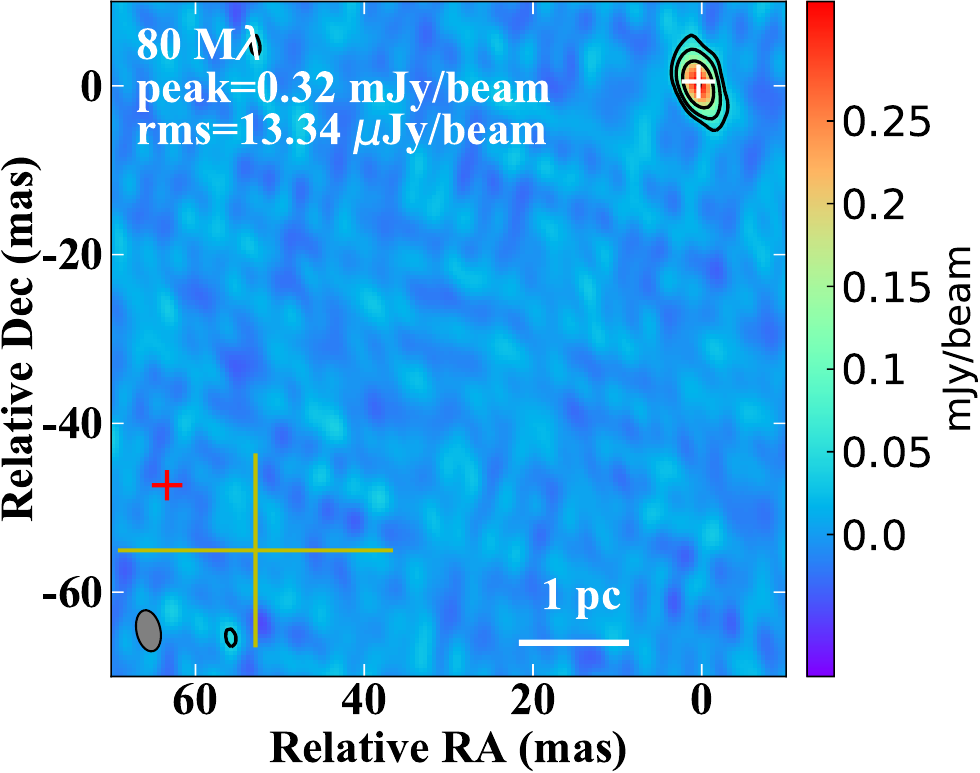}
    \includegraphics[height=3.5cm]{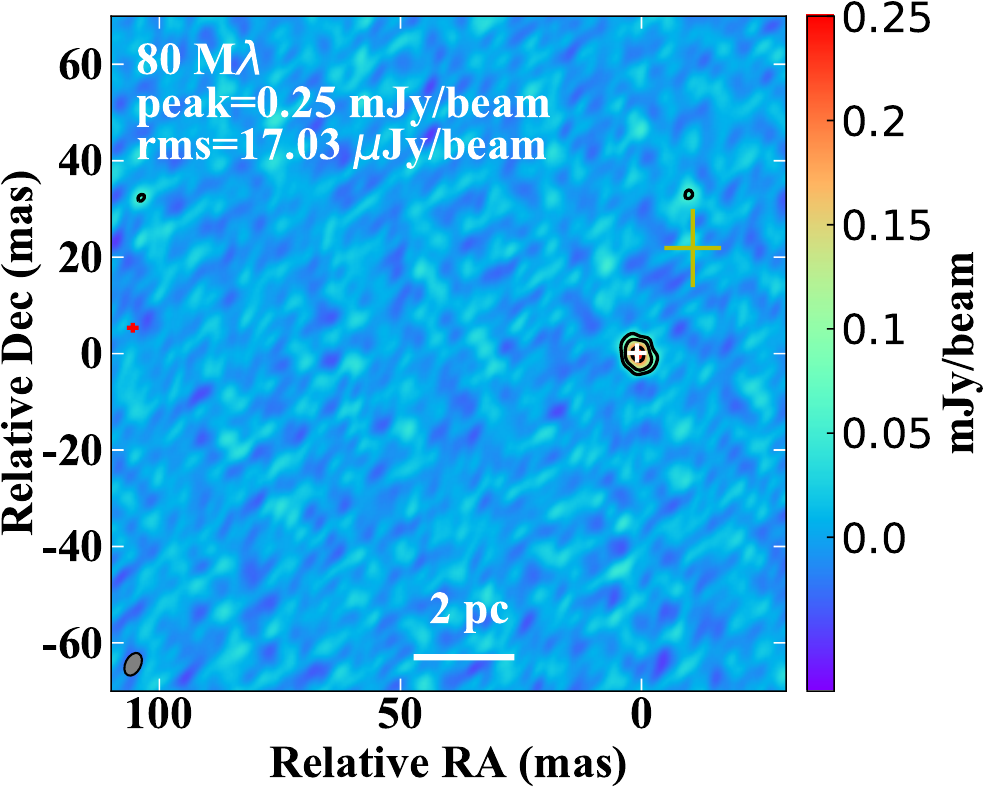}
    \\
    \includegraphics[height=3.5cm]{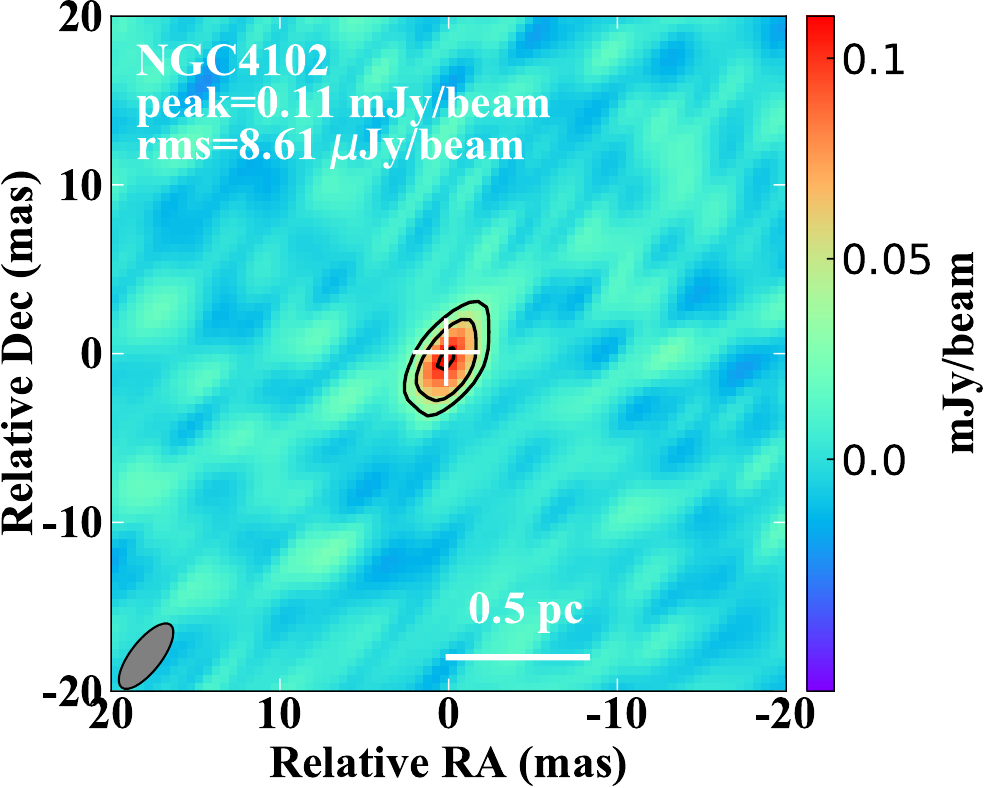}
    \includegraphics[height=3.5cm]{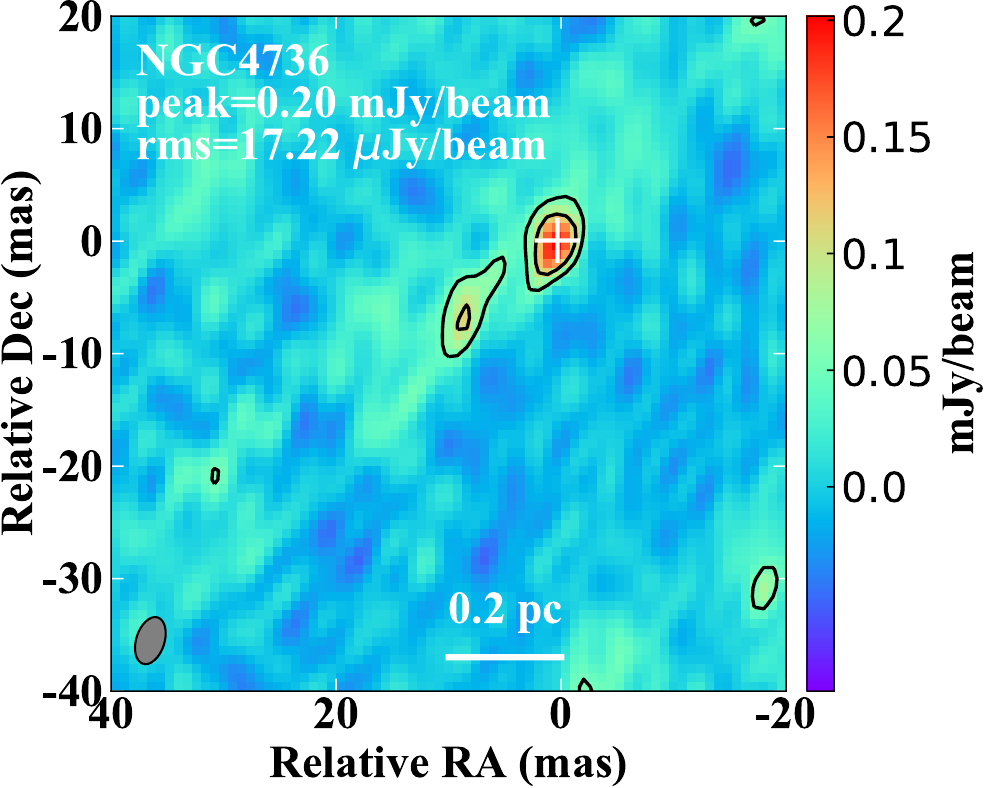}
    \includegraphics[height=3.5cm]{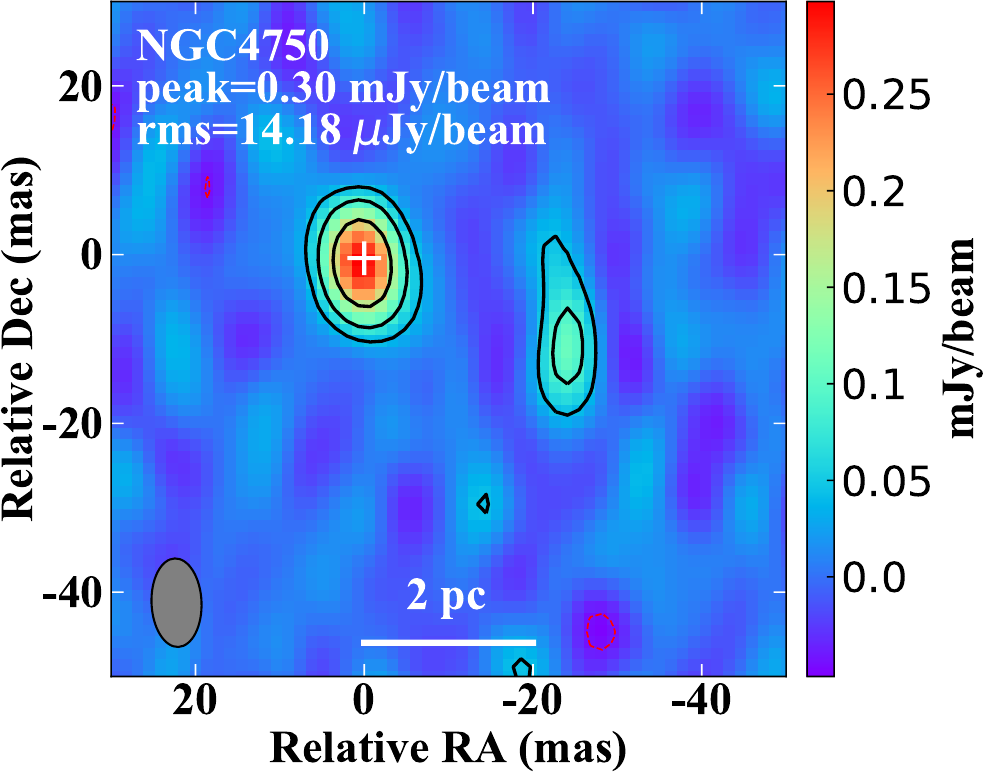}
    \includegraphics[height=3.5cm]{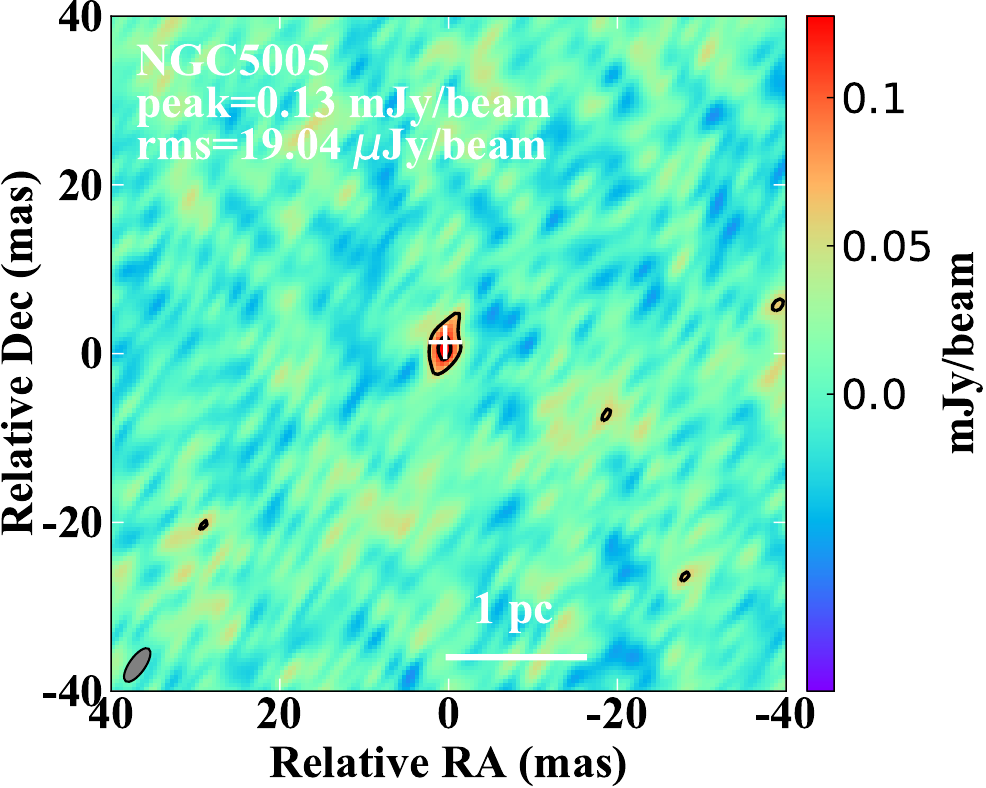}
    \\
    \includegraphics[height=3.5cm]{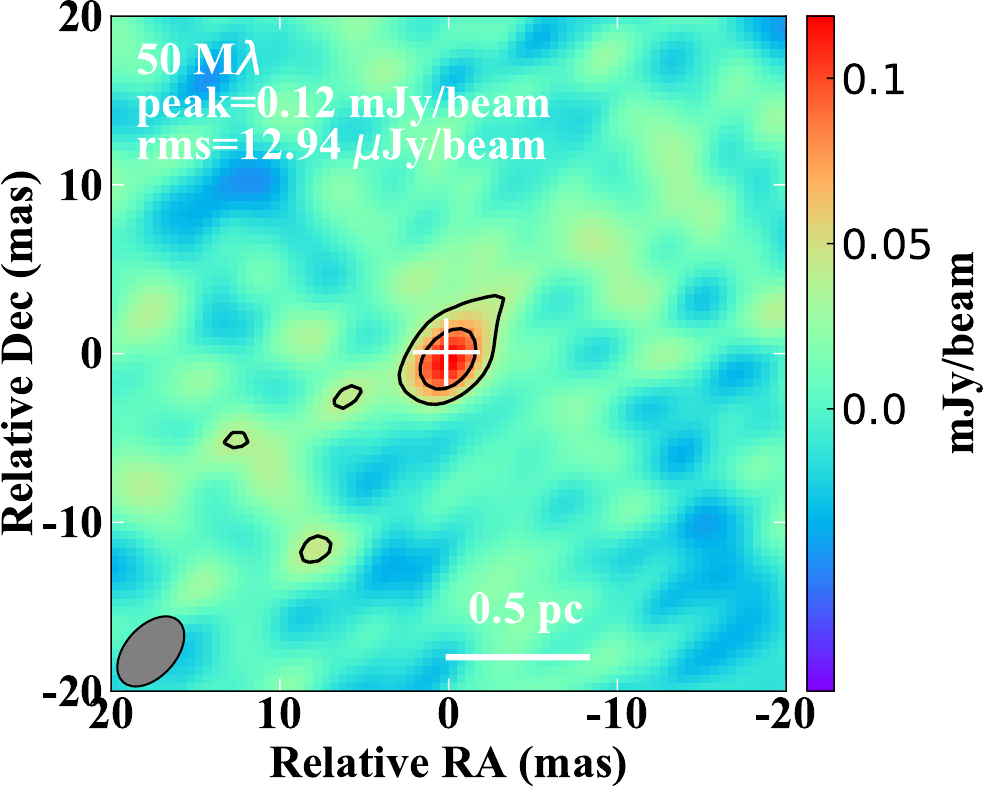}
    \includegraphics[height=3.5cm]{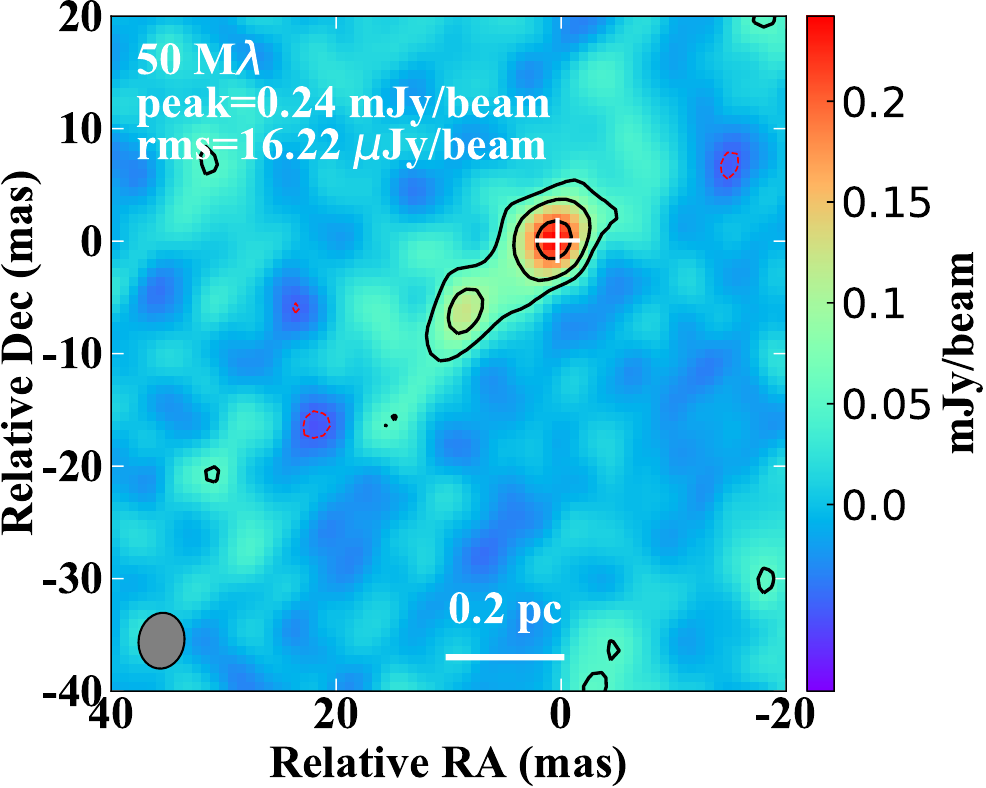}
    \includegraphics[height=3.5cm]{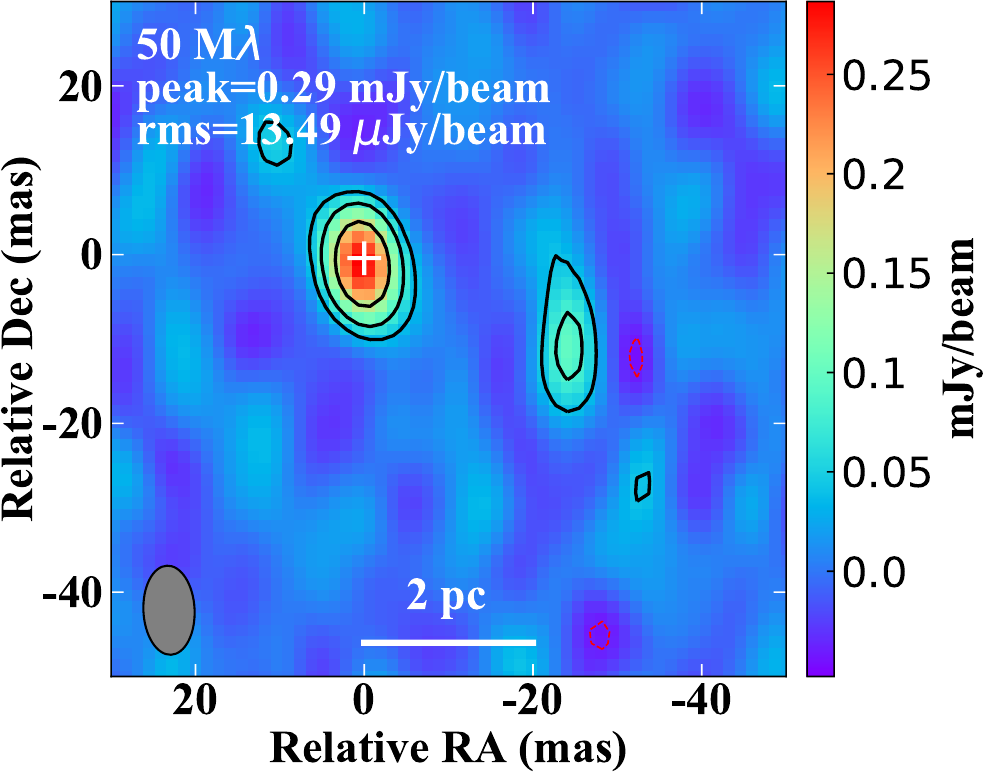}
    \includegraphics[height=3.5cm]{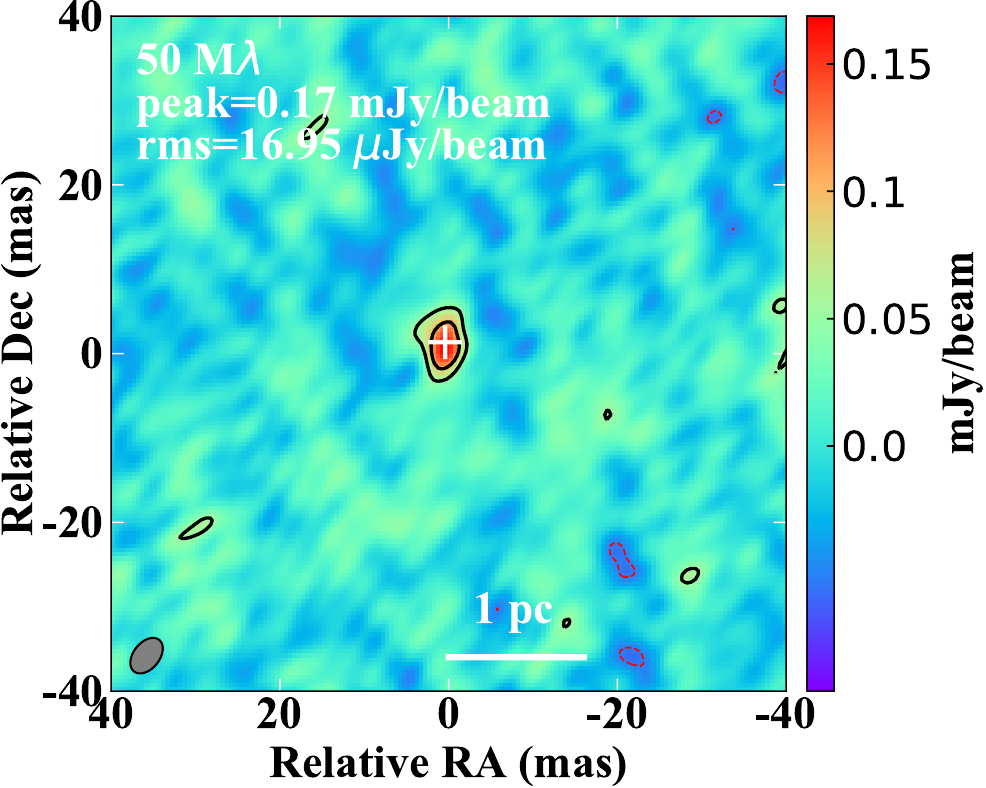}
    \\
    \includegraphics[height=3.5cm]{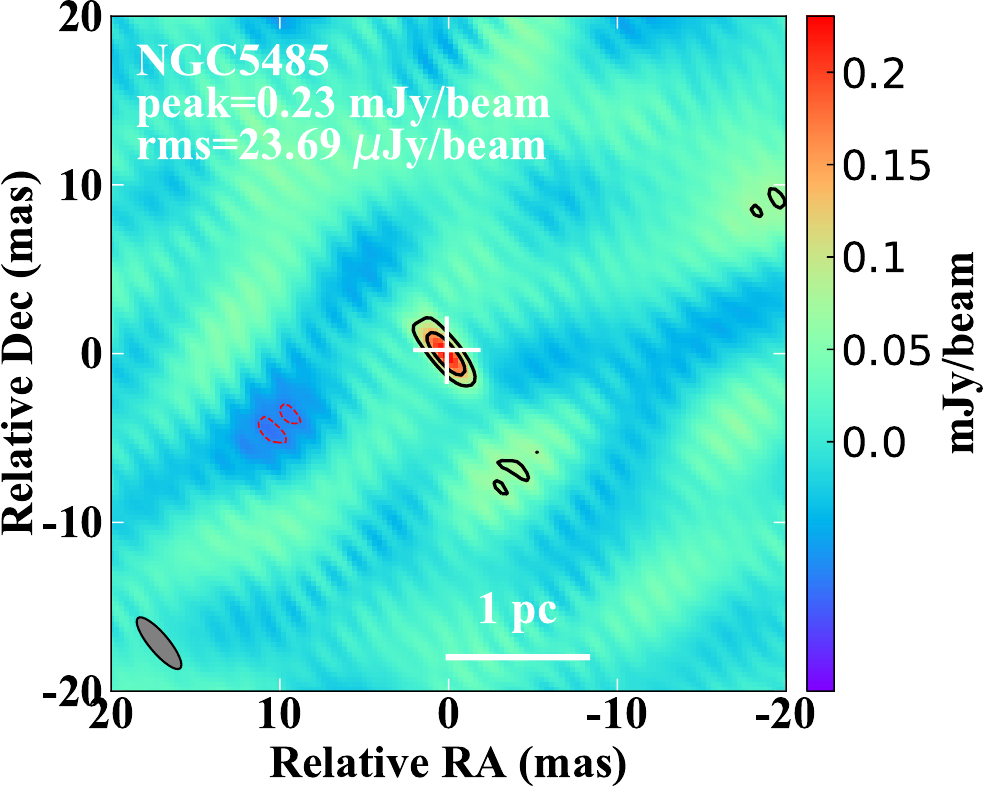}
    \includegraphics[height=3.5cm]{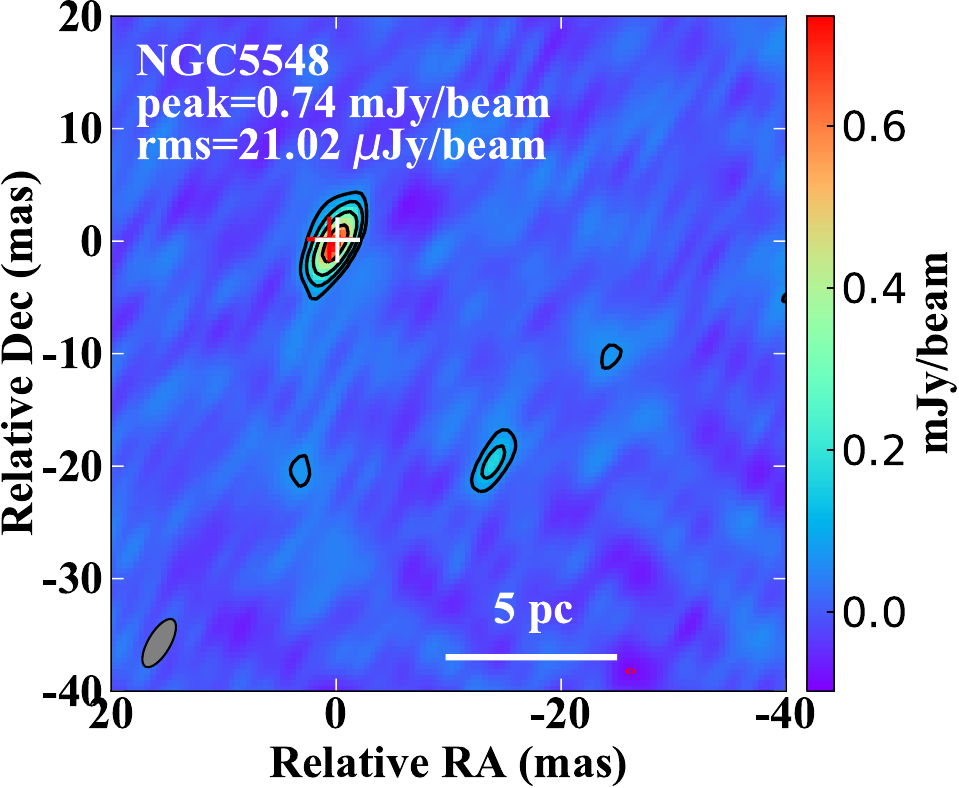}
    \includegraphics[height=3.5cm]{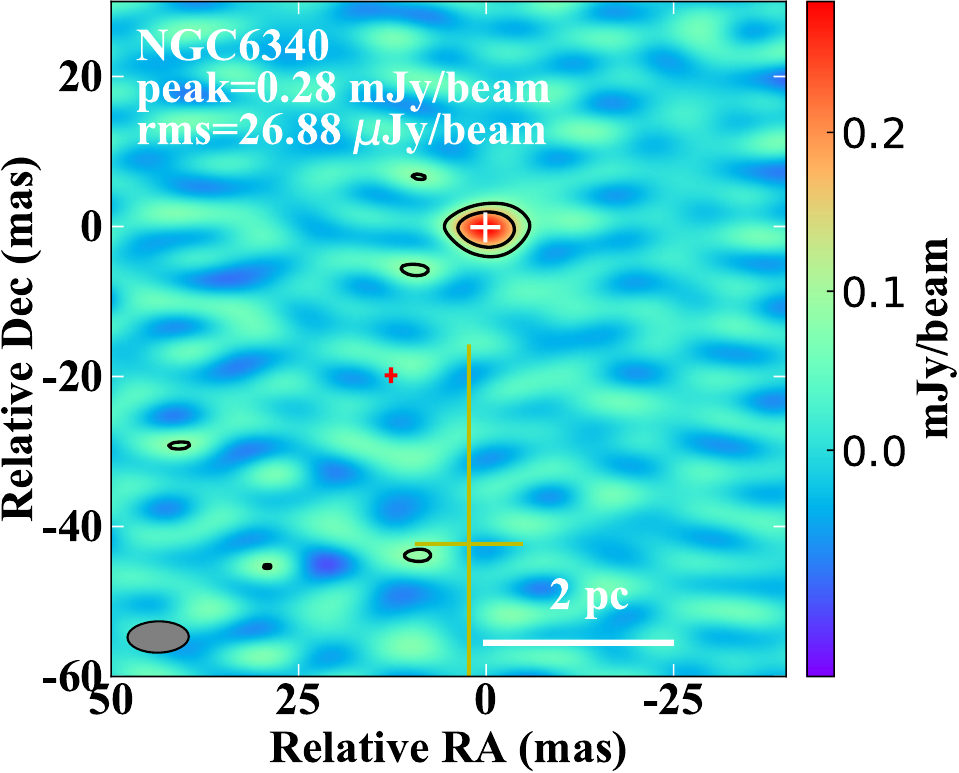}
    \\
    \includegraphics[height=3.5cm]{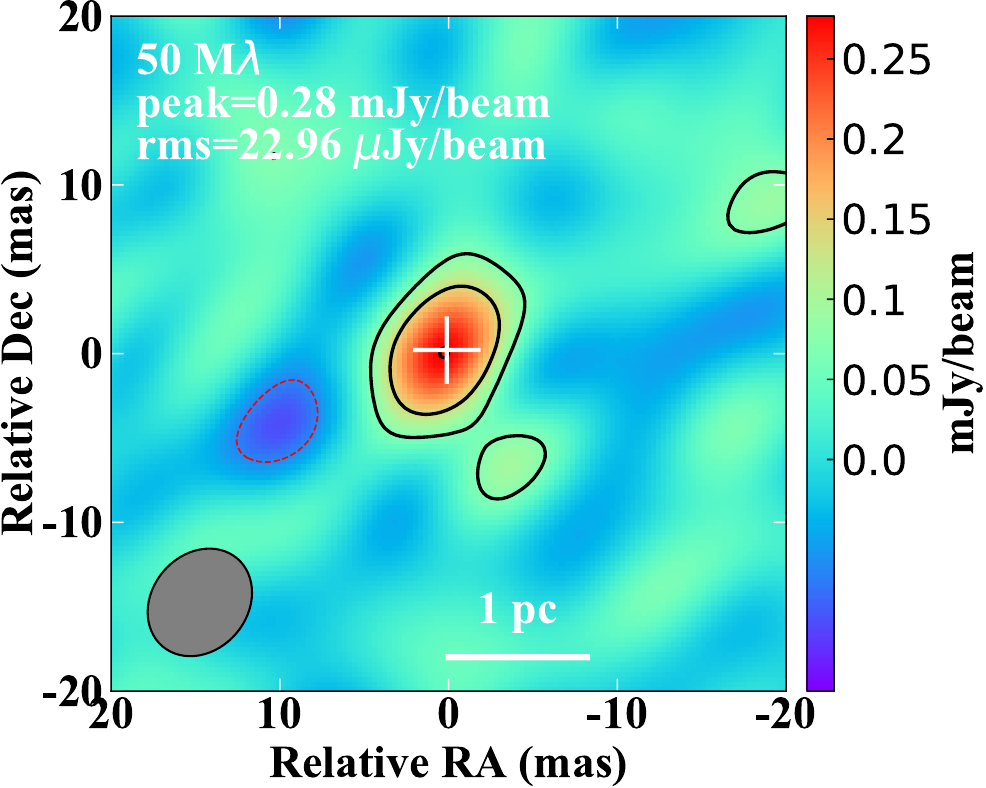}
    \includegraphics[height=3.5cm]{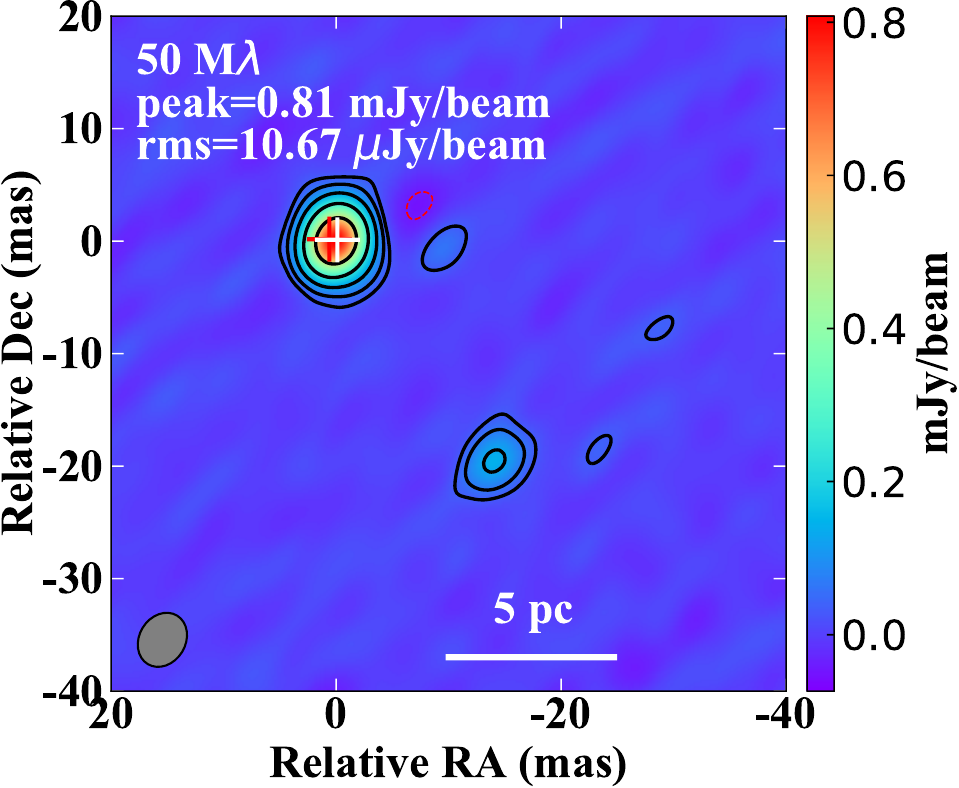}
    \includegraphics[height=3.5cm]{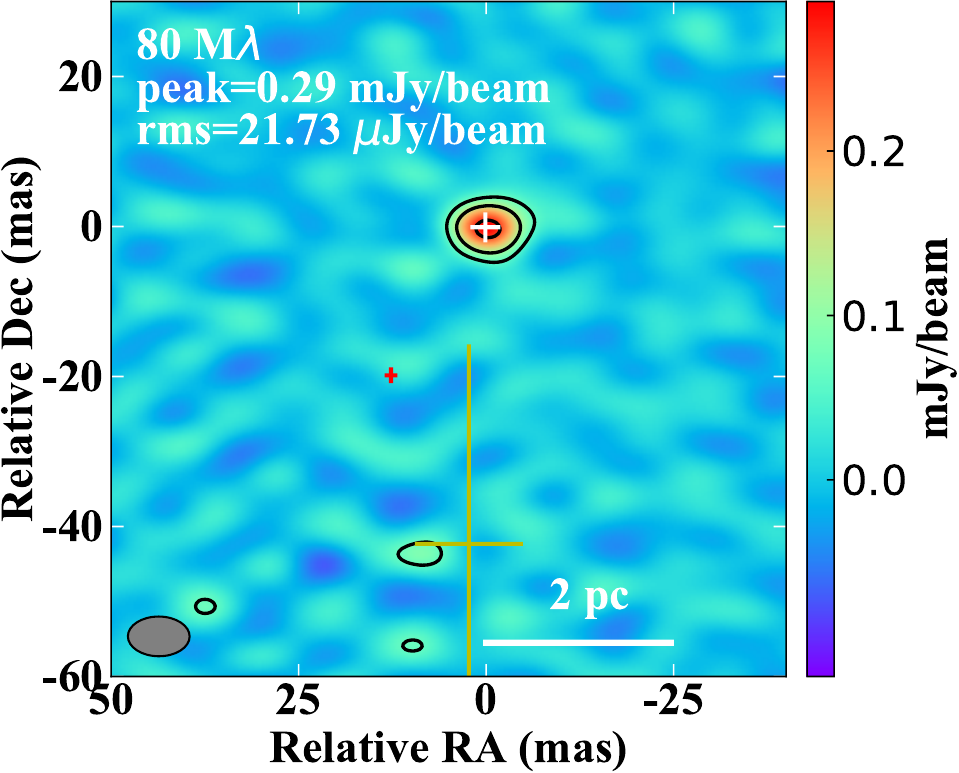}
\caption{Continued}
\end{figure}

\begin{figure}
    \centering
    \includegraphics[height=3.8cm]{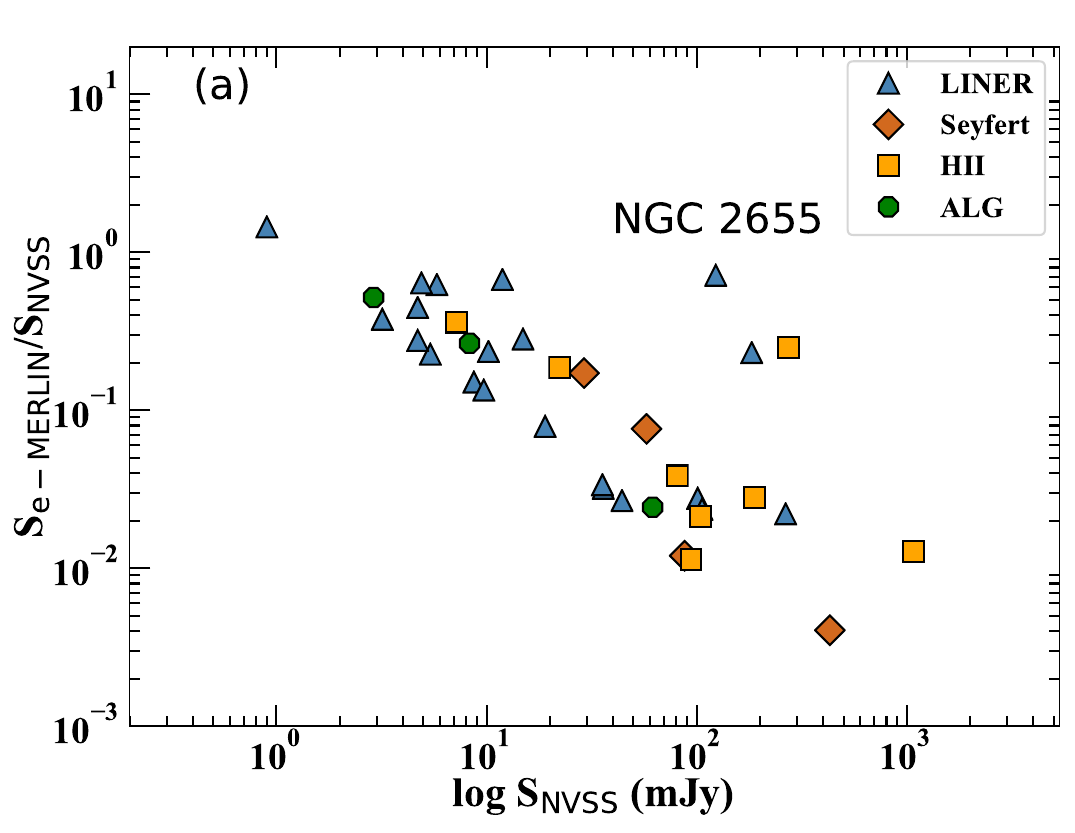}
    \includegraphics[height=3.8cm]{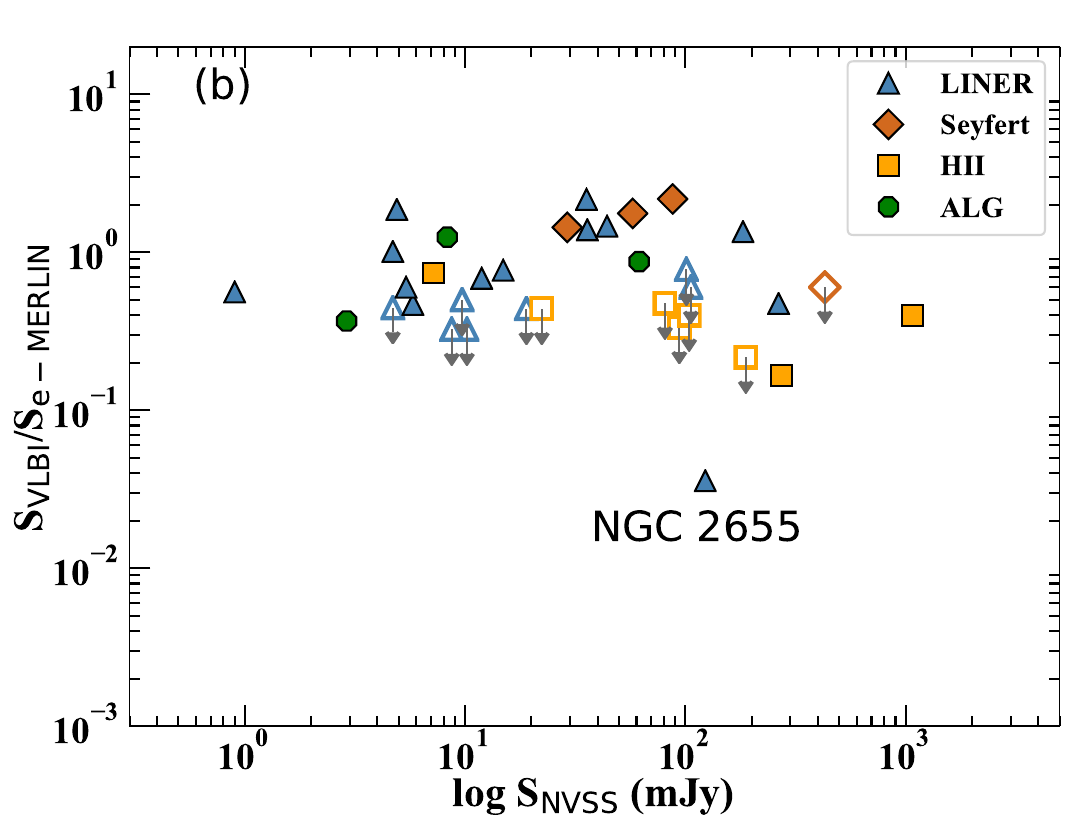}
    \includegraphics[height=3.8cm]{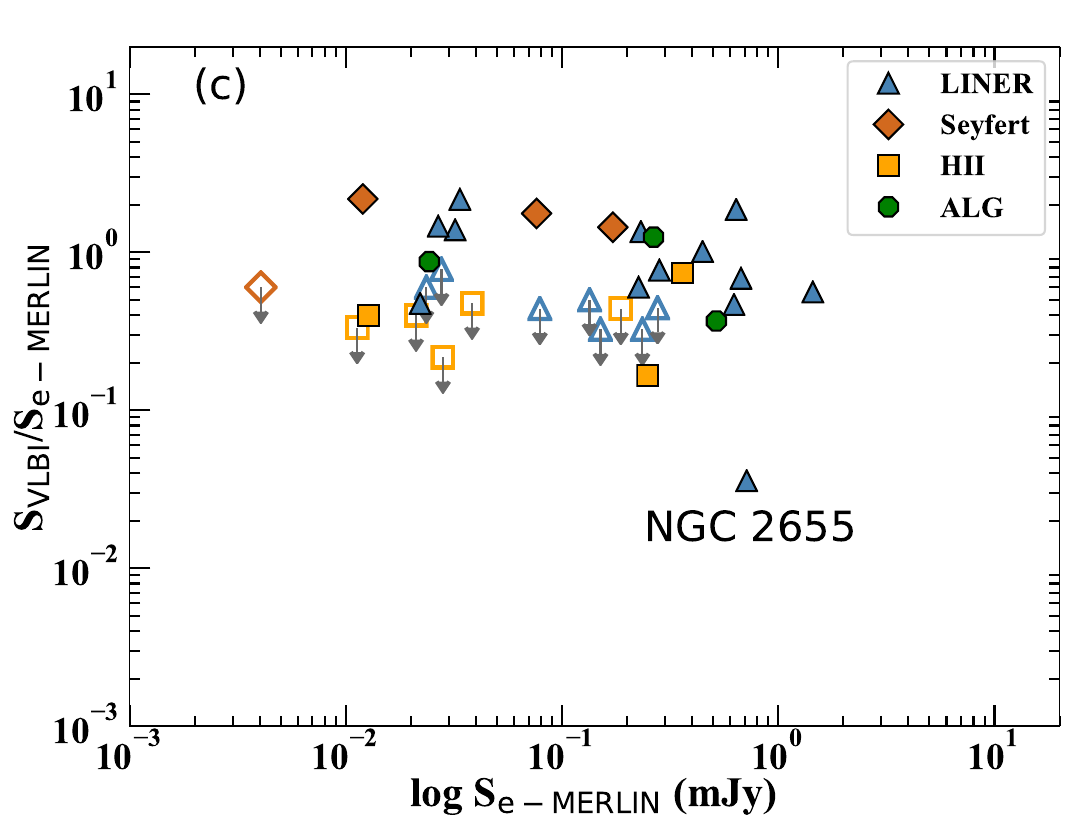}
    \caption{Compactness parameter for all the sources, divided per optical class (color and symbol coded as in the legend). (a) The logarithm of the compactness parameter on arcsecond scales versus the logarithm of the NVSS total flux density. (b) The logarithm of the compactness parameter on mas scales versus the logarithm of the NVSS total flux density. (c) The logarithm of the compactness parameter on arcsecond scales versus the logarithm of the compactness parameter on mas scales.}
    \label{fig:compact}
\end{figure}

\begin{figure}
    \centering
    \includegraphics[height=8cm]{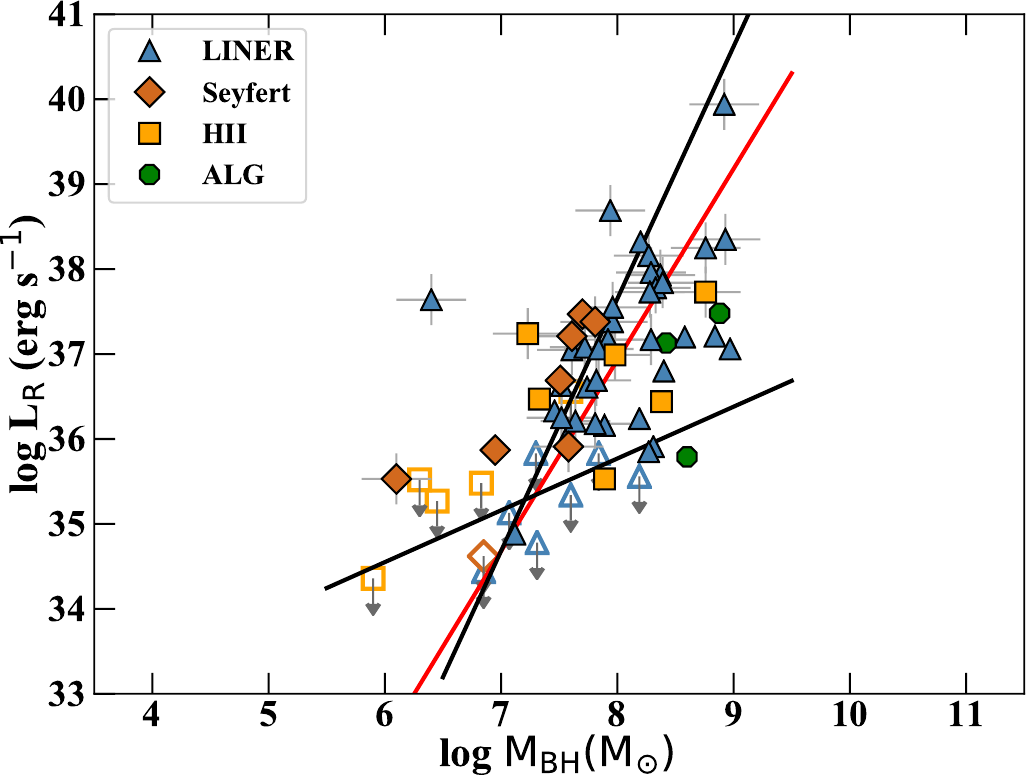}
    \caption{Distribution of the VLBI 5 GHz total radio luminosity as a function of BH mass, divided per optical class (color and symbol coded as in the legend). The data points with gray crosses represent the data from the archive. The red line represents the linear fitting for the whole sample. The black lines represent the two 'broken' scaling relations from \citet{2018MNRAS.476.3478B,2021MNRAS.508.2019B}: L$\rm_{\rm R} \propto L_{\rm BH}^{0.61\pm0.33}$ for non-jetted star-forming galaxies and L$\rm_{\rm R} \propto L_{\rm BH}^{1.65\pm0.25}$ for active galaxies.}
    \label{fig:BH}
\end{figure}

\begin{figure}
    \centering
    \includegraphics[height=5cm]{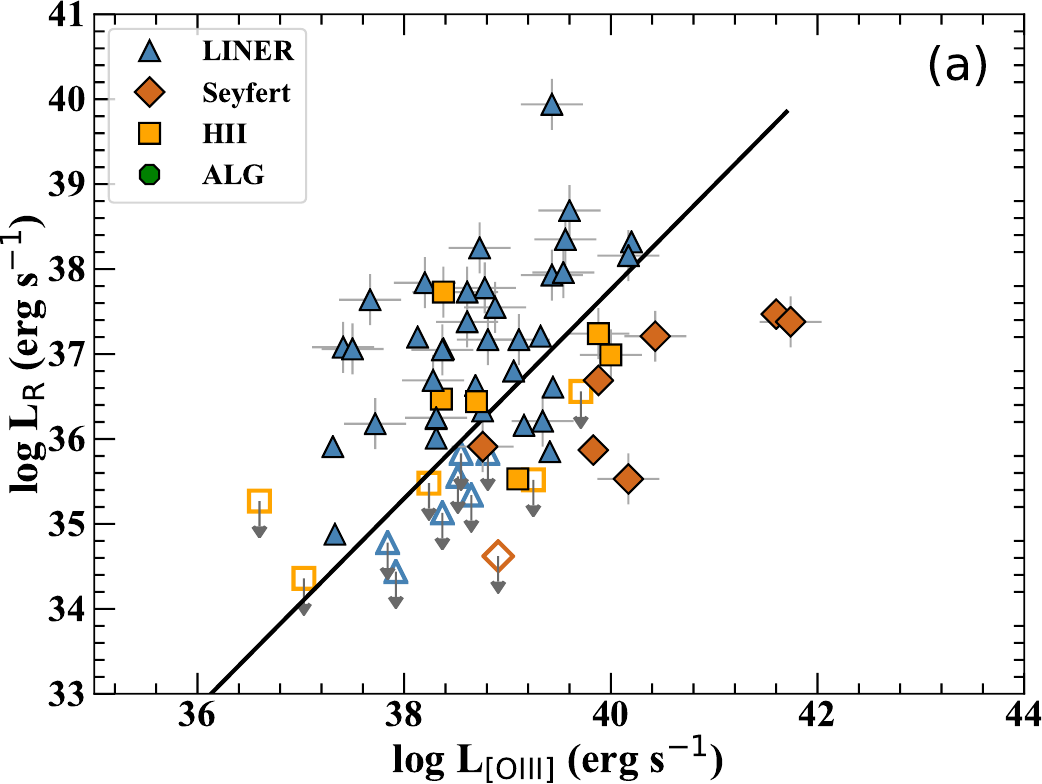}
    \includegraphics[height=5cm]{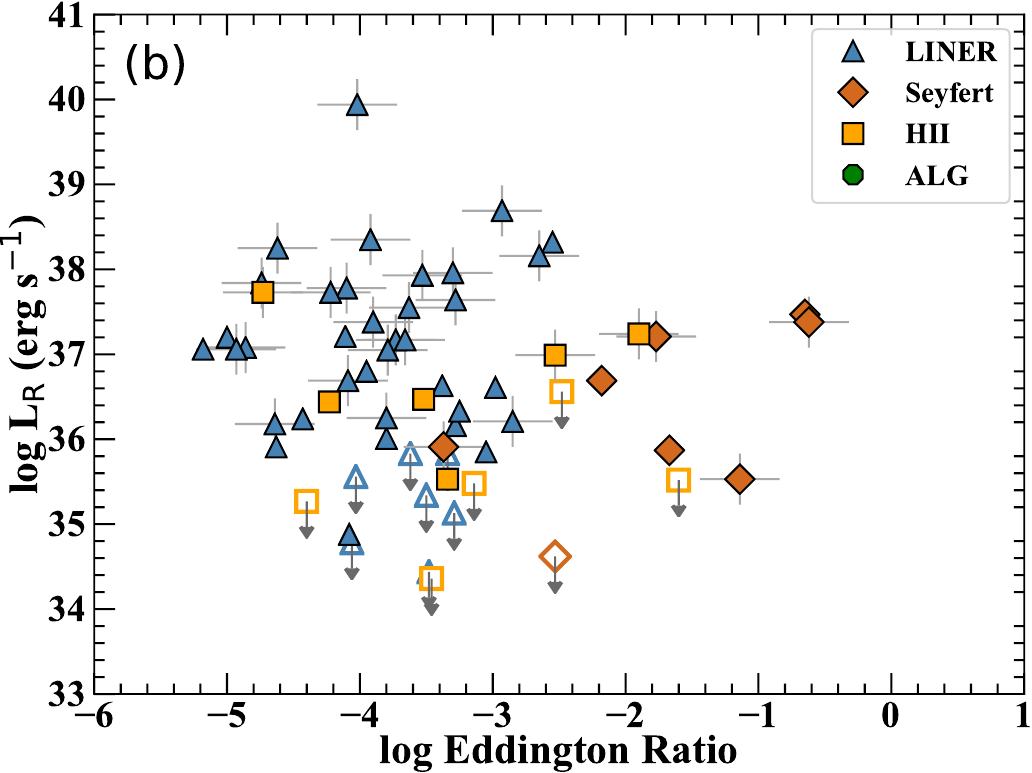}
    \caption{(a) Distribution of the VLBI 5 GHz total radio luminosity as a function of [O III] luminosity, divided per optical class (color and symbol coded as in the legend). The data points with gray crosses represent the data from the archive. The black lines represent the linear fitting for the whole sample. (b) Distribution of the VLBI 5 GHz total radio luminosity as a function of Eddington ratio, divided per optical class (color and symbol coded as in the legend). The data points with gray crosses represent the data from the archive.}
    \label{fig:OIII}
\end{figure}

\begin{figure}
    \centering
    \includegraphics[height=8cm]{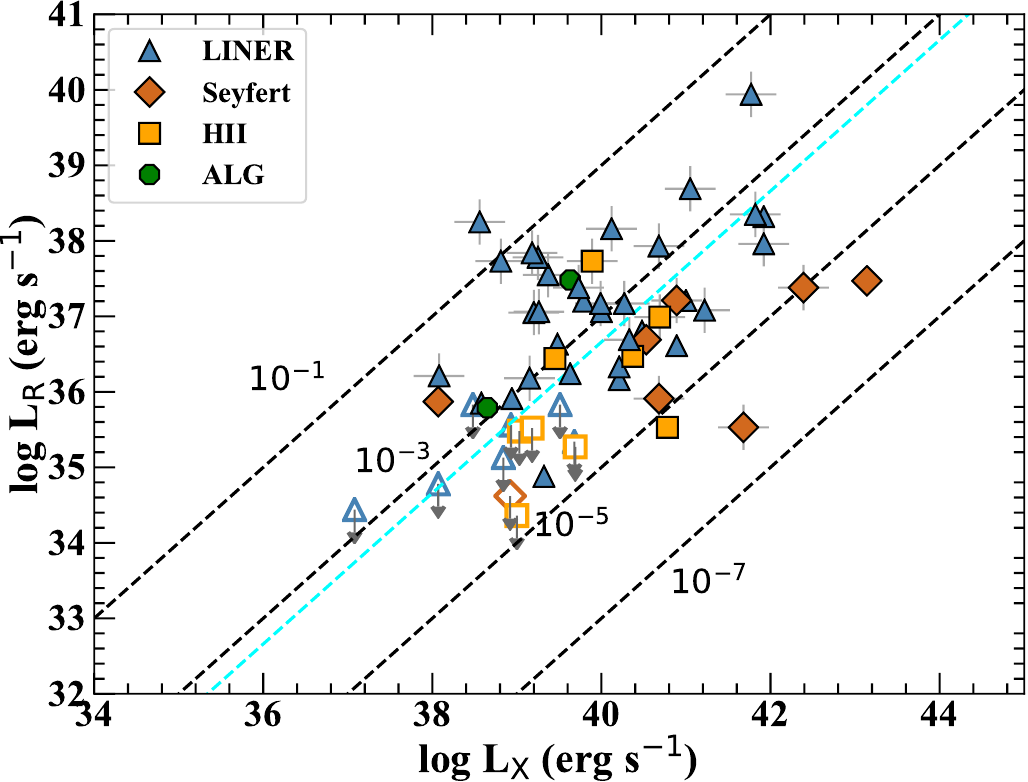}
    \caption{Distribution of the VLBI 5 GHz total radio luminosity as a function of X-ray 2-10 keV luminosity, divided per optical class (color and symbol coded as in the legend). The data points with gray crosses represent the data from the archive. The black dashed lines correspond to L$\rm_{R}$/L$\rm_{X}$ of 10$\rm^{-1}$, 10$\rm^{-3}$, 10$\rm^{-5}$, and 10$\rm^{-7}$. The cyan dashed line is the mean L$\rm_{R}$/L$\rm_{X}$ value of the sample.} 
    \label{fig:Xray}
\end{figure}

\begin{figure}
    \centering
    \includegraphics[height=8cm]{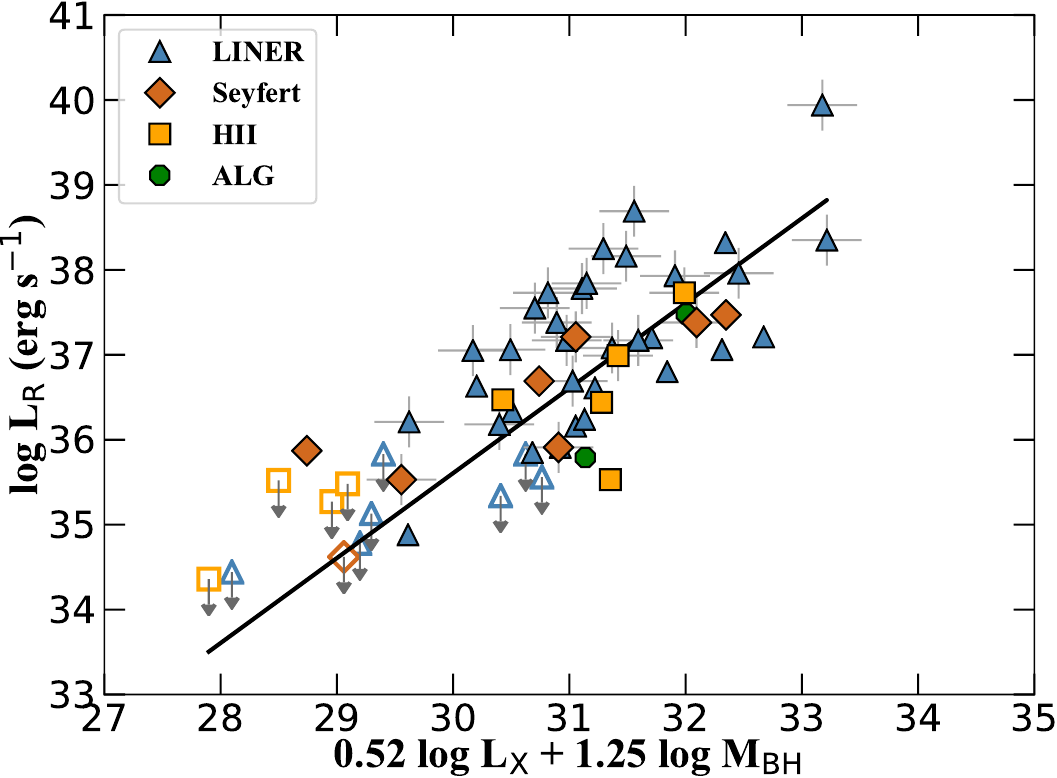}
    \caption{The fundamental plane of BH activity using VLBI 5 GHz luminosity, divided per optical class (color and symbol coded as in the legend). The data points with gray crosses represent the data from the archive. The black line represents the best fitting for the whole sample.} 
    \label{fig:FMP}
\end{figure}

\begin{deluxetable}{cccccccccc}
\tablecolumns{10}
\tabletypesize{\small}
\tablewidth{0pt}
\tablecaption{
Basic properties of the sample
\label{tab:information}}
\tablehead{
\colhead{Name}  & \colhead {R.A.} & \colhead {DEC.} & \colhead {Type} & \colhead {Morphology} & \colhead {M$\rm _{BH}$} & \colhead{L$\rm _{[OIII]}$} & \colhead{L$\rm _{X}$} & \colhead{Morphology} & \colhead{D}   \\
\colhead{}   & \colhead {(J2000)}   & \colhead {(J2000)}      & \colhead {}   & \colhead {(e-MERLIN)} & \colhead {(M$_{\odot}$)}   &  \colhead{(erg s$^{-1}$)} & \colhead{(erg s$^{-1}$)} & \colhead{(VLBI)} & \colhead{(Mpc)}  \\
\colhead{(1)}  & \colhead{(2)}     & \colhead{(3)} & \colhead{(4)}      & \colhead{(5)}      & \colhead{(6)}      & \colhead{(7)}     & \colhead{(8)} & \colhead{(9)} & \colhead{(10)} }
\startdata
NGC 410  & 01 10 58.901 & $+$33 09 06.98 & LINER & core-jet & 8.84 & 39.32 & 41.00$\pm$0.28 & core-jet  & 70.6 \\ 
NGC 507  & 01 23 39.863 & $+$33 15 23.00 & ALG   & core-jet & 8.88 & ...   & 39.63$\pm$0.02 & core      & 65.7 \\
NGC 777  & 02 00 14.908 & $+$31 25 45.84 & LINER & core+components & 8.97 & 38.38 & 40.77   & core-jet  & 66.5 \\
NGC 1161 & 03 01 14.173 & $+$44 53 50.66 & LINER & core     & 8.58 & 38.13 & 39.79$\pm$0.08 & core-jet  & 25.9 \\
NGC 2146 & 06 18 37.589 & $+$78 21 24.25 & HII   & core-jet	& 7.33 & 38.36 & 40.37$\pm$0.30 & twin jets & 17.2 \\
NGC 2300 & 07 32 19.945 & $+$85 42 32.41 & ALG   & core-jet	& 8.6  & ...   & 38.65$\pm$0.19 & core      & 31.0 \\
NGC 2342 & 07 09 18.069 & $+$20 38 09.59 & HII   & complex	& 7.6  & 39.71 & 41.26$^{a}$    &           & 69.5 \\
NGC 2655 & 08 55 37.950 & $+$78 13 23.63 & LINER &S shape jet& 7.74& 39.44 & 40.89$^{b}$    & core-jet  & 24.4 \\
NGC 2681 & 08 53 32.718 & $+$51 18 49.15 & LINER & twin jets& 7.07 & 38.37 & 38.84$\pm$0.12 &           & 13.3 \\
NGC 2841 & 09 22 02.652 & $+$50 58 36.03 & LINER & twin jets& 8.31 & 37.31 & 38.94$\pm$0.10 & core      & 12.0 \\
NGC 2985 & 09 50 22.184 & $+$72 16 44.21 & LINER & core-jet & 7.52 & 38.69 & 39.48$\pm$0.28 & core-jet  & 22.4 \\
NGC 3245 & 10 27 18.386 & $+$28 30 26.63 & HII   &	complex	& 8.38 & 38.70 & 39.45$\pm$0.27 & core      & 22.2 \\
NGC 3348 & 10 47 10.032 & $+$72 50 22.67 & ALG   &	triple source & 8.42 & ...   & ...      & core-jet  & 37.8 \\
NGC 3414 & 10 51 16.211 & $+$27 58 30.28 & LINER & twin jets& 8.40 & 39.06 & 40.48$\pm$0.13 & core      & 23.9 \\
NGC 3729 & 11 33 49.321 & $+$53 07 31.79 & HII   &	core	& 6.45 & 36.60 & 39.69$\pm$0.09 &           & 17.0 \\
NGC 3735 & 11 35 57.216 & $+$70 32 07.81 & Seyfert& core    & 7.51 & 39.88 & 40.53$^{b}$    & core      & 41.0 \\
NGC 3884 & 11 46 12.182 & $+$20 23 29.92 & LINER &	core    & 8.20 & 40.20 & 41.92$\pm$0.14 & core-jet  & 91.6 \\
NGC 3898 & 11 49 15.236 & $+$56 05 04.38 & LINER & 	core    & 8.19 & 38.52 & 38.93$\pm$0.17 &           & 21.9 \\
NGC 3982 & 11 56 28.138 & $+$55 07 30.84 & Seyfert&	core    & 6.95 & 39.83 & 38.07$\pm$0.22 & core      & 17.0 \\
NGC 4036 & 12 01 26.891 & $+$61 53 44.52 & LINER & twin jets& 7.89 & 39.16 & 40.21$\pm$0.08 & core      & 24.6 \\
NGC 4041 & 12 02 12.279 & $+$62 08 12.94 & HII   & complex  & 6.83 & 38.24 & 39.03$\pm$0.44 &           & 22.7 \\
NGC 4102 & 12 06 23.077 & $+$52 42 39.76 & HII &star forming& 7.89 & 39.10 & 40.78$\pm$0.03 & core      & 17.0 \\
NGC 4111 & 12 07 03.136 & $+$43 03 56.45 & LINER & 	core    & 7.60 & 38.65 & 39.68$\pm$0.02 &           & 17.0 \\
NGC 4736 & 12 50 53.018 & $+$41 07 13.53 & LINER & core+components&7.12&37.33& 39.32$\pm$0.04 & core-jet& 4.3  \\
NGC 4750 & 12 50 07.315 & $+$72 52 28.64 & LINER &	core    & 7.46 & 38.76 & 40.21$\pm$0.26 & core-jet  & 26.1 \\
NGC 4826 & 12 56 43.655 & $+$21 40 58.58 & LINER & core+components & 6.85 & 37.92 & 37.08$\pm$0.07 &    & 4.1  \\
NGC 5005 & 13 10 56.260 & $+$37 03 32.74 & LINER & twin jets& 8.27 & 39.41 & 38.58$\pm$0.27 & core      & 21.3 \\
NGC 5194 & 13 29 52.708 & $+$47 11 42.80 & Seyfert&twin jets& 6.85 & 38.91 & 38.92$\pm$0.02 &           & 7.7  \\
NGC 5195 & 13 29 59.535 & $+$47 15 58.37 & LINER & twin jets& 7.31 & 37.84 & 38.07$\pm$0.18 &           & 9.3  \\
NGC 5377 & 13 56 16.631 & $+$47 14 09.16 & LINER & twin jets& 7.84 & 38.81 & 39.51$\pm$0.27 &           & 31.0 \\
NGC 5448 & 14 02 50.035 & $+$49 10 21.75 & LINER &twin jets & 7.30 & 38.55 & 38.48$\pm$0.35 &           & 32.6 \\
NGC 5485 & 14 07 11.345 & $+$55 00 05.97 & LINER &   core   & 8.19 & 38.31 & 39.63$\pm$0.13 & core      & 32.8 \\
NGC 5548 & 14 17 59.540 & $+$25 08 12.61 & Seyfert& core-jet& 7.70 & 41.60 & 43.14$\pm$0.01 & core-jet  & 67.0 \\
NGC 6217 & 16 32 39.267 & $+$78 11 53.35 & HII &core+components& 6.30 & 39.25 & 39.25$^{c}$    &        & 23.9 \\
NGC 6340 & 17 10 24.827 & $+$72 18 15.85 & LINER &  core    & 7.56 & 38.31 & ...            & core      & 22.0 \\
NGC 6946 & 20 34 52.299 & $+$60 09 14.20 & HII  &jet+complex& 5.90 & 37.03 & 39.18          &           & 5.5  \\
\enddata
\tablecomments{(1) Source name; (2)-(3) Right ascension and Declination position (J2000.0) from \citet{2018MNRAS.476.3478B,2021MNRAS.500.4749B}; (4) Optical spectroscopic classification based on \citet{1997ApJS..112..315H}; (5) Radio morphology based on \citet{2018MNRAS.476.3478B,2021MNRAS.500.4749B}; (6) logarithm of the SMBH mass (M$_{\odot}$) determined based on \citet{2002ApJ...574..740T,2016ApJ...831..134V}; (7) logarithm of [OIII] luminosities from \citet{1997ApJS..112..315H}; (8) logarithm of X-ray luminosities in the 2-10 keV from \citet{2022MNRAS.510.4909W}, $^{a}$ the X-ray luminosity in the 0.3-10 keV from \citet{2005MNRAS.357..109J}, $^{b}$ the X-ray luminosity in the 2-10 keV from \citet{Akylas09}, $^{c}$ the X-ray luminosity in the 2-10 keV from \citet{2019MNRAS.486.4962W}; (9) Morphology on parsec-scales in our VLBI observations; (10) Distance (Mpc) from \citet{1997ApJS..112..315H}.}
\end{deluxetable}

\begin{deluxetable}{ccccc}
\tablecolumns{5}
\tabletypesize{\scriptsize}
\tablewidth{0pt}
\tablecaption{ Details of the VLBA and EVN observations. 
\label{tab:obs}} 
\tablehead{
\colhead{Code} & \colhead {$\nu$} & \colhead {Date} & \colhead {Target sources} & \colhead{Participating stations} \\
\colhead{} & \colhead {(GHz)} & \colhead {(yyyy mm dd)} & \colhead {} & \colhead {} \\
\colhead{(1)} & \colhead{(2)} & \colhead{(3)} & \colhead{(4)} & \colhead{(5)} }
\startdata
BA152A & 4.87 & 2022 03 08 & NGC 410, NGC 1161, NGC 2146                                & VLBA \\
BA152B & 4.87 & 2022 03 02 & NGC 2342, NGC 2655, NGC 2681, NGC 3884, NGC 3982           & VLBA \\
BA152C & 4.87 & 2022 03 07 & NGC 3729, NGC 4036, NGC 4102, NGC 5005, NGC 5548, NGC 6946 & VLBA \\
BA152D & 4.87 & 2022 03 16 & NGC 4041, NGC 4736, NGC 5194, NGC 6217                     & VLBA \\
\hline
EC082A & 4.93 & 2022 01 18 & NGC 507, NGC 777, NGC 2300, NGC 2985, NGC 3735 & JbWbEfMcNtT6O8TrYsSvZcBdHhIrCmJlKnPiDe \\
EC082B & 4.93 & 2022 02 16 & NGC 3348, NGC 6340                             & JbWbEfMcNtO8T6TrHhYsIrCmDaKnPiDe \\
EC082C & 4.93 & 2022 03 23 & NGC 4750, NGC 5195, NGC 5448                   & JbWbEfMcNtO8T6TrHhYsIbCmDaKnPiDe \\
EC082D & 4.93 & 2022 04 12 & NGC 3898, NGC 4111, NGC 5377                   & JbWbEfMcNtO8TrHhYsIrDaKnPiDeJl \\
EC082E & 4.93 & 2022 09 20 & NGC 3414, NGC 4826, NGC 5485                   & JbWbEfMcNtO8T6TrYsHhIrCmDaKnPiDe \\
EC082F & 4.93 & 2022 12 06 & NGC 2841, NGC 3245                             & JbWbEfMcO8T6TrYsHhCmDaKnPiDe \\
\enddata
\tablecomments{(1)observing code; (2) observing frequency; (3) observation date; (4) target sources; (5) participating stations. VLBA stations: BR: Brewster, FD: Fort Davis, HN: Hancock, KP: Kitt Peak, LA: Los Alamos, MK: Mauna Kea, NL: North Liberty, OV: Owens Valley, PT: Pie Town, SC: St. Croix. EVN stations: Jb: Jodrell Bank, Wb: Westerbork, Ef: Effelsberg, Mc: Medicina, Nt: Noto, T6: Tianma, O8: Onsala-85, Tr: Torun, Ys: Yebes, Sv: Svetloe, Zc: Zelenchukskaya, Bd: Badary, Hh: Hartebeesthoek, and Ir: Irbene. e-MERLIN stations: Cm: Cambridge, Jl: JB1-MERL, Kn: Knockin, Pi: Pickmere, De: Defford, Da: Darnhall.}
\end{deluxetable}

\begin{deluxetable}{ccccccccccccc}
\tablecolumns{12}
\tabletypesize{\scriptsize}
\tablewidth{0pt}
\tablecaption{
5 GHz VLBI imaging results of the sample 
\label{tab:image}}
\tablehead{
\colhead{Name}  & \colhead{R.A.} & \colhead{DEC.} & \colhead{$\rm \theta_{\mathrm{FWHM}}$} & \colhead{S$_{\rm peak}$}  &  \colhead{$\rm \sigma$} & \colhead{S$_{\rm core}$}  & \colhead{S$_{\rm tot}$} & $\rm \theta_{core}$ & T$_{\rm b}$  & \colhead{L$_{\rm core}$} & \colhead{L$_{\rm tot}$}  & \colhead{Phase calibrator}  \\
\colhead{} & \colhead {(J2000)} & \colhead {(J2000)} & \colhead {(mas$\times$mas, deg)} & \colhead{($\rm \frac{mJy}{beam}$)}   & \colhead{($\rm \frac{\mu Jy}{beam}$)} & \colhead{(mJy)} & \colhead{(mJy)} & \colhead{(mas)} & \colhead{(log(K))} & \colhead{($\rm \frac{erg}{s}$)}  & \colhead{$\rm \frac{erg}{s}$)}  & \colhead{}  \\
\colhead{(1)}  & \colhead{(2)}     & \colhead{(3)} & \colhead{(4)}      & \colhead{(5)}      & \colhead{(6)}      & \colhead{(7)}     & \colhead{(8)} & \colhead{(9)} & \colhead{(10)} & \colhead{(11)} & \colhead{(12)} & \colhead{(13)}  }
\startdata
\multicolumn{12}{c}{VLBI detected LLAGNs} \\
NGC 410  & 01 10 58.9046 & $+$33 09 07.054 & 4.19$\times$1.84, $-$11.8 & 0.17 & 14.68 & 0.35$\pm$0.03 & 0.56 & 1.79$\pm$0.35 & 6.76$\pm$0.43 & 37.00 & 37.21 & J0112$+$3208 \\
NGC 507  & 01 23 39.9318 & $+$33 15 21.832 & 6.45$\times$4.45, $-$3.2  & 0.73 & 19.35 & 1.22$\pm$0.12 & 1.22 & 3.55$\pm$0.16 & 6.70$\pm$0.24 & 37.48 & 37.48 & J0127$+$3305 \\
NGC 777  & 02 00 14.9074 & $+$31 25 45.869 & 6.60$\times$2.42, $-$6.0  & 0.14 & 11.56 & 0.09$\pm$0.01 & 0.45 & 0.75$\pm$0.13 & 6.92$\pm$0.32 & 36.36 & 37.06 & J0159$+$3106 \\
NGC 1161 & 03 01 14.1729 & $+$44 53 50.665 & 4.41$\times$1.69, $-$7.4  & 3.47 & 27.17 & 3.78$\pm$0.38 & 4.10 & 0.51$\pm$0.03 & 8.88$\pm$0.26 & 37.17 & 37.20 & J0304$+$4553 \\
NGC 2146 & 06 18 37.5855 & $+$78 21 24.263 & 5.14$\times$1.91, $-$27.5 & 0.87 & 17.96 & 1.04$\pm$0.10 & 1.71 & 1.14$\pm$0.12 & 7.62$\pm$0.21 & 36.25 & 36.47 & J0617$+$7816 \\
NGC 2300 & 07 32 19.9257 & $+$85 42 32.454 & 6.18$\times$2.29, $-$32.5 & 0.10 & 12.17 & 0.11$\pm$0.01 & 0.11 & 0.97$\pm$0.11 & 6.78$\pm$0.27 & 35.79 & 35.79 & J0702$+$8549 \\
NGC 2655 & 08 55 37.9708 & $+$78 13 23.689 & 5.41$\times$1.82, $-$7.8  & 0.34 & 20.47 & 0.80$\pm$0.08 & 1.18 & 3.23$\pm$0.22 & 6.60$\pm$0.30 & 36.44 & 36.61 & J0713$+$1935 \\
NGC 2841 & 09 22 02.6782 & $+$50 58 35.732 & 6.31$\times$1.16, 11.0    & 0.56 & 22.85 & 0.97$\pm$0.10 & 0.97 & 1.26$\pm$0.18 & 7.50$\pm$0.32 & 35.91 & 35.91 & J0923$+$5136 \\
NGC 2985 & 09 50 22.1843 & $+$72 16 44.178 & 6.73$\times$2.95, 25.8    & 0.87 & 10.93 & 1.04$\pm$0.12 & 1.46 & 0.84$\pm$0.28 & 7.88$\pm$0.42 & 36.48 & 36.63 & J0956$+$7219 \\
NGC 3245 & 10 27 18.3824 & $+$28 30 26.616 & 1.69$\times$1.17, 7.0     & 0.63 & 27.09 & 0.76$\pm$0.10 & 0.76 & 0.19$\pm$0.03 & 9.04$\pm$0.53 & 36.34 & 36.34 & J1023$+$2856 \\
NGC 3348 & 10 47 10.0327 & $+$72 50 22.665 & 13.84$\times$6.04, $-$10.3& 1.29 & 19.11 & 1.29$\pm$0.13 & 1.62 & 3.50$\pm$0.25 & 6.74$\pm$0.27 & 37.03 & 37.13 & J1047$+$7238 \\
NGC 3414 & 10 51 16.2105 & $+$27 58 30.292 & 1.81$\times$1.25, 14.3    & 0.38 & 38.52 & 1.92$\pm$0.21 & 1.92 & 2.65$\pm$0.21 & 7.15$\pm$0.35 & 36.80 & 36.80 & J1047$+$2635 \\
NGC 3735 & 11 35 57.2166 & $+$70 32 07.833 & 12.37$\times$5.65, $-$29.2& 0.12 & 7.59  & 0.50$\pm$0.13 & 0.50 & 14.29$\pm$1.79& 5.10$\pm$0.64 & 36.69 & 36.69 & J1136$+$7009 \\
NGC 3884 & 11 46 12.1819 & $+$20 23 29.919 & 4.06$\times$1.73, $-$10.2 & 3.28 & 26.54 & 3.96$\pm$0.40 & 4.31 & 1.02$\pm$0.10 & 8.30$\pm$0.45 & 38.28 & 38.32 & J1145$+$1936 \\
NGC 3982 & 11 56 28.1395 & $+$55 07 30.916 & 4.77$\times$1.78, $-$13.6 & 0.26 & 21.09 & 0.44$\pm$0.04 & 0.44 & 2.04$\pm$0.03 & 6.74$\pm$0.32 & 35.87 & 35.87 & J1157$+$5527 \\
NGC 4036 & 12 01 26.7534 & $+$61 53 44.621 & 4.69$\times$1.78, $-$25.8 & 0.19 & 20.57 & 0.41$\pm$0.04 & 0.41 & 2.67$\pm$0.36 & 6.47$\pm$0.40 & 36.16 & 36.16 & J1203$+$6031 \\
NGC 4102 & 12 06 23.0650 & $+$52 42 39.515 & 4.71$\times$1.83, $-$37.9 & 0.11 & 19.25 & 0.20$\pm$0.02 & 0.20 & 2.19$\pm$0.22 & 6.33$\pm$0.20 & 35.53 & 35.53 & J1204$+$5228 \\
NGC 4736 & 12 50 53.0803 & $+$41 07 13.056 & 4.34$\times$1.80, $-$12.1 & 0.14 & 17.82 & 0.42$\pm$0.03 & 0.71 & 3.46$\pm$0.06 & 6.26$\pm$0.26 & 34.66 & 34.88 & J1244$+$4048 \\
NGC 4750 & 12 50 07.3208 & $+$72 52 28.634 & 10.50$\times$2.36, $-$2.4 & 0.30 & 15.20 & 0.42$\pm$0.04 & 0.54 & 4.61$\pm$0.47 & 6.01$\pm$0.34 & 36.22 & 36.44 & J1243$+$7315 \\
NGC 5005 & 13 10 56.2602 & $+$37 03 32.728 & 4.73$\times$1.88, $-$35.3 & 0.13 & 18.82 & 0.27$\pm$0.04 & 0.29 & 2.70$\pm$0.26 & 6.28$\pm$0.18 & 35.85 & 35.85 & J1308$+$3546 \\
NGC 5485 & 14 07 11.3472 & $+$55 00 06.017 & 3.89$\times$1.15, 40.0    & 0.23 & 18.02 & 0.28$\pm$0.04 & 0.28 & 0.87$\pm$0.05 & 7.28$\pm$0.36 & 36.24 & 36.24 & J1408$+$5613 \\ 
NGC 5548 & 14 17 59.5398 & $+$25 08 12.604 & 4.85$\times$1.97, $-$30.6 & 0.74 & 20.92 & 0.80$\pm$0.08 & 1.15 & 0.74$\pm$0.11 & 8.04$\pm$0.34 & 37.47 & 37.32 & J1419$+$2706 \\
NGC 6340 & 17 10 24.8338 & $+$72 18 15.940 & 8.18$\times$4.18, $-$89.0 & 0.28 & 26.60 & 0.36$\pm$0.07 & 0.36 & 2.84$\pm$0.96 & 6.36$\pm$0.62 & 36.01 & 36.01 & J1711$+$7334 \\
\hline 
\multicolumn{13}{c}{VLBI non-detected LLAGNs} \\
NGC 2342 & ...           & ...             & 3.92$\times$1.52, $-$2.7  & ...  & 25.78 & ...         &  $<$ 0.13    & ...   & ...  & ...   & $<$ 36.56 & J0919$+$7825        \\
NGC 2681 & ...           & ...             & 4.58$\times$1.53, $-$6.3  & ...  & 25.92 & ...         &  $<$ 0.13    & ...   & ...  & ...   & $<$ 36.13 & J0849$+$5108        \\
NGC 3729 & ...           & ...             & 4.49$\times$1.53, $-$13.2 & ...  & 22.01 & ...         &  $<$ 0.11    & ...   & ...  & ...   & $<$ 36.27 & J1131$+$5128        \\
NGC 3898 & ...           & ...             & 7.94$\times$5.31,    44.1 & ...  & 26.04 & ...         &  $<$ 0.13    & ...   & ...  & ...   & $<$ 35.56 & J1150$+$5632        \\
NGC 4041 & ...           & ...             & 4.67$\times$1.55,    8.3  & ...  & 19.94 & ...         &  $<$ 0.10    & ...   & ...  & ...   & $<$ 35.48 & J1203$+$6031        \\
NGC 4111 & ...           & ...             & 7.66$\times$5.38,   52.4  & ...  & 25.74 & ...         &  $<$ 0.13    & ...   & ...  & ...   & $<$ 35.34 & J1211$+$4234        \\
NGC 4826 & ...           & ...             & 6.21$\times$1.67,   73.2  & ...  & 35.85 & ...         &  $<$ 0.18    & ...   & ...  & ...   & $<$ 34.44 & J1302$+$2039        \\
NGC 5194 & ...           & ...             & 4.46$\times$1.59, $-$20.5 & ...  & 24.15 & ...         &  $<$ 0.12    & ...   & ...  & ...   & $<$ 34.62 & J1332$+$4722        \\
NGC 5195 & ...           & ...             & 8.21$\times$4.45,   57.4  & ...  & 24.67 & ...         &  $<$ 0.12    & ...   & ...  & ...   & $<$ 34.78 & J1332$+$4722        \\
NGC 5377 & ...           & ...             & 6.19$\times$4.82,   71.1  & ...  & 24.54 & ...         &  $<$ 0.12    & ...   & ...  & ...   & $<$ 35.83 & J1358$+$4737        \\
NGC 5448 & ...           & ...             & 12.10$\times$4.54,  40.6  & ...  & 22.02 & ...         &  $<$ 0.11    & ...   & ...  & ...   & $<$ 35.83 & J1356$+$4831        \\
NGC 6217 & ...           & ...             & 4.93$\times$1.64, $-$0.2  & ...  & 19.82 & ...         &  $<$ 0.10    & ...   & ...  & ...   & $<$ 35.52 & J1610$+$7809        \\
NGC 6946 & ...           & ...             & 4.79$\times$2.13,   29.0  & ...  & 26.72 & ...         &  $<$ 0.13    & ...   & ...  & ...   & $<$ 34.36 & J2030$+$5957        \\
\enddata
\tablecomments{(1) Source name; (2)-(3) Right ascension and Declination coordinates of the peak radio emission; (4) major axis and minor axis of the restoring beam, and the position angle of the major axis, measured from the north to east; (5) peak intensity; (6) root-mean-square noise of the image, measured in off-source regions; (7)  integrated flux density of the core; (8) total integrated radio flux density; (9) component size; (10) core brightness temperature; (11)-(12) logarithm of the radio core and total luminosity; (13) Phase-reference calibrator name.}
\end{deluxetable}

\begin{table}[!ht]
    \centering
    \small
    \caption{Basic properties for 29 archival sources}
	\begin{tabular}{cccccccccccccc}
	\hline \hline
Name & $z$ & Class  & Array & Freq. & $\rm S_{tot}$ & L$\rm _{tot}$      & Mor. & M$\rm _{BH}$  & L$\rm _{[OIII]}$    & L$\rm _{X}$              & M$\rm _{G}$ & Ref. \\
     &     &        &       & (GHz) &      (mJy)    & (erg s$\rm ^{-1}$) &    & (M$_{\odot}$) &  (erg s$\rm ^{-1}$) &  (erg s$\rm ^{-1}$) & (mag)       &      \\
 (1) & (2) & (3)    & (4)   &  (5)  &        (6)    &        (7)         &(8)&      (9)   &       (10)    &       (11)          &         
 (12)&       (13)    \\
\hline
NGC 266  & 0.016 & LINER & VLBA & 5 & 3.8 & 37.93 & core     & 8.37 & 39.43 & 40.68$\pm$0.10 & 18.76 &  1 \\
NGC 315  & 0.016 & LINER & VLBA & 5 & 350 & 39.94 & core-jet & 8.92 & 39.43 & 41.77$\pm$0.04 & 20.00 &  2 \\
NGC 1167 & 0.016 & LINER & VLBA & 5 & 5.9 & 38.16 & core     & 8.27 & 40.17 & 40.12$\pm$0.45 & 19.51 &  1 \\
NGC 1961 & 0.013 & LINER & EVN  & 5 & 0.9 & 37.17 & core     & 8.29 & 39.11 & 40.27$\pm$0.22 & 19.56 &  6 \\
NGC 2273 & 0.006 &Seyfert& VLBA & 5 & 3.5 & 37.21 & core-jet & 7.61 & 40.43 & 40.89$\pm$0.17 & 18.41 &  1 \\
NGC 2639 & 0.011 & LINER & VLBA &8.4& 47  & 38.69 & core-jet & 7.94 & 39.60 & 39.72$\pm$0.32 & 18.71 &  3 \\
NGC 2768 & 0.005 & LINER & VLBA & 5 & 7.3 & 37.38 & core     & 7.96 & 38.61 & 39.73$\pm$0.15 & 17.84 &  1 \\
NGC 2782 & 0.008 &  HII  & EVN  &1.6& 1.2 & 36.99 & twin jets& 7.98 & 40.00 & 40.69$\pm$0.02 & 17.98 &  6 \\
NGC 2787 & 0.002 & LINER & VLBA & 5 & 11.5& 37.05 & core     & 7.61 & 38.37 & 39.20$\pm$0.17 & 17.40 &  1 \\
NGC 3031 & 0.001 & LINER & VLBA & 5 &132.0& 36.18 & core-jet & 7.81 & 37.72 & 39.15$\pm$0.01 & 15.13 &  1 \\
NGC 3079 & 0.004 & LINER & VLBA & 5 & 18.0& 37.64 & core-jet & 6.40 & 37.67 & 40.05$\pm$0.16 & 21.52 &  1 \\
NGC 3147 & 0.009 & LINER & EVN  & 5 & 9.3 & 37.96 & core     & 8.29 & 39.54 & 41.92$\pm$0.09 & 17.72 &  6 \\
NGC 3504 & 0.005 &  HII  & VLBA &1.4& 4.2 & 37.24 & core-jet & 7.23 & 39.88 & ...   & 18.03 &  7 \\
NGC 3665 & 0.007 &  HII  & VLBA & 5 & 8.8 & 37.73 & core     & 8.76 & 38.38 & 39.89$\pm$0.39 & ...   &  4 \\
NGC 3718 & 0.003 & LINER & EVN  & 5 & 7.1 & 37.08 & core     & 7.72 & 37.41 & 41.22$\pm$0.10 & 18.23 &  1 \\
NGC 3998 & 0.004 & LINER & VLBA & 5 & 83.0& 38.35 & core-jet & 8.93 & 39.56 & 41.82$\pm$0.02 & 15.55 &  8 \\
NGC 4051 & 0.002 &Seyfert& EVN  &1.6& 0.5 & 35.53 &twin jets & 6.10 & 40.17 & 41.68$\pm$0.01 & 14.72 &  5 \\
NGC 4143 & 0.003 & LINER & VLBA & 5 & 8.7 & 37.17 & core-jet & 7.92 & 38.81 & 39.99$\pm$0.27 & 17.32 &  1 \\
NGC 4151 & 0.003 &Seyfert& VLBA & 5 & 10.0& 37.38 &twin jets & 7.81 & 41.74 & 42.39$\pm$0.01 & 13.38 &  1 \\
NGC 4203 & 0.004 & LINER & VLBA & 5 & 8.9 & 36.69 & core-jet & 7.82 & 38.28 & 40.33$\pm$0.03 & 16.60 &  1 \\
NGC 4258 & 0.001 &Seyfert& VLBA & 5 & 3.0 & 35.91 & core     & 7.58 & 38.76 & 40.68$\pm$0.01 & 16.73 &  1 \\
NGC 4278 & 0.002 & LINER & VLBA & 5 &65.2 & 37.55 & core-jet & 7.96 & 38.88 & 39.37$\pm$0.04 & ...   &  1 \\
NGC 4589 & 0.007 & LINER & VLBA & 5 &11.5 & 37.78 &twin jets & 8.33 & 38.78 & 39.28$\pm$0.38 & 18.22 &  1 \\
NGC 5033 & 0.003 & LINER & EVN  & 5 & 0.8 & 36.21 & core     & 7.64 & 39.34 & 38.08$\pm$0.41 & 16.63 &  5 \\
NGC 5322 & 0.006 & LINER & VLBA & 5 & 12.0& 37.84 & core     & 8.39 & 38.20 & 39.18$\pm$0.26 & 18.32 &  1 \\
NGC 5353 & 0.008 & LINER & VLBA & 5 &21.6 & 38.25 & core     & 8.76 & 38.73 & 38.56$\pm$0.37 & 19.86 &  1 \\
NGC 5354 & 0.009 & LINER & VLBA & 5 & 8.6 & 37.73 & core-jet & 8.28 & 38.61 & 38.81$\pm$0.36 & 19.24 &  1 \\
NGC 5866 & 0.003 & LINER & VLBA & 5 & 8.4 & 37.06 & core     & 7.84 & 37.50 & 39.26$\pm$0.02 & ...   &  1 \\
NGC 7217 & 0.003 & LINER & EVN  & 5 & 1.2 & 36.25 & core-jet & 7.52 & 38.31 & ...   & 19.02 &  6 \\
\hline
	\end{tabular}
\\[0.1cm]
\label{archival}
Notes-Col: {(1) Source name; (2) redshift; (3) Optical spectroscopic classification based on \citet{1997ApJS..112..315H}; (4) VLBI observation Array; (5) observing frequency; (6) VLBI integrated flux density at 5 GHz. For NGC 2639, NGC 2782, NGC 3504, and NGC 4051, we calculated the total integrated flux density at 5 GHz based on the assumption of spectral index $-$0.5; (7) logarithm of the radio total luminosity; (8) the radio morphology in pc scales; (9) logarithm of the SMBH mass (M$_{\odot}$) determined based on \citet{2002ApJ...574..740T,2016ApJ...831..134V}; (10) logarithm of [OIII] luminosities from \citet{1997ApJS..112..315H}; (11) logarithm of X-ray luminosities in the 2-10 keV from \citet{2022MNRAS.510.4909W}; (12) Absolute G magnitude of the nucleus from \citet{2020yCat.1350....0G}; (13) References for the VLBI data: (1) \citet{2005A&A...435..521N}, (2) \citet{2021A&A...647A..67B}, (3) \citet{2005ApJ...627..674A}, (4) \citet{2009A&A...505..509L}, (5) \citet{2009ApJ...706L.260G}, (6) \citet{2007A&A...464..553K}, (7) \citet{2014AJ....147...14D}, (8) \citet{2007ApJ...658..203H}; }
\end{table}

\end{document}